\documentclass[twocolumn,twoside]{IEEEtran}
\usepackage[utf8]{inputenc}
\usepackage{comment}
\usepackage{xcolor,soul,framed} 
\usepackage{amsmath}
\usepackage{bbding}
\usepackage{pifont}
\colorlet{shadecolor}{yellow}
\usepackage[pdftex]{graphicx}
\graphicspath{{../pdf/}{../jpeg/}}
\DeclareGraphicsExtensions{.pdf,.jpeg,.png}

\usepackage{array}
\usepackage{tabularx}
\usepackage{mdwmath}
\usepackage{mdwtab}
\usepackage{eqparbox}
\usepackage{url}
\usepackage{flushend,cuted}
\usepackage[noend]{algpseudocode}
\usepackage{algorithmicx,algorithm}
\usepackage{subfigure}
\usepackage{caption}
\usepackage{cite}
\usepackage{xcolor}
\usepackage{epstopdf}
\usepackage{cases}
\usepackage{amscd,textcomp,amssymb,epsfig,graphics,epsf,color,amsmath,balance,cite}
\usepackage{multirow}
\usepackage{booktabs}
\usepackage{stfloats}
\usepackage{setspace}
\usepackage{graphicx}
\usepackage{amsmath, amsthm, amsfonts, amssymb, amsbsy}
\usepackage{dsfont}
\usepackage{color}
\usepackage{lipsum}
\usepackage{bm}
\usepackage{setspace}
\usepackage{tabu}                     
\usepackage{multirow}                 
\usepackage{multicol}                 
\usepackage{multirow}                
\usepackage{float}                    
\usepackage{makecell}                 
\usepackage{booktabs} 

\usepackage{amsmath}
\allowdisplaybreaks[4] 

\begin{document}

\title{Channel Estimation for 6G Near-Field Wireless Communications: A Comprehensive Survey}

\author{Wen-Xuan Long, \emph{Member, IEEE}, Shengyu Ye, \emph{Graduate Student Member, IEEE},\\ Marco Moretti, \emph{Member, IEEE}, Michele Morelli, \emph{Senior Member, IEEE}, \\ Luca Sanguinetti, \emph{Fellow, IEEE},  Rui Chen, \emph{Member, IEEE}, and Cheng-Xiang Wang, \emph{Fellow, IEEE}

\thanks{The work of Wen-Xuan Long, Marco Moretti, and Luca Sanguinetti has been performed in the framework of the HORIZON-JUSNS-2022 EU project TIMES under grant no. 101096307, by the European
Union under the Italian National Recovery and Resilience Plan (NRRP) of NextGenerationEU, partnership on ``Telecommunications of the Future'' (PE00000001 – Program ``RESTART'', Cascade Call SMART), and also supported in part by Italian Ministry of Education and Research (MUR) in the Framework of the FoReLab Project (Departments of Excellence) and the Project GARDEN funded by EU in NextGenerationEU Plan through Italian ``Bando Prin 2022-D.D.1409 del 14-09-2022''. The work of Rui Chen and Shengyu Ye was supported in part by the National Natural Science Foundation of China (NSFC) under Grant 62271376, and in part by the Guangdong Natural Science Fund for Distinguished Young Scholar under Grant 2023B1515020079. The work of Cheng-Xiang Wang was supported by the Major Science and Technology Project of Jiangsu Province under Grant BG2025039 and the Research Fund of National Mobile Communications Research Laboratory, Southeast University, under Grant 2026A05.}

\thanks{W.-X. Long and M. Morelli are with the Dipartimento di Ingegneria dell’Informazione, University of Pisa, 56126 Pisa, Italy (e-mail: wenxuan.long@ing.unipi.it, michele.morelli@unipi.it).}
\thanks{S. Ye and R. Chen are with the State Key Laboratory of Integrated Service Networks, Xidian University, Xi’an 710071, Shaanxi, China (e-mail: 24011211130@stu.xidian.edu.cn, rchen@xidian.edu.cn).}
\thanks{M. Moretti and L. Sanguinetti are with the Dipartimento di Ingegneria dell’Informazione, University of Pisa, 56126 Pisa, Italy, and also with National Inter-University Consortium for Telecommunications (CNIT), 43124 Parma, Italy (e-mail: marco.moretti@unipi.it, luca.sanguinetti@unipi.it).}
\thanks{C.-X. Wang is with the National Mobile Communications Research Laboratory, School of Information Science and Engineering, Southeast University, Nanjing, 211189, China, and also with the Purple Mountain Laboratories, Nanjing 211111, China. (e-mail: chxwang@seu.edu.cn).}
}

\maketitle

\begin{abstract}
Sixth-generation (6G) wireless communication systems are expected to embrace extremely large aperture arrays (ELAAs), novel antenna architectures, and operation in high-frequency bands to meet the rapidly growing demand for data transmission. By increasing the number of antenna elements, ELAAs enable finer spatial resolution and enhanced beamforming capabilities. At these high operating frequencies, an ELAA aperture may span tens or even hundreds of wavelengths, causing the propagation conditions to gradually depart from the conventional far-field assumption, and making spherical-wave effects increasingly prominent. More generally, near-field behavior is jointly determined by the array size relative to the wavelength and the link distance, and it may also arise in short-range deployments. Consequently, near-field propagation is expected to play a key role in 6G systems. In the near-field region, the electromagnetic field impinges on the array with a non-negligible wavefront curvature. Consequently, the channel is parameterized by both angular information and the propagation distance (range) between transmitter and receiver. This additional distance-dependent degree of freedom increases the dimensionality of the channel parameters and alters the structural properties exploited by far-field channel estimators. As a result, straightforward extensions of conventional far-field channel estimation techniques, typically designed to exploit only angular information, to near-field scenarios may lead to significant high computational complexity. These challenges motivate the development of estimation methods tailored to the distinctive characteristics of near-field propagation.

This paper provides a comprehensive overview of recent advances in near-field channel estimation. From an electromagnetic-wave perspective, we first delineate the boundary between near- and far-field regions and highlight the fundamental differences in their propagation mechanisms. We then summarize representative ELAA architectures, spherical-wavefront control techniques, emerging near-field applications, and ongoing standardization efforts related to near-field communications. Next, we introduce widely used near-field channel models and contrast them with their far-field counterparts. Finally, we systematically review major estimation techniques under various system configurations, including single- and multi-user, as well as single- and multi-carrier settings, covering both direct channel estimation between the base station and user equipments, and cascaded channel estimation assisted by a reconfigurable intelligent surface. The surveyed techniques reflect different trade-offs among estimation accuracy, complexity, and robustness. Overall, this survey aims to provide technical insights and theoretical foundations for  efficient and scalable near-field channel estimation in 6G systems, while also highlighting key challenges and promising future research directions.
\end{abstract}


\section{Introduction}

\begin{figure}[t!]
\setlength{\abovecaptionskip}{-0.00cm}
\setlength{\belowcaptionskip}{-0.40cm}
\centering
\includegraphics[scale=1.22]{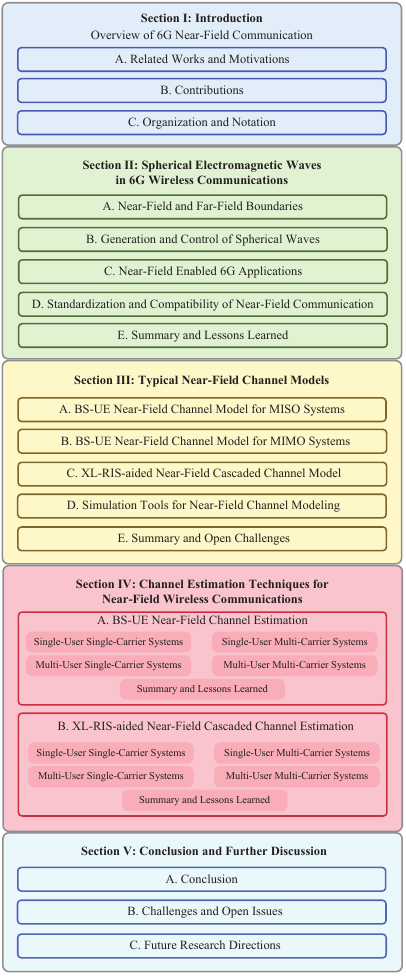}
\caption{The structure of this survey.}
\label{Fig.Org}
\end{figure}

With the global commercialization and deployment of fifth-generation (5G) wireless communication systems, both academia and industry are actively paving the way for the development of sixth-generation (6G) networks \cite{Chen2025Channel}. The aim is to meet the requirements of emerging applications such as  intelligent industrial systems, metaverse, and fully autonomous vehicles \cite{Liu2023Near, lu2024tutorial, An2024Near}. In June 2023, the International Telecommunication Union (ITU) released its official vision for 6G networks \cite{ITU2023IMT}, outlining a recommended system architecture, key enabling technologies, and a comprehensive development roadmap. This publication marked the formal initiation of 6G standardization efforts. Notably, the ITU also identifies six representative usage scenarios for 6G systems. These include enhancements to existing 5G paradigms such as enhanced mobile broadband (eMBB+), massive machine-type communications (mMTC+), and ultra-reliable low-latency communications (URLLC+). They also encompass three transformative new use cases: integrated sensing and communication (ISAC), integrated artificial intelligence (AI) and communication, and ubiquitous connectivity. To support these advanced use cases, 6G systems are expected to achieve peak data rates of at least 1 Tb/s, air interface latency as low as 0.01 - 0.1 ms, and device densities up to  $10^7$ devices/$\text{km}^2$ \cite{Long2021A, Wang2023On}. These ambitious performance targets exceed the capabilities of current 5G technologies, necessitating the development of novel and potentially disruptive communication paradigms.

In 6G wireless communications, extremely large aperture arrays (ELAAs) are regarded as one of the key enabling technologies \cite{bjornson2019massive, Ye2024Extremely}.
As a natural evolution of massive multiple-input multiple-output (MIMO) in 5G systems, ELAAs typically include extremely large MIMO (XL-MIMO) \cite{bjornson2019massive,Wang20206G} and extremely large reconfigurable intelligent surface (XL-RIS) \cite{lee2024near}, which typically integrate hundreds or even thousands of antenna elements. The enhanced spatial multiplexing and beamforming capabilities of these solutions lead to substantial improvements in spectral efficiency (SE) and spatial resolution.
With higher operating frequencies \cite{Han2022Terahertz, Jiang2024Terahertz, Thomas2025A}, the aperture of an ELAA can span tens or even hundreds of wavelengths, thereby substantially expanding the near-field region \cite{lu2024tutorial}. For instance, an array operating at 24 GHz with a physical aperture of 0.5 m $\times$ 0.5 m yields a Rayleigh distance of 80 m, which covers a substantial portion of a typical cell. Meanwhile, near-field propagation may also arise at lower frequency bands when the communication range is sufficiently short. As a result, near-field wireless communications are anticipated to become an operating mode in future networks \cite{Liu2023Near,Cui2023Near2, liu2024near,An2024Near,Liu2025Near}.

\begin{table*}[t]
\centering
\renewcommand{\arraystretch}{1.0} 
\caption{Comparison of this work with existing magazine articles, surveys, and tutorials.}
\resizebox{1.0\linewidth}{!}{
\begin{tabular}{|cccccccccccccccccccc|}
\hline
\multicolumn{3}{|c|}{\multirow{2}{*}{Existing Work}}                                                                                                                                                                                                                                          & \multicolumn{8}{c|}{Magazine Articles}                                                                                                                                                                                                                                                                                                                                                                                                                                                                                  & \multicolumn{3}{c|}{Surveys}                                                                                                                                                                & \multicolumn{5}{c|}{Tutorials}                                                                                                                                                                                                                                                                                     & \multirow{2}{*}{This Work} \\ \cline{4-19}
\multicolumn{3}{|c|}{}                                                                                                                                                                                                                                                                        & \multicolumn{1}{c|}{\cite{Zhang2022Near}} & \multicolumn{1}{c|}{\cite{Cui2023Near2}} & \multicolumn{1}{c|}{\cite{Zhang20236G}} & \multicolumn{1}{c|}{\cite{An2024Near}} & \multicolumn{1}{c|}{\cite{Mu2024Reconfigurable}} & \multicolumn{1}{c|}{\cite{Wang2024Rethinking}} & \multicolumn{1}{c|}{\cite{Liu2025Near}} & \multicolumn{1}{c|}{\cite{Lei2025}} & \multicolumn{1}{c|}{\cite{Gong2024Holographic}} & \multicolumn{1}{c|}{\cite{lU2024Integrated}} & \multicolumn{1}{c|}{\cite{liu2024near}} & \multicolumn{1}{c|}{\cite{Liu2023Near}} & \multicolumn{1}{c|}{\cite{han2023toward}} & \multicolumn{1}{c|}{\cite{lu2024tutorial}} & \multicolumn{1}{c|}{\cite{wang2024tutorial}} & \multicolumn{1}{c|}{\cite{Emil2024Towards}} &                            \\ \hline
\multicolumn{3}{|c|}{Year}                                                                                                                                                                                                                                                                    & \multicolumn{1}{c|}{2022}                                  & \multicolumn{1}{c|}{2023}                                & \multicolumn{1}{c|}{2023}                                & \multicolumn{1}{c|}{2024}                                 & \multicolumn{1}{c|}{2024}                                         & \multicolumn{1}{c|}{2024}                                       & \multicolumn{1}{c|}{2025}                                             & \multicolumn{1}{c|}{2025}                                  & \multicolumn{1}{c|}{2024}                                        & \multicolumn{1}{c|}{2024}                                     & \multicolumn{1}{c|}{2025}                                & \multicolumn{1}{c|}{2023}                                & \multicolumn{1}{c|}{2023}                                  & \multicolumn{1}{c|}{2024}                                   & \multicolumn{1}{c|}{2024}                                     & \multicolumn{1}{c|}{2024}                                    & 2025                       \\ \hline
\multicolumn{3}{|c|}{Core concepts}                                                                                                                                                                                                                                                           & \multicolumn{1}{c|}{$\triangle$}                           & \multicolumn{1}{c|}{$*$}                                  & \multicolumn{1}{c|}{$\Box$}                              & \multicolumn{1}{c|}{$\Box$}                             & \multicolumn{1}{c|}{$\Box$}                                       & \multicolumn{1}{c|}{}                                           & \multicolumn{1}{c|}{$\Box$}                              & \multicolumn{1}{c|}{$\Box$}                                          & \multicolumn{1}{c|}{}                                            & \multicolumn{1}{c|}{}                                         & \multicolumn{1}{c|}{$*$}                                 & \multicolumn{1}{c|}{$*$}                                 & \multicolumn{1}{c|}{$*$}                                   & \multicolumn{1}{c|}{$\Box$}                                 & \multicolumn{1}{c|}{$*$}                                      & \multicolumn{1}{c|}{$\Box$}                                  & $*$                        \\ \hline
\multicolumn{1}{|c|}{\multirow{2}{*}{\begin{tabular}[c]{@{}c@{}}Spherical-wave \\ generation and control\end{tabular}}} & \multicolumn{2}{c|}{Novel antenna types}                                                                                                                            & \multicolumn{1}{c|}{$\Box$}                                & \multicolumn{1}{c|}{}                                     & \multicolumn{1}{c|}{}                                    & \multicolumn{1}{c|}{$\triangle$}                        & \multicolumn{1}{c|}{}                                             & \multicolumn{1}{c|}{}                                           & \multicolumn{1}{c|}{}                                    & \multicolumn{1}{c|}{}                                                & \multicolumn{1}{c|}{$*$}                                         & \multicolumn{1}{c|}{}                                         & \multicolumn{1}{c|}{$\Box$}                              & \multicolumn{1}{c|}{$\Box$}                              & \multicolumn{1}{c|}{$*$}                                   & \multicolumn{1}{c|}{$\Box$}                                 & \multicolumn{1}{c|}{$*$}                                      & \multicolumn{1}{c|}{$\Box$}                                  & $\Box$                     \\ \cline{2-20}
\multicolumn{1}{|c|}{}                                                                                                  & \multicolumn{2}{c|}{Beamforming design}                                                                                                                             & \multicolumn{1}{c|}{$\Box$}                                & \multicolumn{1}{c|}{$\triangle$}                          & \multicolumn{1}{c|}{$*$}                                 & \multicolumn{1}{c|}{$\Box$}                             & \multicolumn{1}{c|}{$\Box$}                                       & \multicolumn{1}{c|}{$\Box$}                                     & \multicolumn{1}{c|}{$\Box$}                              & \multicolumn{1}{c|}{}                                                & \multicolumn{1}{c|}{$\Box$}                                      & \multicolumn{1}{c|}{}                                         & \multicolumn{1}{c|}{$\Box$}                              & \multicolumn{1}{c|}{$*$}                                 & \multicolumn{1}{c|}{$\Box$}                                & \multicolumn{1}{c|}{$*$}                                    & \multicolumn{1}{c|}{$*$}                                      & \multicolumn{1}{c|}{}                                        & $\Box$                     \\ \hline
\multicolumn{1}{|c|}{\multirow{4}{*}{Channel modeling}}                                                                 & \multicolumn{1}{c|}{\multirow{2}{*}{BS-UE channels}}                                                                   & \multicolumn{1}{c|}{LoS}                   & \multicolumn{1}{c|}{}                                      & \multicolumn{1}{c|}{}                                     & \multicolumn{1}{c|}{$\triangle$}                         & \multicolumn{1}{c|}{$\triangle$}                        & \multicolumn{1}{c|}{}                                             & \multicolumn{1}{c|}{$\triangle$}                                & \multicolumn{1}{c|}{}                                    & \multicolumn{1}{c|}{}                                                & \multicolumn{1}{c|}{$*$}                                         & \multicolumn{1}{c|}{}                                         & \multicolumn{1}{c|}{$*$}                                 & \multicolumn{1}{c|}{$*$}                                 & \multicolumn{1}{c|}{$*$}                                   & \multicolumn{1}{c|}{$*$}                                    & \multicolumn{1}{c|}{$*$}                                      & \multicolumn{1}{c|}{$\Box$}                                  & $*$                        \\ \cline{3-20}
\multicolumn{1}{|c|}{}                                                                                                  & \multicolumn{1}{c|}{}                                                                                                  & \multicolumn{1}{c|}{NLoS}                  & \multicolumn{1}{c|}{}                                      & \multicolumn{1}{c|}{}                                     & \multicolumn{1}{c|}{$\triangle$}                         & \multicolumn{1}{c|}{$\triangle$}                        & \multicolumn{1}{c|}{}                                             & \multicolumn{1}{c|}{}                                           & \multicolumn{1}{c|}{$\triangle$}                         & \multicolumn{1}{c|}{}                                                & \multicolumn{1}{c|}{$*$}                                         & \multicolumn{1}{c|}{}                                         & \multicolumn{1}{c|}{$*$}                                 & \multicolumn{1}{c|}{$*$}                                 & \multicolumn{1}{c|}{$*$}                                   & \multicolumn{1}{c|}{$*$}                                    & \multicolumn{1}{c|}{$*$}                                      & \multicolumn{1}{c|}{$\Box$}                                  & $*$                        \\ \cline{2-20}
\multicolumn{1}{|c|}{}                                                                                                  & \multicolumn{1}{c|}{\multirow{2}{*}{\begin{tabular}[c]{@{}c@{}}RIS-aided\\ cascaded channel\end{tabular}}}             & \multicolumn{1}{c|}{LoS}                   & \multicolumn{1}{c|}{}                                      & \multicolumn{1}{c|}{}                                     & \multicolumn{1}{c|}{}                                    & \multicolumn{1}{c|}{}                                   & \multicolumn{1}{c|}{$\Box$}                                       & \multicolumn{1}{c|}{}                                           & \multicolumn{1}{c|}{$\triangle$}                         & \multicolumn{1}{c|}{}                                                & \multicolumn{1}{c|}{}                                            & \multicolumn{1}{c|}{}                                         & \multicolumn{1}{c|}{}                                    & \multicolumn{1}{c|}{}                                    & \multicolumn{1}{c|}{$\triangle$}                           & \multicolumn{1}{c|}{}                                       & \multicolumn{1}{c|}{}                                         & \multicolumn{1}{c|}{}                                        & $*$                        \\ \cline{3-20}
\multicolumn{1}{|c|}{}                                                                                                  & \multicolumn{1}{c|}{}                                                                                                  & \multicolumn{1}{c|}{NLoS}                  & \multicolumn{1}{c|}{}                                      & \multicolumn{1}{c|}{}                                     & \multicolumn{1}{c|}{}                                    & \multicolumn{1}{c|}{}                                   & \multicolumn{1}{c|}{}                                             & \multicolumn{1}{c|}{}                                           & \multicolumn{1}{c|}{}                                    & \multicolumn{1}{c|}{}                                                & \multicolumn{1}{c|}{}                                            & \multicolumn{1}{c|}{}                                         & \multicolumn{1}{c|}{}                                    & \multicolumn{1}{c|}{}                                    & \multicolumn{1}{c|}{$\triangle$}                           & \multicolumn{1}{c|}{}                                       & \multicolumn{1}{c|}{}                                         & \multicolumn{1}{c|}{}                                        & $*$                        \\ \hline
\multicolumn{1}{|c|}{\multirow{4}{*}{Channel estimation}}                                                               & \multicolumn{1}{c|}{\multirow{2}{*}{BS-UE channel estimation}}                                                         & \multicolumn{1}{c|}{Single-carrier system} & \multicolumn{1}{c|}{}                                      & \multicolumn{1}{c|}{$\triangle$}                          & \multicolumn{1}{c|}{$\triangle$}                         & \multicolumn{1}{c|}{}                                   & \multicolumn{1}{c|}{}                                             & \multicolumn{1}{c|}{}                                           & \multicolumn{1}{c|}{}                                    & \multicolumn{1}{c|}{$\Box$}                                          & \multicolumn{1}{c|}{$\triangle$}                                 & \multicolumn{1}{c|}{}                                         & \multicolumn{1}{c|}{$\Box$}                              & \multicolumn{1}{c|}{}                                    & \multicolumn{1}{c|}{$\Box$}                                & \multicolumn{1}{c|}{$\Box$}                                 & \multicolumn{1}{c|}{$\Box$}                                   & \multicolumn{1}{c|}{$\Box$}                                  & $*$                        \\ \cline{3-20}
\multicolumn{1}{|c|}{}                                                                                                  & \multicolumn{1}{c|}{}                                                                                                  & \multicolumn{1}{c|}{Wideband system}       & \multicolumn{1}{c|}{}                                      & \multicolumn{1}{c|}{}                                     & \multicolumn{1}{c|}{}                                    & \multicolumn{1}{c|}{$\triangle$}                        & \multicolumn{1}{c|}{}                                             & \multicolumn{1}{c|}{}                                           & \multicolumn{1}{c|}{}                                    & \multicolumn{1}{c|}{$\Box$}                                          & \multicolumn{1}{c|}{$\triangle$}                                 & \multicolumn{1}{c|}{}                                         & \multicolumn{1}{c|}{$\triangle$}                         & \multicolumn{1}{c|}{}                                    & \multicolumn{1}{c|}{}                                      & \multicolumn{1}{c|}{$\triangle$}                            & \multicolumn{1}{c|}{$\triangle$}                              & \multicolumn{1}{c|}{}                                        & $*$                        \\ \cline{2-20}
\multicolumn{1}{|c|}{}                                                                                                  & \multicolumn{1}{c|}{\multirow{2}{*}{\begin{tabular}[c]{@{}c@{}}RIS-aided cascaded \\ channel estimation\end{tabular}}} & \multicolumn{1}{c|}{Single-carrier system} & \multicolumn{1}{c|}{}                                      & \multicolumn{1}{c|}{}                                     & \multicolumn{1}{c|}{}                                    & \multicolumn{1}{c|}{}                                   & \multicolumn{1}{c|}{}                                             & \multicolumn{1}{c|}{}                                           & \multicolumn{1}{c|}{}                                    & \multicolumn{1}{c|}{}                                                & \multicolumn{1}{c|}{$\triangle$}                                 & \multicolumn{1}{c|}{}                                         & \multicolumn{1}{c|}{}                                    & \multicolumn{1}{c|}{}                                    & \multicolumn{1}{c|}{$\triangle$}                           & \multicolumn{1}{c|}{}                                       & \multicolumn{1}{c|}{}                                         & \multicolumn{1}{c|}{}                                        & $*$                        \\ \cline{3-20}
\multicolumn{1}{|c|}{}                                                                                                  & \multicolumn{1}{c|}{}                                                                                                  & \multicolumn{1}{c|}{Wideband system}       & \multicolumn{1}{c|}{}                                      & \multicolumn{1}{c|}{}                                     & \multicolumn{1}{c|}{}                                    & \multicolumn{1}{c|}{$\triangle$}                        & \multicolumn{1}{c|}{}                                             & \multicolumn{1}{c|}{}                                           & \multicolumn{1}{c|}{}                                    & \multicolumn{1}{c|}{}                                                & \multicolumn{1}{c|}{$\triangle$}                                 & \multicolumn{1}{c|}{}                                         & \multicolumn{1}{c|}{}                                    & \multicolumn{1}{c|}{}                                    & \multicolumn{1}{c|}{}                                      & \multicolumn{1}{c|}{}                                       & \multicolumn{1}{c|}{}                                         & \multicolumn{1}{c|}{}                                        & $*$                        \\ \hline
\multicolumn{3}{|c|}{Applications}                                                                                                                                                                                                                                                            & \multicolumn{1}{c|}{}                                      & \multicolumn{1}{c|}{}                                     & \multicolumn{1}{c|}{}                                    & \multicolumn{1}{c|}{$\Box$}                             & \multicolumn{1}{c|}{}                                             & \multicolumn{1}{c|}{}                                           & \multicolumn{1}{c|}{$\Box$}                              & \multicolumn{1}{c|}{}                                                & \multicolumn{1}{c|}{$\Box$}                                      & \multicolumn{1}{c|}{$\Box$}                                   & \multicolumn{1}{c|}{$\Box$}                              & \multicolumn{1}{c|}{}                                    & \multicolumn{1}{c|}{}                                      & \multicolumn{1}{c|}{}                                       & \multicolumn{1}{c|}{$*$}                                      & \multicolumn{1}{c|}{}                                        & $\Box$                     \\ \hline
\multicolumn{3}{|c|}{Standardization and Compatibility}                                                                                                                                                                                                                                                            & \multicolumn{1}{c|}{}                                      & \multicolumn{1}{c|}{}                                     & \multicolumn{1}{c|}{}                                    & \multicolumn{1}{c|}{}                             & \multicolumn{1}{c|}{}                                             & \multicolumn{1}{c|}{}                                           & \multicolumn{1}{c|}{}                              & \multicolumn{1}{c|}{}                                                & \multicolumn{1}{c|}{}                                      & \multicolumn{1}{c|}{}                                   & \multicolumn{1}{c|}{}                              & \multicolumn{1}{c|}{}                                    & \multicolumn{1}{c|}{}                                      & \multicolumn{1}{c|}{}                                       & \multicolumn{1}{c|}{}                                      & \multicolumn{1}{c|}{}                                        & $\Box$                     \\ \hline
\multicolumn{20}{|l|}{Here, $*$, $\Box$, and $\triangle$ denote ``discussed in detail'', ``partially discussed'', and ``only mentioned'', respectively. The existing literature is listed in chronological order.}                                                                                                                                                             \\ \hline
\end{tabular}
}
\label{TableVIII}
\end{table*}

Due to the inherent spherical wavefront of near-field propagation, the channel response depends on the three-dimensional (3-D) position of the receiver rather than solely on the direction of arrival (DoA), as in far-field models. This dual-domain capability enables fine-grained beam focusing and interference suppression, thereby enhancing spatial resolution, particularly for high-density access and ISAC scenarios requiring precise spatial discrimination.
However, under the conventional planar-wave approximation, angular-only channel representations become inadequate, necessitating refined models that jointly account for both angular and distance characteristics \cite{Zheng2023Measurements}.  Existing near-field channel modeling approaches can be classified into three categories. The first one comprises geometry-based spherical wavefront models \cite{zhang2022beam}, which parameterize the channel in terms of array geometry and transceiver positions, capturing wavefront curvature while often neglecting other physical effects. In line-of-sight (LoS) scenarios, the channel response is typically an explicit function of the transmitter-receiver distance. The second category includes multipath spherical wavefront models \cite{cui2022channel}, tailored to environments with a finite number of propagation paths, where the channel is expressed as a superposition of spherical-wave components, each corresponding to an individual path characterized by different propagation distances and DoAs. The third category consists of spatial correlation models \cite{dong2022near,long2025parametric}, which describe non-line-of-sight (NLoS) channels through second-order statistics, usually in the form of spatial correlation matrices derived under spherical-wave assumptions.
Together, these approaches offer complementary views of near-field propagation and serve as the foundation for key tasks such as channel estimation and beamforming.

Furthermore, near-field channels associated with ELAAs typically exhibit increased dimensionality, which leads to higher computational burden. The conventional far-field estimators such as the linear minimum mean-square error (LMMSE) can become prohibitively expensive, as it typically require large-scale matrix operations. This highlights the need for high accuracy and low complexity algorithms that leverage the unique structure of spherical wavefronts in near-field propagation. When the near-field channel can be accurately characterized by a limited set of physical parameters, such as the relative distance and DoA between the transmitter and receiver in a LoS scenario, the channel can  be reconstructed by estimating these parameters. This leads to the development of \textit{parametric channel estimation} techniques. Moreover, due to the inherent sparsity of near-field channels in the polar domain, \textit{sparsity-aware} methods have also been widely adopted to jointly recover angular and distance information, enabling accurate reconstruction of the complete near-field channel. Finally, \textit{deep learning} approaches are emerging as a promising alternative, providing an effective means to achieve low-complexity and high-efficiency solutions.

The purpose of this survey is to provide a comprehensive review and analysis of channel estimation techniques tailored for near-field communication scenarios. We cover both direct base-station (BS)-to-user equipment (UE) estimation and reconfigurable intelligent surface (RIS)-assisted cascaded estimation across a range of deployment configurations. The goal is to synthesize recent advances in near-field channel inference and enable accurate channel state information (CSI) acquisition, thereby supporting reliable and high-performance near-field communications.

\begin{table}[t]
\centering
\renewcommand{\arraystretch}{1.0} 
\caption{List of acronyms.}
\resizebox{1.0\linewidth}{!}{
\begin{tabular}{|c|c|}
\hline
1-D          & One-dimensional                                       \\ \hline
2-D          & Two-dimensional                                       \\ \hline
3-D          & Three-dimensional                                     \\ \hline
5G           & Fifth-generation                                      \\ \hline
6G           & Sixth-generation                                      \\ \hline
ADC          & Analog-to-digital converter                             \\ \hline
AI           & Artificial intelligence                               \\ \hline
AGSBL        & Adaptive group sparse Bayesian learning               \\ \hline
B5G          & Beyond fifth-generation                               \\ \hline
BER          & Bit error rate                                        \\ \hline
BPD          & Bilinear pattern detection                            \\ \hline
BS           & Base station                                          \\ \hline
CAP          & Continuous aperture                                   \\ \hline
CLRA         & Collaborative low-rank approximation                  \\ \hline
CNN          & Convolutional neural network                          \\ \hline
CRLB         & Cram\'er-Rao lower bound                              \\ \hline
CSI          & Channel state information                             \\ \hline
DMA          & Dynamic metasurface antenna                           \\ \hline
DNN          & Deep neural network                                   \\ \hline
DoA          & Direction of arrival                                  \\ \hline
ELAA         & Extremely large aperture array                        \\ \hline
GBSM         & Geometry-based stochastic model                                             \\ \hline
GHz          & Gigahertz                                             \\ \hline
HMIMO        & Holographic multiple-input multiple-output             \\ \hline
ITU          & International Telecommunication Union                 \\ \hline
ISAC         & Integrated sensing and communication                  \\ \hline
LIS          & Large intelligent surface                             \\ \hline
LMMSE        & Linear minimum mean square error                      \\ \hline
LoS          & Line-of-sight                                         \\ \hline
LS           & Least squares                                         \\ \hline
MIMO         & Multiple-input multiple-output                         \\ \hline
MISO         & Multiple-input single-output                           \\ \hline
mmWave       & Millimeter-wave                                       \\ \hline
MMSE         & Minimum mean square error                             \\ \hline
MUSIC        & Multiple signal classification                        \\ \hline
NMSE         & Normalized mean square error                          \\ \hline
NLoS         & Non-line-of-sight                                     \\ \hline
OFDM         & Orthogonal frequency division multiplexing            \\ \hline
OMP          & Orthogonal matching pursuit                           \\ \hline
PIN          & Positive intrinsic-negative                           \\ \hline
P-OMP        & Polar-domain orthogonal matching pursuit              \\ \hline
P-SIGW       & Polar-domain simultaneous iterative gridless weighted \\ \hline
P-SOMP       & Polar-domain sparse orthogonal matching pursuit       \\ \hline
RF           & Radio-frequency                                       \\ \hline
RIS          & Reconfigurable intelligent surface                    \\ \hline
SE           & Spectral efficiency                                   \\ \hline
SNR          & Signal-to-noise ratio                                 \\ \hline
SOMP         & Sparse orthogonal matching pursuit                    \\ \hline
SPD          & Spatially discrete                                    \\ \hline
THz          & Terahertz                                             \\ \hline
Turbo-CoSaMP & Turbo-type compressive sampling matching pursuit      \\ \hline
UCA          & Uniform circular array                                \\ \hline
UE           & User equipment                                        \\ \hline
ULA          & Uniform linear array                                  \\ \hline
UPA          & Uniform planar array                                  \\ \hline
VBL          & Variational Bayesian learning                         \\ \hline
VR           & Visibility region                                     \\ \hline
XL-MIMO      & Extremely large multiple-input multiple-output         \\ \hline
XL-RIS       & Extremely large reconfigurable intelligent surface    \\ \hline
\end{tabular}
}
\label{TableIX}
\end{table}

\subsection{Related Works and Motivations}

To date, several surveys \cite{liu2024near},\cite{Gong2024Holographic,lU2024Integrated}, tutorials \cite{Liu2023Near,lu2024tutorial},\cite{han2023toward,wang2024tutorial,Emil2024Towards} and magazine articles \cite{Cui2023Near2,Liu2025Near,An2024Near,Lei2025},\cite{Zhang2022Near,Zhang20236G,Mu2024Reconfigurable,Wang2024Rethinking} have introduced the concepts, operating principles, and potential applications of near-field communications, offering complementary perspectives with varying depth and scope.

\subsubsection{Magazine Articles}

Among the earliest efforts, \cite{Cui2023Near2} introduced near-field communications by describing the distinction between near- and far-field regions and discussing key concepts, technical challenges, and envisioned applications. It also briefly outlined the idea of near-field channel estimation based on suitably designed codebooks. Following this, \cite{Liu2025Near} offered a focused exploration of the fundamental differences between spherical wavefront-based near-field communications and conventional planar wavefront-based far-field communications, particularly in the context of channel modeling, beamforming, system performance, and integration with next-generation technologies. Complementing the above efforts, \cite{An2024Near} discussed several core technical challenges in near-field communications, spanning channel modeling, estimation, beamforming, hardware design, and sensing, and briefly highlighted recent research efforts in each area. Of particular note, it summarized the limitations faced by codebook-based compressive sensing algorithms in near-field channel estimation and reviewed three representative solutions to address these challenges. In addition, several recent magazine articles have focused on specific research directions in near-field wireless communications. For example, \cite{Zhang2022Near} provided a concise overview of the opportunities and challenges associated with radiative near-field wireless power transfer (WPT). In another line of research, \cite{Zhang20236G} offered an in-depth discussion on the potential advantages of spherical-wave-based beam focusing in 6G networks, including multi-user interference suppression, high-precision localization, and focused sensing. Furthermore, \cite{Mu2024Reconfigurable} presented a comprehensive analysis of the performance gains enabled by XL-RISs in near-field communications, and briefly discussed related challenges in beam training and beamforming design. Meanwhile, \cite{Wang2024Rethinking} provided a system-level overview of wideband near-field ISAC systems, covering key issues such as spectrum allocation, antenna array deployment, transceiver architecture, and waveform design. Lastly, \cite{Lei2025} offered an introductory perspective on the challenges posed by spherical wave propagation in the context of user localization and channel estimation, and outlined a few representative algorithmic strategies designed to mitigate these issues.

\subsubsection{Surveys}

Among existing review papers on near-field communications, \cite{Gong2024Holographic} considers holographic MIMO (HMIMO) systems enabled by continuous-aperture (CAP) arrays as a case study and offers a systematic overview of their integration with near-field wireless communications. Besides, \cite{lU2024Integrated} focuses on the latest advances and open challenges in near-field ISAC, whereas \cite{liu2024near} offers a comprehensive overview of near-field wireless communications, covering fundamental concepts, basic operating principles, channel modeling, performance analysis, signal processing techniques, and emerging applications. Notably, \cite{liu2024near} also summarizes more than 20 representative works on near-field channel estimation published prior to 2023.

\subsubsection{Tutorials}

Building upon the above surveys, several tutorial articles have provided systematic and in-depth treatment of the concepts, methodologies, and key technologies of near-field wireless communications. Among them, \cite{Liu2023Near} presented a comprehensive tutorial that systematically reviewed near-field channel modeling, antenna architectures, beamforming techniques, and system performance analysis. Furthermore, from the perspective of extremely large-scale MIMO (XL-MIMO), the tutorial \cite{lu2024tutorial} offered a holistic overview of near-field modeling, system performance, and practical design considerations, providing valuable guidance for researchers working on near-field propagation, particularly in the context of array design. Subsequently, the tutorial \cite{Liu2023Near} was further extended in \cite{han2023toward, Emil2024Towards, wang2024tutorial}, with additional emphasis on the physical-layer characteristics and electromagnetic properties of spherical wave propagation \cite{han2023toward}, stochastic channel modeling \cite{wang2024tutorial}, and low-complexity signal processing algorithms \cite{Emil2024Towards}.

Despite all  these valuable contributions, the existing literature lacks a dedicated, systematic, and in-depth survey focusing specifically on near-field channel estimation. To bridge this gap, this paper provides a comprehensive and structured review of the recent advances in near-field channel estimation, aiming to establish a clear knowledge framework and highlight promising directions for future research. In addition, Table~\ref{TableVIII} compares this survey with the existing literature in terms of scope and research focus.

\subsection{Contributions}

The main contributions of this survey are summarized as:

\textbullet\ We first clarify the boundary criteria between the far-field, radiative near-field, and reactive near-field regions for different array configurations, based on Rayleigh- and Fresnel-distance definitions. This provides the theoretical foundation for the subsequent analysis of near-field communication characteristics. We then review existing ELAA architectures capable of generating extended near-field regions, including CAP arrays and spatially discrete (SPD) arrays, along with their corresponding beamforming and wavefront-control techniques.

\textbullet\ We provide an overview of representative near-field applications and briefly discuss standardization progress and compatibility issues for near-field communications in 6G, thereby clarifying the role of near-field technologies and their evolution within the 6G ecosystem.

\textbullet\ We discuss representative near-field channel modeling frameworks, including array-response formulations for different geometries, LoS and multipath representations, and spatial-correlation descriptions that capture near-field statistics. We then contrast them with far-field counterparts to highlight key differences and introduce simulation platforms that provide unified support for near-field modeling and algorithm design.

\textbullet\ As the core part of this survey, we provide a systematic summary of recent advances in near-field channel estimation under various system configurations, including single-user and multi-user systems, as well as single-carrier and multi-carrier transmission scenarios. The review covers not only the direct channel estimations between the BS and UEs, but also the cascaded channel estimation techniques in the presence of RIS. For these different types of channel structures, we categorize the main estimation approaches proposed in the literatures, including parametric estimation methods, sparsity-aware techniques, low-rank algorithm, and deep learning-based solutions. These methods exhibit different trade-offs among estimation accuracy, online complexity, and robustness. This part aims to offer a comprehensive reference and technical roadmap for researchers and engineers working in the field of near-field communications.

\subsection{Organization and Notation}

\begin{figure*}[t!]
\setlength{\abovecaptionskip}{0.05cm}
\setlength{\belowcaptionskip}{-0.2cm}
\centering
\includegraphics[scale=0.35]{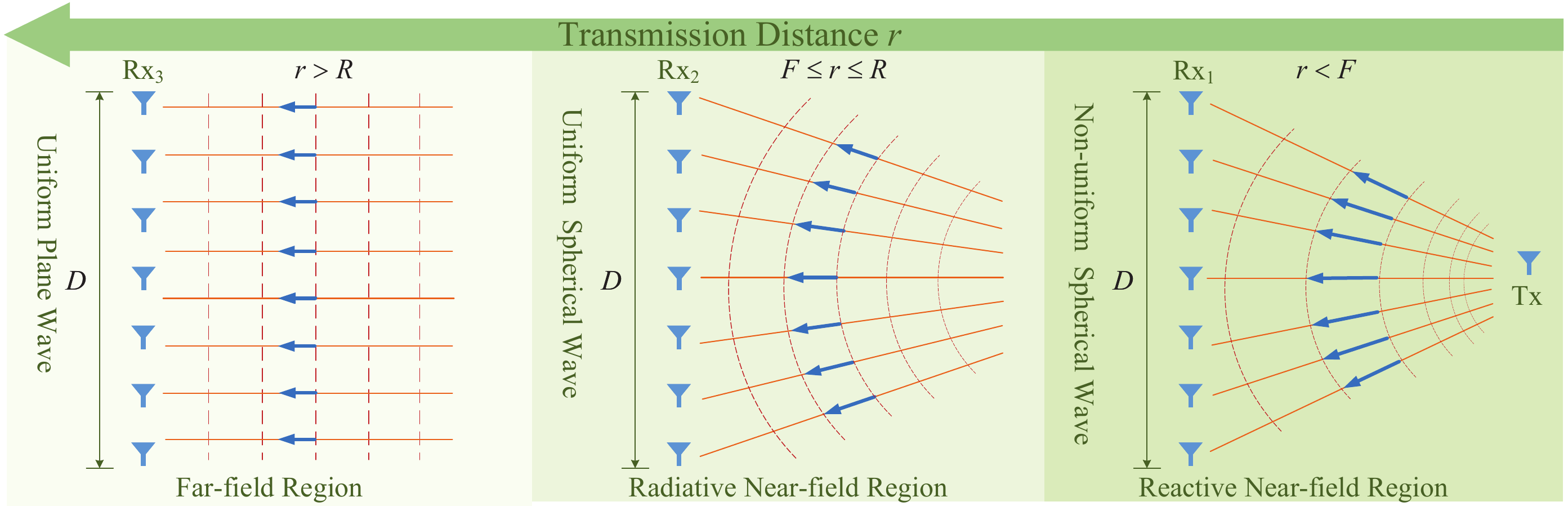}
\caption{Near-field vs. far-field boundary in a ULA scenario.}
\label{Fig.1}
\end{figure*}

The remainder of this survey is structured as follows. Section II first reviews the boundary between the far-field and near-field regions,  summarizes the mainstream antenna architectures that can enable an expanded near-field region together with their corresponding wavefront manipulation techniques. It then summarizes the potential 6G applications enabled by near-field spherical-wave propagation, as well as the ongoing standardization progress related to 6G near-field technologies. Section III presents the fundamental models for near-field channels. Section IV provides a systematic overview of existing studies on near-field channel estimation under various scenarios. Finally, Section V concludes this paper by summarizing the open challenges and future research directions in near-field channel estimation. The organizational structure of the paper is illustrated in Fig.~\ref{Fig.Org}, and the key abbreviations used throughout the paper are listed in Table~\ref{TableIX}.

Throughout this paper, scalars are denoted by italic letters (i.e., $x$), vectors by bold lowercase letters (i.e., $\mathbf{x}$), and matrices by bold uppercase letters (i.e., $\mathbf{X}$). The operators $(\cdot)^{\mathrm{T}}$ and $(\cdot)^{\mathrm{H}}$ denote the transpose and Hermitian transpose, respectively. The Euclidean norm of a vector $\mathbf{x}$ is written as $||\mathbf{x}||$, while $\mathbb{E}\{\cdot\}$ denotes the expectation operator. The notation $\text{diag}(\mathbf{x})$ refers to a diagonal matrix formed by placing the elements of $\mathbf{x}$ along the main diagonal. Moreover, $\mod{(\cdot,\cdot)}$ represents the modulo operation, and $ \lfloor\cdot\rfloor$ denotes the floor function.

\section{Spherical Electromagnetic Waves \\in 6G Wireless Communications}

This section first reviews the fundamental criteria that distinguish near-field and far-field regions, and then examines array architectures that can extend the near-field range along with their beam/wavefront control mechanisms. Building on these foundations, it highlights how spherical-wave characteristics in the near-field support emerging 6G applications and discusses related standardization progress and compatibility considerations.

\subsection{Near-Field and Far-Field Boundaries}

As illustrated in Fig. \ref{Fig.1}, the electromagnetic field around an antenna array is divided into near-field and far-field regions, separated by the Rayleigh distance. The near-field region can be further partitioned into reactive and radiative subregions by the Fresnel distance. When the source is located closer to the array than the Fresnel distance, the electromagnetic wavefront exhibits pronounced curvature and strong spatial non-uniformity, resulting in non-uniform spherical-wave behavior. As the distance increases beyond the Fresnel region, the wavefront becomes more uniform across the aperture but still retains  some curvature, resulting in a quasi-uniform spherical shape. Eventually, in the far-field region, the electromagnetic radiation is well approximated by a uniform plane~wave.

\subsubsection{Rayleigh Distance}

The Rayleigh distance is commonly used to characterize the transition between the radiative near-field and far-field regions of an antenna array. Within this range, differences in propagation path lengths lead to a distance-dependent phase profile and wavefront curvature must be accounted for. Beyond it, propagation paths are approximately parallel over the array aperture, and the received field can be approximated as a plane wave whose spatial response depends only on the angle of arrival (AoA). The Rayleigh distance is given by
\begin{align}\label{eq1}
R=\frac{2D^2}{\lambda },
\end{align}
where $D$ is the maximum aperture of the array and $\lambda$ is the wavelength. From \eqref{eq1}, the Rayleigh distance scales quadratically with the array size and inversely with the wavelength. When the distance between the array and the source exceeds $R$, the planar-wave assumption is valid. Otherwise, the nonlinear phase variation across the aperture calls for a spherical-wave propagation model.

According to \cite{cui2022channel,lee2022intelligent,wu2022near,cui2024near}, for a multiple-input single-output (MISO) system equipped with a uniform linear array (ULA) and an inter-element spacing of $d_t =\lambda/2$, the array aperture $D$ can be expressed as $D= N d_t = N\lambda/2$, where $N$ denotes the number of ULA elements. Accordingly, the Rayleigh distance for the ULA-based MISO system is calculated as
\begin{align}\label{eq2}
R_{\text{ULA}}^{\text{MISO}}=\frac{1}{2} N^{2} \lambda.
\end{align}
For a ULA-based MIMO system with $N_t$ transmit and $N_r$ receive elements, and an inter-element spacing of $d_t = d_r = \lambda/2$, the Rayleigh distance is given by \cite{lu2023near1}
\begin{align}\label{eq3}
R_{\text{ULA}}^{\text{MIMO}} =\frac{\lambda  \left ( N_{t} +N_{r} \right ) ^{2}}{2}.
\end{align}

As stated in \cite{demir2024spatial},  the Rayleigh distance for a MISO system equipped with a uniform planar array (UPA), where $N_H$ and $N_V$ denote the numbers of antenna elements per row and per column, respectively, and the inter-element spacing is $d_t$, can be expressed as
\begin{align}\label{eq5}
R_{\text{UPA}}^{\text{MISO}}=\frac{2d_t^2\left( N_H^2+N_V^2 \right)}{\lambda}.
\end{align}
When the inter-element spacing $d_t = \lambda/2$, \eqref{eq5} can be further simplified as \cite{hu2023design,han2023toward,lee2024near}:
\begin{align}\label{eq6}
\bar{R}_{\text{UPA}}^{\text{MISO}}=\frac{\left( N_H^{2}+N_V^{2} \right)\lambda }{2}.
\end{align}
Besides, an explicit closed-form expression of the Rayleigh distance for UPAs in the MIMO setting has not been clearly stated in the references surveyed in this work.

Finally, for a MISO system employing a uniform circular array (UCA) with radius $r_{\text{UCA}}$, the Rayleigh distance is expressed in \cite{lemaitre2013circular,xie2023near} as
\begin{align}\label{eq7}
\bar{R}_{\text{UCA}}^{\text{MISO}}=\frac{8r_{\text{UCA}}^{2}}{\lambda }.
\end{align}
Similarly, an expression for the Rayleigh distance of UCAs in the MIMO configuration is not reported in the surveyed literature.

\subsubsection{Fresnel Distance}

The Fresnel distance further partitions the near-field region into radiative and reactive subregions. Below this threshold, the electric and magnetic fields are out of phase, leading to localized energy oscillations rather than effective radiation, and resulting in a highly non-uniform spherical wavefront. Due to their rapid spatial attenuation, such waves are confined to a very limited region near the antenna. Beyond the Fresnel distance, the wavefront becomes progressively regular and can be approximated as a uniform spherical wave. Originally introduced by C. Polk in \cite{polk1956optical}, the general expression for the Fresnel distance is given by
\begin{align}\label{eq8}
F=0.62\sqrt{\frac{D^{3}}{\lambda }}.
\end{align}
According to \eqref{eq8}, the Fresnel distance increases with a larger array size $D$ and with a shorter wavelength $\lambda$.
For a MISO system employing a ULA, the Fresnel distance can be expressed as \cite{li2020doa}
\begin{align}\label{eq9}
F_{\text{ULA}}^{\text{MISO}}=0.62\sqrt{\frac{\left(Nd_{t}\right)^{3}}{\lambda}},
\end{align}
while in the case of a UPA-based MISO system, where the array aperture is given by $D = d_t\sqrt{\left(N_H^2+N_V^2\right)}$, we have
\begin{align}\label{eq10}
F_{\text{UPA}}^{\text{MISO}}=0.62\sqrt{\frac{d_{t}^{3}\left(N_H^{2}+N_V^{2}\right)^{\frac{3}{2}}}{\lambda}}.
\end{align}
Finally, for a MISO system utilizing a UCA with radius $r_{\text{UCA}}$, the Fresnel distance is given by
\begin{align}\label{eq11}
F_{\text{UCA}}^{\text{MISO}}=0.62\sqrt{\frac{8r_{\text{UCA}}^{3}}{\lambda }}.
\end{align}
It is further noted that, for the above arrays, closed-form expressions for the Fresnel distance under MIMO configurations have not been reported in the literature surveyed in this work.

\subsection{Generation and Control of Spherical Waves}

With the growing demand for wireless communications, massive MIMO in 5G is evolving toward ELAAs to support the higher capacity and data rates required in Beyond 5G (B5G).
When the array aperture becomes tens or even hundreds of times larger than the operating wavelength, spherical-wave effects become increasingly pronounced, making near-field propagation a non-negligible transmission paradigm in B5G networks. Typically, ELAAs are classified into CAP arrays, whose radiating surface is approximately spatially continuous, and SAP arrays, which are composed of discrete antenna elements. This classification is based on the ratio between the antenna element size $A$ and the square of the inter-element spacing $d$ \cite{lu2021communicating}:
\begin{align}\label{CAP_SDP}
\xi \stackrel{\triangle}{=} \frac{A}{d^2}.
\end{align}
%
In the extreme case when $\xi =1$, the discrete array becomes a continuous holographic surface.
Notably, the Fresnel distance of SPD arrays is typically negligible in practical communication scenarios, since non-uniform propagation is confined to a very limited region around each antenna element. In contrast, CAP arrays generally exhibit a non-negligible Fresnel distance due to their large continuous radiating surface \cite{Liu2025Near}.

\subsubsection{Continuous-Aperture Array}

A CAP array consists of densely packed elements with infinitesimal inter-element spacing \cite{liu2024near}. Due to the practical difficulty of integrating a large number of antennas within a finite aperture, CAP implementations often rely on emerging metamaterial technologies \cite{wang2024tutorial}. Metamaterial-based CAP arrays can be broadly categorized into four types: (i) dynamic metasurface antennas (DMAs) \cite{shlezinger2021dynamic,wang2019dynamic, shlezinger2019dynamic,yoo2018enhancing}, (ii) RISs \cite{huang2019reconfigurable,dai2020reconfigurable,venkatesh2020high,araghi2022reconfigurable}, (iii) continuous HMIMO \cite{yurduseven2017design,yurduseven2018dynamically,badawe2016true,yoo2021holographic} and (iv) large intelligent surfaces (LISs) \cite{foo2017liquid, hu2020spherical}. In the following, we briefly introduce these four categories, starting with DMAs.

Specifically, a DMA is a planar architecture comprising numerous sub-wavelength controllable elements, whose electromagnetic responses, such as amplitude and phase, can be dynamically tuned in an electronic manner  \cite{shlezinger2021dynamic}. Unlike traditional phased arrays, DMAs integrate part of the radio-frequency (RF) processing functionalities, such as analog weighting and combining, directly into the antenna structure, enabling beamforming and spatial multiplexing with reduced complexity and hardware cost, as well as improved energy efficiency. The power consumption of DMAs mainly stems from driving the tunable elements and the associated control/feeding networks. However, since they do not require a dedicated and fully independent RF chain per element, the resulting system-level power budget typically remains moderate. Notably, DMAs often support joint amplitude-phase control, offering a richer set of controllable degrees of freedom than most RISs, which are commonly limited to phase-only tuning. In terms of deployment, DMAs can be used at the BS as a low-complexity alternative to conventional massive MIMO to reduce power consumption and hardware cost \cite{wang2019dynamic, shlezinger2019dynamic}. They can also be integrated into mobile terminals to enable compact near-field beam focusing through reconfigurability, thereby further enhancing system capacity  \cite{yoo2018enhancing}.

\begin{table*}[]
\caption{Comparison of characteristics of different ELAA architectures}
\resizebox{\textwidth}{!}{
\begin{tabular}{|c|cc|c|c|c|c|c|}
\hline
\textbf{Type}                                                                  & \multicolumn{2}{c|}{\textbf{Structure}}                                                                                                                          & \begin{tabular}[c]{@{}c@{}}\textbf{Transmission}\\ \textbf{Mode}\end{tabular} & \begin{tabular}[c]{@{}c@{}}\textbf{Power} \\ \textbf{Budget}\end{tabular} & \begin{tabular}[c]{@{}c@{}}\textbf{Element} \\ \textbf{Density}\end{tabular}                        & \textbf{Control Type}                                                               & \begin{tabular}[c]{@{}c@{}}\textbf{Main}\\ \textbf{Functions}\end{tabular} \\ \hline
\multirow{4}{*}{\begin{tabular}[c]{@{}c@{}}\textbf{CAP} \\ \textbf{Array}\end{tabular}} & \multicolumn{2}{c|}{DMA}                                                                                                                                & Active or semi-passive                                                     & Medium                                                  & \begin{tabular}[c]{@{}c@{}}Dense \\or extremely dense\end{tabular}             & \begin{tabular}[c]{@{}c@{}}Dynamic amplitude\\ /phase control\end{tabular} & BS and Mobile terminals                                  \\ \cline{2-8}
                                                                      & \multicolumn{2}{c|}{\begin{tabular}[c]{@{}c@{}}Continuous\\ HMIMO\end{tabular}}                                                                         & Active                                                      & Medium-High                                             & Near-continuous                                                                   & Continuous holographic control                                             & Future XL-MIMO systems                                   \\ \cline{2-8}
                                                                      & \multicolumn{2}{c|}{RIS}                                                                                                                                & Passive or quasi-passive                                                    & Low                                                     & \begin{tabular}[c]{@{}c@{}}Dense \\ or extremely dense\end{tabular}             & Dynamic phase control                                                      & Assisted communications                                  \\ \cline{2-8}
                                                                      & \multicolumn{2}{c|}{LIS}                                                                                                                                & Active or quasi-passive                                           & Medium-High                                                    & Continuous                                                                        & Continuous surface control                                                 & Future indoor/macro coverage                             \\ \hline
\multirow{4}{*}{\begin{tabular}[c]{@{}c@{}}\textbf{SPD} \\ \textbf{Array}\end{tabular}} & \multicolumn{2}{c|}{\begin{tabular}[c]{@{}c@{}}Massive \\ MIMO\end{tabular}}                                                                            & Active                                                      & Medium-High                                             & Sparse                                                                            & \multirow{2}{*}{Hybrid beamforming}                                        & 5G BS                                                    \\ \cline{2-6} \cline{8-8}
                                                                      & \multicolumn{2}{c|}{XL-MIMO}                                                                                                                            & Active                                                      & High                                                    & \begin{tabular}[c]{@{}c@{}}Sparse\\ or relatively dense\end{tabular}                                                                  &                                                                            & B5G/6G BS                                                \\ \cline{2-8}
                                                                      & \multicolumn{1}{c|}{\multirow{2}{*}{\begin{tabular}[c]{@{}c@{}}Discrete\\ HMIMO\end{tabular}}} & \begin{tabular}[c]{@{}c@{}}Discrete\\ RIS\end{tabular} & Passive or quasi-passive                                                   & Low                                                     & \begin{tabular}[c]{@{}c@{}}Relatively dense \\ or quasi-continuous\end{tabular} & Dynamic phase control                                                      & Assisted communications                                  \\ \cline{3-8}
                                                                      & \multicolumn{1}{c|}{}                                                                          & \begin{tabular}[c]{@{}c@{}}Discrete\\ LIS\end{tabular} & Active or quasi-passive                                           & Medium-High                                                    & \begin{tabular}[c]{@{}c@{}}Relatively dense \\or quasi-continuous\end{tabular} & Dynamic amplitude/phase control                                            & Future indoor/macro coverage                             \\ \hline
\end{tabular}
}
\label{TableII}
\end{table*}

As the second category, the RIS is a planar metasurface comprising numerous passive or quasi-passive reflecting elements \cite{huang2019reconfigurable}. Each element is typically implemented by integrating positive intrinsic-negative (PIN) diodes \cite{dai2020reconfigurable} or leveraging complementary metal-oxide-semiconductor (CMOS) technology \cite{venkatesh2020high}, enabling independent phase control of incident electromagnetic waves and, in turn, reconfiguring the wireless propagation environment without active RF transmission or reception. RISs do not actively radiate electromagnetic energy. Instead, they redirect impinging signals via programmable phase control, and therefore generally operate with a very low power budget. Owing to their passive nature, RISs are most commonly deployed as assisting surfaces in wireless communication systems and are typically mounted on walls, building facades, ceilings, corridors and tunnels to facilitate link establishment, improve coverage under blockage, and enhance signal quality in targeted areas.

As the third category, continuous HMIMO can be regarded as a representative realization of the ideal CAP model.
By leveraging a dense or quasi-continuous distribution of antenna elements over a surface, it can generate, manipulate, and detect electromagnetic wavefronts under holographic principles, thereby enabling extremely fine-grained beamforming and potentially ultra-high spatial multiplexing \cite{yurduseven2017design, yurduseven2018dynamically}. Since a continuous aperture requires active excitation and precise control of the radiated fields over a finite surface, continuous HMIMO is typically considered an active transmission architecture \cite{badawe2016true}. As a result, the associated feeding/control networks often entail non-negligible implementation complexity and power consumption. In terms of deployment, continuous HMIMO is commonly envisioned at the BS side of future XL-MIMO systems to support stronger near-field beam focusing and further exploit the resulting spatial multiplexing gains.

Finally, as the fourth category, the LIS typically consists of large surfaces embedded within the environment, such as walls and ceilings \cite{foo2017liquid}. By electromagnetically activating these surfaces, the LIS can operate in either active or passive modes to transmit, receive, reflect, or modulate electromagnetic signals. Since it can be viewed as a form of massively distributed antenna array, such electrically activated surfaces are expected to substantially enhance the capacity, coverage, and link reliability of wireless communication systems \cite{hu2020spherical}. It is worth emphasizing that an LIS can either be deployed as a passive assisting reflective surface, akin to an RIS, or be equipped with active elements and feeding networks to operate as a transceiver-enabled extremely large aperture. Given the large surface area and the potentially massive number of elements, even when the per-element power consumption is low, the system-level power budget can still be moderate to high. From a deployment perspective, LISs are commonly envisioned as an important component of future indoor and macro-coverage infrastructures. For example, they can be distributedly integrated into entire indoor walls or ceilings, or embedded into outdoor building facades, so as to realize ``environment-as-a-node'' coverage enhancement and capacity improvement. Based on the above discussion, the different CAP arrays are summarized in Table \ref{TableII} for comparison.

Owing to their spatial continuity, CAP arrays support smoother beamforming over broader frequency ranges and enable more flexible wavefront shaping, offering enhanced directionality and high gain in near-field communications. This comes at the cost of higher structural complexity and increased signal-processing overhead. Recent works have investigated beam focusing and beamforming design across different CAP realizations. For DMA-based systems, \cite{zhang2022beam} proposed a beam focusing scheme to maximize the achievable sum rate in multi-user near-field communications. In RIS-aided MIMO, joint active-passive beamforming designs were developed in \cite{li2019joint,wu2019intelligent} by optimizing the BS precoding vector and RIS phase-shifts in order to maximize the total received power or minimize the signal-to-interference-plus-noise ratio (SINR). A hybrid beamforming framework was further introduced in \cite{di2020hybrid}, combining digital beamforming at the BS with analog beamforming at the RIS for sum-rate maximization. For continuous HMIMO, \cite{xu2024near} proposed a near-field receive beamforming method for uplink transmissions that mitigates degradations caused by near-field propagation, frequency selectivity, and spatial wideband effects, achieving higher transmission rates than conventional hybrid combiners. Finally, for LIS-based systems, \cite{yan2019passive} formulated a passive beamforming design for signal-to-noise ratio (SNR) maximization in a passive beamforming and information transfer (PBIT)-enhanced single-input multiple-output (SIMO) system and solved the resulting optimization problem using the semidefinite relaxation (SDR) approach.

\subsubsection{Spatially-Discrete Array}

In contrast to CAP arrays, SPD arrays comprise a finite number of antenna elements with fixed inter-element spacing, forming a discrete aperture. They are widely used in practical wireless communications, including MIMO in 4G and massive MIMO in 5G \cite{zhang2015massive, zhang2017hybrid}. With growing traffic demands, SPD arrays are now evolving toward XL-MIMO for B5G \cite{arun2020rfocus}.
Relative to massive MIMO, XL-MIMO features a substantially larger number of antenna elements and a much wider array aperture (potentially spanning tens or even hundreds of wavelengths), which makes user terminals more likely to fall within the radiative near-field region. The resulting spherical-wavefront effects and angle-distance coupling make near-field beam focusing a key capability of XL-MIMO. At the same time, XL-MIMO largely retains the active transceiver architecture of massive MIMO and still relies on RF chains, power amplifiers, and baseband processing to support multi-stream transmission and beamforming. As a result, it typically entails a higher power budget. From a deployment perspective, massive MIMO is mainly adopted at 5G BSs to support multiuser access and coverage enhancement, whereas XL-MIMO is more commonly envisioned for B5G/6G BSs and high-capacity hotspot access points to enable finer near-field beam focusing, higher spatial multiplexing potential, and more flexible coverage shaping.

Owing to the large number of antenna elements, SPD arrays, such as massive MIMO and XL-MIMO, can steer signal energy toward desired locations through adaptive beamforming, following environmental variations and user demands. However, fully digital beamforming is often impractical due to its high power consumption, elevated cost, and increased hardware and signal-processing complexity. As a result, most massive MIMO and XL-MIMO deployments rely on hybrid beamforming architectures to generate near-field beams \cite{ahmed2018survey}. For massive MIMO, the authors of \cite{nwalozie2023near} compared several beamforming strategies in mmWave near-field communications, including maximum ratio transmission (MRT), zero-forcing (ZF), minimum mean square error (MMSE), and signal-to-leakage-plus-noise ratio (SLNR), and reported that SLNR achieves the best performance among the considered schemes. Moving to XL-MIMO systems, the authors of \cite{el2014spatially} exploited the structural characteristics of mmWave channels and designed an orthogonal matching pursuit (OMP)-based beamforming approach, attaining near-optimal hybrid precoding performance. Beyond OMP-based designs, manifold-optimization methods using alternating minimization \cite{yu2015hybrid,yu2016alternating} have also been shown to achieve near-optimal precoding performance. In \cite{sohrabi2015hybrid}, a two-stage hybrid beamforming structure was adopted at both the transmitter and receiver to reduce the number of RF chains while maximizing the SE of XL-MIMO systems. To address the near-field beam-split issue in XL-MIMO, the authors of \cite{cui2024near} developed a hybrid beamforming scheme incorporating phase-delay focusing (PDF). Additionally, \cite{zhang2024dynamic} proposed a dynamic hybrid beamforming architecture for near-field XL-MIMO that employs a switching module to maximize the sum rate, while minimizing hardware power consumption.

By packing a large number of antenna elements with sub-wavelength spacing into a compact area, SPD arrays can approximate CAP arrays in terms of electromagnetic radiation characteristics while remaining physically discrete. This configuration gives rise to quasi-continuous aperture antennas, also known as discrete HMIMO \cite{wan2021terahertz}, which typically includes discrete RIS \cite{yang2016programmable,zhang2018space,zhang2020optically} and discrete LIS \cite{sanchez2022distributed}. A discrete RIS is usually composed of a large number of passive or quasi-passive controllable elements and mainly relies on dynamic phase control to programmably redirect impinging signals. Hence, it typically operates with a low power budget and is more often deployed as an assisting device in wireless systems. In contrast, a discrete LIS admits both quasi-passive and active realizations. The quasi-passive option leverages passive elements to shape the incident wavefront, whereas the active realization distributedly integrates a certain number of active antenna elements, thereby enabling joint amplitude-phase control over an extremely large aperture. As a result, discrete LIS implementations generally incur a moderate-to-high power budget. Similar to their continuous counterparts, discrete LISs are also envisioned as a key enabler for future indoor and macro-coverage infrastructures.

From a near-field beam manipulation perspective, \cite{wang2024wideband} proposed a deep learning-based end-to-end framework for wideband beamforming in discrete RIS-aided MIMO systems, aiming to maximize SE while mitigating the double beam split effect. For discrete LIS-aided near-field communications, \cite{hu2023design} developed a two-step 3-D beamforming scheme: it first applies conventional two-dimensional (2-D) far-field beamforming to compensate azimuth and elevation phase variations, and then uses one-dimensional (1-D) near-field beamforming to correct residual distance-induced phase errors. The characteristics of different SPD arrays are summarized in Table \ref{TableII} for intuitive comparison.

\subsubsection{Hardware Challenges and Solutions for ELAAs}

\setlength{\abovecaptionskip}{-0.0cm}
\setlength{\belowcaptionskip}{-0.40cm}
\begin{figure*}[t!]
\centering
\includegraphics[width=0.78\textwidth]{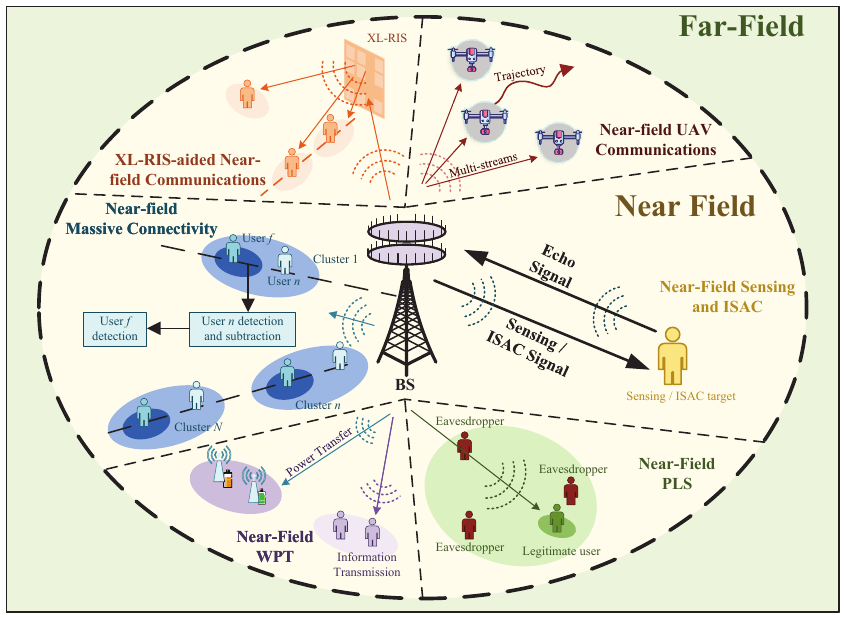}
\caption{Potential spherical-wave-enabled applications for 6G networks.}
\label{Fig14}
\end{figure*}
%

Although ELAAs offer substantial potential for spatial gain and capacity enhancement from a theoretical perspective \cite{Cui2023Near2, Wang2024Extremely}, their large-scale practical deployment still faces significant engineering challenges. Integrating a very large number of elements within a limited physical aperture makes fully digital architectures difficult to sustain. In particular, assigning a dedicated RF chain and a high-resolution ADC to each element, as in conventional MIMO or massive MIMO systems, quickly becomes impractical and gives rise to several challenges. On the one hand, the layout and routing burden increases sharply at the hardware integration level. In addition, limited routing space may push the inter-element spacing below half a wavelength, exacerbating mutual coupling. This, in turn, can distort element radiation patterns and input impedances, causing the realized array gain and beam patterns to deviate from their intended designs. On the other hand, dedicating an RF chain and a high-resolution ADC to each element typically leads to hardware cost and power consumption that scale approximately linearly with the number of elements. Since deployment cost and energy efficiency are key practical metrics, such power-hungry and expensive fully digital architectures are often challenging to realize in real-world systems.

To mitigate the above bottlenecks, hardware-efficient ELAA implementations have been widely explored. Among them, hybrid beamforming architectures, introduced in Section II-B(2), adopt a hierarchical design that combines a limited number of RF chains with an extremely large-scale array. The key idea is to implement an analog beamforming network in the RF front end using phase shifters, power splitters/combiners, and switch matrices, so as to provide coarse-grained directional gain and energy focusing. The digital baseband then operates on the resulting low-dimensional signals from the limited RF chains to perform fine-grained multi-stream precoding/combining, interference suppression, and multi-user separation. In this way, the number of required RF chains and high-resolution ADCs can be substantially reduced, while spatial multiplexing capability and beamforming flexibility remain close to those of fully digital architectures. Consequently, hybrid architectures can significantly alleviate hardware integration burden, deployment cost, and power consumption. Representative realizations include the fully connected and partially connected hybrid architectures in \cite{Ye2024Extremely}, as well as the dynamic hybrid architectures presented in \cite{zhang2024dynamic} and \cite{Park2017Dynamic}.

Complementary to hybrid architectures, metasurface-based arrays provide another hardware-efficient realization of ELAAs. In particular, the passive or semi-passive RISs, LISs, and DMAs reviewed in Sections II-B(1) and II-B(2) can form quasi-continuous apertures by leveraging a large number of low-power, low-cost programmable metamaterial elements or waveguide-fed structures. By using mode control or effective-aperture modulation in place of ultra-dense discrete feeding and per-element active RF chains, these arrays enable more flexible mutual-coupling suppression and electromagnetic-field manipulation at the physical-structure level. As a result, this metasurface-based paradigm inherently mitigates the hardware integration burden of ELAAs. Moreover, it can substantially reduce the reliance on active RF networks (e.g., large numbers of phase shifters, power splitters/combiners, and switches), further lowering deployment cost and power consumption.

In summary, ELAAs are a key enabler for B5G wireless communications and substantially extend the applicability of near-field propagation. However, their large-scale deployment hinges on carefully addressing hardware constraints, deployment cost, and power consumption. At the same time, the massive number of antenna elements in ELAAs significantly increases the complexity of signal processing tasks such as channel estimation and beamforming. Therefore, developing low-complexity algorithms tailored to ELAA-enabled near-field channels remains a key requirement for practical deployment.

\subsection{Near-Field Enabled 6G Applications}

\setlength{\abovecaptionskip}{-0.0cm}
\setlength{\belowcaptionskip}{-0.40cm}
\begin{figure}[t!]
\centering
\includegraphics[scale=0.56]{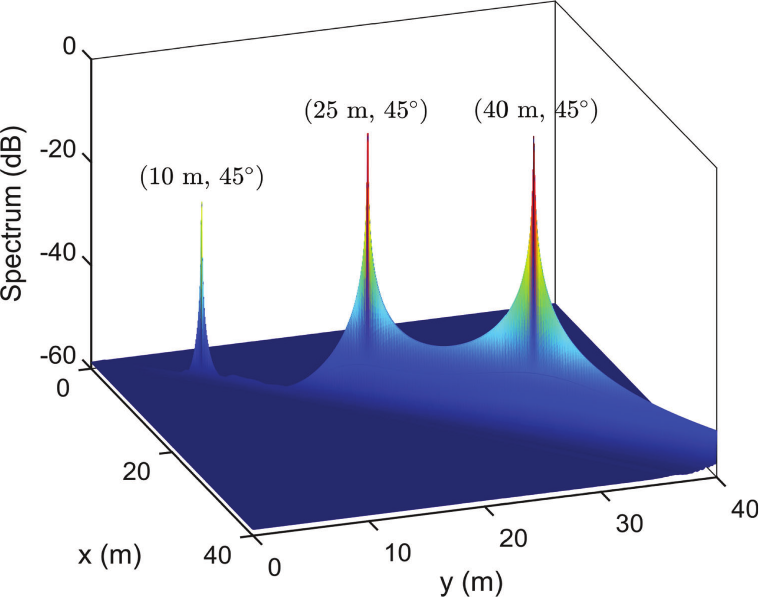}
\caption{The near-field sensing using MUSIC algorithm, reproduced from~\cite{Liu2025Near}, \textcopyright\ IEEE.}
\label{Fig13b}
\end{figure}

As discussed earlier, spherical-wave propagation in the near-field region introduces additional distance degrees of freedom and enables finer spatial resolution, allowing conventional DoA-based far-field beamforming to evolve into location-aware near-field beam focusing. Consequently, near-field beam manipulation extends beyond angular-domain control to joint angle-distance control, substantially improving spatial energy concentration and user separability. As illustrated in Fig. \ref{Fig14}, this capability enables a range of emerging applications, including near-field sensing and ISAC, near-field  physical layer security (PLS), near-field WPT, near-field massive connectivity, XL-RIS-aided near-field communications, and near-field unmanned aerial vehicle (UAV) communications. The key distinctions between these near-field applications and their conventional far-field counterparts are summarized as follows:

\textbullet\ \textbf{Near-field Sensing and ISAC}: Since spherical waves are jointly characterized by both angular and distance parameters, near-field sensing signals exhibit structured features that enable high-resolution holographic sensing \cite{Liu2025Near,Lei2025}. As illustrated in Fig. \ref{Fig13b}, near-field sensing based on the multiple signal classification (MUSIC) algorithm achieves joint angle-distance estimation \cite{Wang2023Near}. Owing to spherical-wave propagation, the locations of multiple targets can be accurately distinguished even when they share the same angular direction, a capability that is generally unattainable in conventional far-field sensing. The resulting high-precision angle-distance estimates can be directly exploited in near-field ISAC systems for accurate CSI acquisition and dynamic updates. By feeding sensing results into the communication process, refined spatial features and user location information can be leveraged to enhance link performance. This integration lays the foundation for joint near-field ISAC optimization, enabling communication and sensing functions to operate collaboratively within a unified architecture and further unlocking the potential performance gains of near-field ISAC systems \cite{lU2024Integrated}.

\begin{figure}[t!]
\centering
\includegraphics[width=0.385\textwidth]{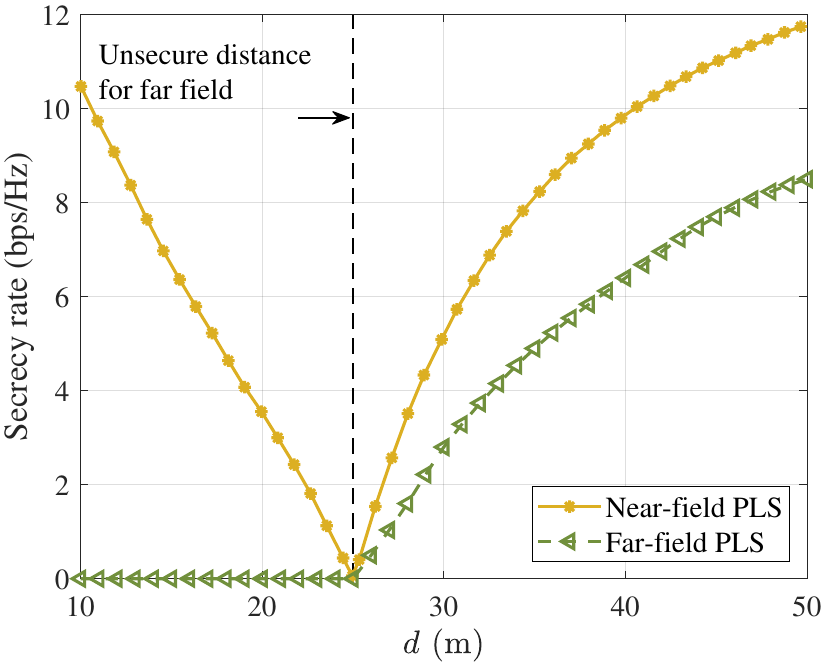}
\caption{Secrecy rate vs. the distance between the eavesdropper and the BS~\cite{Liu2025Near}.}
\label{Fig13a}
\end{figure}

\textbullet\ \textbf{Near-field Physical Layer Security}: Spherical-wave propagation can offer substantial PLS benefits, since near-field beam focusing can effectively mitigate eavesdropping threats from users located at similar angular positions, an issue that often compromises security under plane-wave propagation \cite{wang2024tutorial}. Fig.~\ref{Fig13a} illustrates how the distance between the eavesdropper and the BS affects the secrecy rate in both near-field and far-field communications \cite{Liu2025Near}. Notably, even when the eavesdropper is closer to the BS than the legitimate user, near-field operation can still improve the secrecy rate by beam focusing, highlighting a fundamental difference from the PLS mechanisms in far-field communications. More importantly, Fig.~\ref{Fig13a} shows that near-field PLS performance is primarily governed by the separation between the eavesdropper and the legitimate user, rather than by the eavesdropper-BS distance, as is typically the case in far-field settings.

\textbullet\ \textbf{Near-field Wireless Power Transfer}: Near-field beam focusing enables the BS to concentrate radiated energy on specific spatial locations with high precision, thereby significantly improving spatial resolution and transfer efficiency, which in turn creates new opportunities for near-field WPT \cite{Liu2025Near,Zhang2022Near}. Unlike far-field WPT, where energy spreads over the angular domain, near-field WPT achieves stronger energy focusing and higher localization accuracy. Since RF signals can simultaneously convey information and energy, near-field simultaneous wireless information and power transfer (SWIPT) also emerges as a promising 6G application. By exploiting the joint angle-distance controllability enabled by spherical wave propagation, near-field SWIPT can exploit optimized beam focusing to mitigate interference between energy and information transmission, thereby enhancing both functionalities \cite{Zhang2022Near}.

\setlength{\abovecaptionskip}{-0.0cm}
\setlength{\belowcaptionskip}{-0.40cm}

\textbullet\ \textbf{Near-field Massive Connectivity}:
Owing to the joint angle-distance controllability of spherical-wave propagation, near-field NOMA broadens the design space for multiuser resource allocation and interference management in 6G networks \cite{liu2024near,Liu2025Near}, providing a more flexible paradigm for massive connectivity. In particular, near-field beam focusing can concentrate signal energy toward users located farther from the BS, alleviating the far-field NOMA limitation whereby the effective channel gain decreases with distance, and potentially allowing far users to achieve higher gains. This property enables a ``far-to-near'' channel-gain-based successive interference cancellation (SIC) decoding order, which is especially beneficial when far users have higher communication demands than near users.

\textbullet\ \textbf{XL-RIS-aided Near-field Communications}: Owing to its substantially enlarged aperture, an XL-RIS can place both the signal source and the user terminal within the near-field region.
Unlike small-scale RISs operating primarily in the far-field, an XL-RIS can jointly design phase compensation in both the angular and distance domains to achieve spatial focusing of spherical wavefronts. This capability enables the simultaneous service of multiple users sharing the same angular direction but located at different distances, thereby improving spatial resolution and enhancing the energy efficiency of the communication system.

\textbullet\ \textbf{Near-field UAV Communications}: In aerial communications, near-field beam focusing can effectively mitigate strong interference among different LoS channels \cite{Liu2025Near}. By designing dedicated beamfocusing vectors for each UAV based on its spatial position, the BS can simultaneously serve multiple UAVs while significantly suppressing inter-UAV interference. Moreover, whereas conventional far-field LoS channels are often low-rank, near-field LoS channels can exhibit higher ranks, enabling the transmission of multiple data streams to each UAV.

Overall, spherical-wave-enabled near-field communications introduce an additional spatial degree of freedom for 6G, enabling refined wavefront control, improved energy efficiency, and enhanced spatial multiplexing. These gains over conventional far-field solutions fundamentally stem from the higher resolvability of spherical wavefronts in the joint angle-distance domain, which allows spatially selective beam focusing. Such focusing, in turn, requires real-time acquisition and updating of location-dependent near-field CSI. Therefore, the practical realization of these applications hinges on continued advances in accurate near-field channel modeling, efficient CSI acquisition, and intelligent beam focusing strategies tailored to spherical-wave propagation.

\subsection{Near-Field Standardization and Compatibility}

As discussed earlier, near-field communications are widely recognized as a key transmission paradigm for 6G networks. In response, international standardization bodies and industry alliances have initiated preliminary efforts on channel modeling, performance metrics, and backward compatibility considerations.

\subsubsection{Standardization and Industry Alliance Activities}

From a standardization perspective, the Third Generation Partnership Project (3GPP) has incorporated near-field propagation and spatial non-stationarities into Release 19 (Rel-19) channel modeling enhancements for the FR3 band (6-24 GHz) and XL-MIMO systems \cite{poddar2025overview3gpprelease19}. Building on the far-field statistical framework of TR 38.901, Rel-19 introduces a unified XL-MIMO channel model that explicitly captures spherical-wave propagation and element-level parameter variations. This candidate FR3 model for 6G supports both near- and far-field channel generation within a common framework. In parallel, 3GPP has launched a dedicated work item on near-field channel characteristics, addressing region definition, channel parameterization, and representative deployment scenarios (e.g., UMa, UMi, indoor office, and factory) \cite{wang2025newparadigmunifiednearfield}. A preliminary consensus on region definition, parameterization, and overall model structure has been reached within the 3GPP Technical Specification Group Radio Access Network (TSG RAN).

At the industry-alliance level, the RIS Alliance has published the 6G near-field technologies white paper and its 2.0 version \cite{Zhao2024NearFieldWP}, which systematically review key near-field scenarios, typical XL-MIMO and XL-RIS architectures, and related standardization requirements in channel modeling, testing and evaluation, and protocol design. The report further emphasizes the need to align near-field development with ongoing and forthcoming 3GPP standardization activities in spectrum planning, channel modeling, and evaluation methodologies, thereby facilitating integration into future 6G standards.

\subsubsection{Backward Compatibility}

The above activities indicate that near-field communications are evolving from a primarily theoretical topic into a concrete direction for 6G standardization. Nevertheless, 5G new radio (NR) has only recently entered large-scale commercial deployment, with substantial infrastructure already in place and still expanding. Replacing the current 5G ecosystem with a fully near-field-centric architecture is therefore impractical from both engineering and economic perspectives. The key challenge is to incorporate near-field enhancements into ongoing 5G-Advanced and future 6G evolution in a backward-compatible manner, while preserving the existing 5G NR framework and interface architecture.

Within the current 5G NR framework, the geometry-based stochastic model (GBSM) in 3GPP TR 38.901 is built on far-field plane-wave and spatial-stationarity assumptions, limiting its ability to capture spherical-wave and spatially non-stationary effects. To support near-field behavior while preserving established modeling frameworks and simulation platforms, the 3GPP Rel-19 FR3 study adopts an XL-MIMO channel model that preserves the original cluster-ray structure and parameter interfaces while incorporating BS/UE-to-spherical-wave-source distance information \cite{xu2025near}. This extension allows unified generation of near- and far-field channels within a common framework. In parallel, \cite{wang2025newparadigmunifiednearfield} advocates a `unified near-/far-field communication paradigm', in which near-field features are introduced through procedures compatible with TR 38.901. This framework supports both conventional far-field cellular operation and emerging near-field services (e.g., beam focusing and high-precision sensing), facilitating a smooth evolution toward 6G near-field communications based on existing evaluation platforms.

Beyond channel modeling, near-field operation raises backward compatibility challenges for the physical layer and across the protocol stack. The current 5G NR frame structure, reference-signal design, and limited-feedback codebooks are tailored to far-field plane-wave channels and angular-domain beamforming. In contrast, near-field operation requires incorporating range information in addition to angular parameters to extend beamforming to beam focusing, which increases codebook dimensionality and feedback overhead \cite{Lu2024A}. A central standardization question is therefore how to introduce polar-domain or 3-D focal-point codebooks while largely preserving the existing NR frame and reference-signal structure. In this direction, \cite{hu2023design} proposes a hierarchical near-field codebook, where 2-D far-field beamforming first compensates angular phase variations, and 1-D near-field beam focusing then corrects residual distance-induced phase errors, enabling 3-D beam focusing while remaining compatible with 5G NR \cite{An2024Near}. Moreover, in wideband systems, near-field propagation can also induce beam split due to stronger frequency-spatial coupling. As a result, pilot designs and channel estimation methods derived under narrowband plane-wave assumptions may suffer from substantial model mismatch. Standardization efforts must therefore balance continuity with the orthogonal frequency-division multiplexing (OFDM) framework and the explicit inclusion of spherical wavefronts, so that near-field capabilities can be introduced as NR physical-layer extensions rather than as a wholesale redesign.

Overall, near-field standardization remains largely focused on channel modeling and requirement analysis. As formal 6G work items are launched and FR3-based systems move toward deployment, a key objective is to integrate near-field transmission into a unified 6G framework while maintaining backward compatibility with existing 5G NR standards.

\subsection{Summary and Lessons Learned}

This section provided an integrated overview of near-field spherical-wave communications from four perspectives: electromagnetic region partitioning, array architectures and beam control, representative applications, and standardization. The Rayleigh and Fresnel distances define the boundaries between near-/far-field regions and reactive-/radiative near-field subregions, highlighting that for ELAAs the near-field region expands substantially, rendering plane-wave assumptions invalid over much of the practical range and necessitating spherical-wave channel models. CAP and SPD architectures were contrasted to clarify differences among DMA, RIS, LIS, HMIMO, and XL-MIMO in their structural properties and typical realizations.
%
Representative applications, including sensing and ISAC, PLS, WPT/SWIPT, near-field NOMA, XL-RIS-aided communications, and UAV links, demonstrate the advantages of spherical-wave propagation in joint angle-distance resolution, spatial focusing, and interference management, all fundamentally dependent on accurate near-field CSI acquisition. From a standardization perspective, we summarized 3GPP Rel-19 progress in FR3 and XL-MIMO channel modeling and highlighted key backward-compatibility requirements: unified near-/far-field modeling compatible with TR 38.901 and the continued evolvability of physical-layer protocols and reference-signal designs.

Building on the above discussion, the lessons learned can be summarized as follows:

\textbullet\ \textit{Near-field propagation will no longer be a ``marginal effect'' but one of the primary operating regimes in 6G networks}: In 6G networks, the aperture of an ELAA may reach tens or even hundreds of operating wavelengths, which significantly enlarges the Rayleigh distance and allows the near-field region to extend deep into, and potentially cover, the main service area of conventional cellular cells. Consequently, spherical-wave characteristics will evolve from a negligible effect into a core design premise. This shift calls for channel models, signal processing algorithms, and reference-signal designs that explicitly incorporate near-field effects, rather than relying on the conventional plane-wave assumption.

\textbullet\ \textit{The practical deployment of ELAAs requires hybrid architectures and passive or semi-passive metasurfaces}: Full-duplex active ELAAs typically face the dual challenges of substantial hardware cost and power consumption in practical deployments. Hybrid beamforming, together with passive or semi-passive metasurface-based structures such as DMAs, RISs, and LISs, provides a viable path toward low-cost and energy-efficient ELAA implementations. However, these architectures also introduce new constraints on the design of subsequent signal processing algorithms.

\textbullet\ \textit{Spherical-wave-enabled near-field applications fundamentally rely on the availability of high-quality near-field CSI}: Emerging near-field applications fundamentally rely on accurate angle-distance information to enable fine-grained control of the spherical wavefront. Consequently, near-field channel estimation and real-time CSI tracking constitute fundamental prerequisites for realizing these applications, thereby placing more stringent requirements on near-field estimation algorithms in terms of accuracy, computational complexity, and robustness.

\textbullet\ \textit{The key to standardization is compatibility-driven evolution within a unified framework}: For near-field technologies to be truly deployed, it is essential, without overturning the existing 5G NR evaluation framework and protocol stack, to gradually incorporate spherical-wave characteristics, near-field scenarios, and new service types into a unified standardization framework. For many key 6G use cases, such as XL-RIS and near-field ISAC, communication and high-precision sensing performance should be evaluated under a common set of scenarios and channel models, so that beamforming strategies and reference-signal designs remain reusable and consistent across different applications.


\section{Near-Field Channel Models}

When an ELAA is deployed at the BS, its array aperture may reach several tens or even hundreds of wavelengths. As a result, the extension of the Rayleigh distance makes it highly likely that UEs will be located within the near-field region of the BS. As an example, consider a BS array with a physical size of $0.8\times 0.4\text{m}^2$ and inter-element spacing $d=\lambda/2$, operating at $24 \,\text{GHz}$ (corresponding to $\lambda = 6.25\,\text{mm}$). This configuration yields a total of $N=8192$ antennas arranged in a $128\times 64$ layout. For this array, the Rayleigh distance is $R =128\,\text{m}$, which spans a substantial portion of a typical cell. As a result, it is highly likely that the UEs operate within the radiative near-field of the BS.

Owing to the large aperture of ELAAs, the propagation distances and angles between individual antenna elements and the UE or scatterers can vary significantly when these objects reside within the array's near-field region. Such variations lead to pronounced amplitude and phase fluctuations across the array, posing considerable challenges for accurate near-field channel modeling. To this end, this section introduces a representative channel modeling framework grounded in the spherical wavefront assumption.

\subsection{BS-UE Near-Field Channel Model for MISO Systems}

We begin by considering a MISO system where the UE is equipped with a single antenna and the BS employs an antenna array with $N$ elements. The array response vector at the BS is given by:
\begin{align}\label{eq12}
\mathbf{a}(\theta,\phi,r)=
    \frac{1}{\sqrt{N}}
\left[1,\cdots ,e^{-j\frac{2\pi}{\lambda}\left ( r_{n}-r\right)},\cdots,e^{-j\frac{2\pi}{\lambda}\left (r _{N}- r \right ) }\right]^{\mathrm{T}},
\end{align}
where $r$ denotes the distance from the UE to the BS reference element, $\theta$ and $\phi$ are the elevation and azimuth angles of the UE with respect to this element, while $r_{n}$ is the distance from the $n$-th BS array element to the UE. Denote $\mathbf{u}_{n} = [x_n,y_n,z_n]^{\mathrm{T}}$ as the coordinates of the $n$-th BS element relative to the reference element. Then, a general distance model from the $n$-th BS element to the UE can be expressed as
\begin{equation}
\begin{aligned}\label{eq13}
r_{n}&=\left \| \mathbf{r}-\mathbf{u}_{n} \right \|\\&=r\sqrt{1 -\frac{2\mathbf{k}^{\mathrm{T}}\left (\theta,\phi\right)\mathbf{u}_{n}}{r}+\frac{\left\|\mathbf{u}_{n}\right\|^{2}}{r^2}}\\
&\overset{(a)}{=} r-\mathbf{k}^{\mathrm{T}}\left (\theta,\phi\right)\mathbf{u}_{n}+\frac{\left\|\mathbf{u}_{n}\right\|^{2}-\left ( \mathbf{k}^{\mathrm{T}}\left (\theta,\phi\right)\mathbf{u}_{n} \right )^{2}}{2r}\\
& \quad+\frac{\mathbf{k}^{\mathrm{T}}\left (\theta,\phi\right)\mathbf{u}_{n}\left\|\mathbf{u}_{n}\right\|^{2}}{2r^{2}}-\frac{\left\|\mathbf{u}_{n}\right\|^{4}}{8r^{3}}+\cdots ,
\end{aligned}
\end{equation}
where $\mathbf{k}(\theta,\phi)=[\cos\theta\cos\phi,\cos\theta\sin\phi,\sin\theta]^{\mathrm{T}}$ denotes the propagation direction vector from the UE to the BS, and (a) utilizes the Taylor series expansion $\sqrt{1+x} = 1+\frac{1}{2}x-\frac{1}{8} x^{2} + \ldots$. Then, the distance difference in \eqref{eq12} can be expressed as:
\begin{align}\label{eq14}
\triangle r_n &= r_{n}-r=-\mathbf{k}^{\mathrm{T}}\left (\theta,\phi\right)\mathbf{u}_{n}+\frac{\left\|\mathbf{u}_{n}\right\|^{2}-\left ( \mathbf{k}^{\mathrm{T}}\left (\theta,\phi\right)\mathbf{u}_{n} \right )^{2}}{2r}\nonumber\\
&\quad +\frac{\mathbf{k}^{\mathrm{T}}\left (\theta,\phi\right)\mathbf{u}_{n}\left\|\mathbf{u}_{n}\right\|^{2}}{2r^{2}}-\frac{\left\|\mathbf{u}_{n}\right\|^{4}}{8r^{3}}+\cdots .
\end{align}
When the distance $r$ between the BS and the UE exceeds the Rayleigh distance, the Taylor approximation $\sqrt{1+x}\approx 1+\frac{1}{2}x$ under the far-field condition is applied, where only the first term in \eqref{eq14} is considered, resulting in
\begin{equation}
\begin{aligned}\label{eq14FF}
\triangle r_n =-\mathbf{k}^{\mathrm{T}}\left (\theta,\phi\right)\mathbf{u}_{n}.
\end{aligned}
\end{equation}
Furthermore, when the distance $r$ lies between the Rayleigh distance and the Fresnel distance, the following Fresnel approximation can be obtained:
\begin{equation}
\begin{aligned}\label{eq14UNF}
\triangle r_n =-\mathbf{k}^{\mathrm{T}}\left (\theta,\phi\right)\mathbf{u}_{n}+\frac{\left\|\mathbf{u}_{n}\right\|^{2}-\left ( \mathbf{k}^{\mathrm{T}}\left (\theta,\phi\right)\mathbf{u}_{n} \right )^{2}}{2r}.
\end{aligned}
\end{equation}
Unlike the far-field approximation in \eqref{eq14FF}, the near-field formulations in \eqref{eq14} and \eqref{eq14UNF} indicate that the array response vector depends not only on the DoA from the UE to the BS, but also on their separation distance. For different BS array configurations, such as ULA, UPA, and UCA, the explicit expressions of the above distance differences will be derived in detail in Section III-A(1)-(3).

Under the quasi-static channel assumption, the UE-BS channel with a single LoS path can be modeled as \cite{Liu2023Near,lu2023near1,cui2022near}:
%
\begin{align}\label{eq15}
\mathbf{h}_{\mathrm{NF}}^{\mathrm{LoS}}
=g_{\mathrm{NF}}^{\mathrm{LoS}}\mathbf{a}(\theta,\phi,r),
\end{align}
where $g_{\mathrm{NF}}^{\mathrm{LoS}}$ is the attenuation factor with zero mean and variance $\beta_{\mathrm{LoS}}$. On the other hand, when multiple transmission paths are present between the UE and the BS, the multipath channel model can be expressed as \cite{Saleh1987A}:
\begin{align}\label{eq161}
\mathbf{h}_{\mathrm{NF}}^{\mathrm{NLoS}}(t)
=\sqrt{\frac{N}{L}}\sum_{l=1}^{L}g_{l}\mathbf{a}(\theta_l,\phi_l,r_l)\delta(t-\tau_l),
\end{align}
where $g_{l}$ is the attenuation factor of the $l$-th path with zero mean and variance $\beta_{\text{NLoS}}$, while $(\theta_l,\phi_l,r_l)$ is the position of the $l$-th scatterer relative to the reference element of the BS array, $\tau_l$ is the propagation delay of the $l$-th path, and $\delta(\cdot)$ is the Dirac delta function, $l =1,\cdots,L$. Nevertheless, to simplify the parametric representation and facilitate channel estimation, most existing works neglect the inter-path delays \cite{Liu2023Near,lu2023near1,cui2022near}, i.e.,
\begin{align}\label{eq16}
\mathbf{h}_{\mathrm{NF}}^{\mathrm{NLoS}}
=\sqrt{\frac{N}{L}}\sum_{l=1}^{L}g_{l}\mathbf{a}(\theta_l,\phi_l,r_l).
\end{align}
In this case, as $L\rightarrow \infty$, we obtain the spatially correlated Rayleigh fading model \cite{demir2024spatial,kosasih2024roles}
\begin{align}\label{eq17}
\mathbf{h}_{\mathrm{NF}}^{\mathrm{Ray}}\sim \mathcal{CN}\left(\mathbf{0}_N,\mathbf{R}_{\mathrm{Ray}}\right),
\end{align}
where the spatial correlation matrix $\mathbf{R}_{\mathrm{Ray}}$ is given by:
\begin{align}\label{RhNF1}
&\mathbf{R}_{\mathrm{Ray}}=\mathbb{E}\left\{\mathbf{h}\mathbf{h}^{\mathrm{H}}\right\}=\beta_{\text{Ray}}\cdot\nonumber\\
&\!\int_{r_1}^{r_2}\!\int_{\varphi_1}^{\varphi_2} \!\int_{\theta_1}^{\theta_2}\!f(\tilde{\theta},\tilde{\phi},\tilde{r})\mathbf{a}(\tilde{\theta},\tilde{\phi},\tilde{r})\mathbf{a}^{\mathrm{H}} (\tilde{\theta},\tilde{\phi},\tilde{r}) d\tilde{\theta} d\tilde{\phi} d\tilde{r}.
\end{align}
In the above equation, $\beta_{\text{Ray}}$ is the average channel power, $\tilde{\theta}\in[\theta_1,\theta_2]$, $\tilde{\phi}\in[\phi_1,\phi_2]$ and $\tilde{r}\in[r_1,r_2]$ define the spatial distribution range of the UE-BS channel, while $f(\tilde{\theta},\tilde{\phi},\tilde{r})$ is the normalized spatial scattering function satisfying $\int_{r_1}^{r_2}\int_{\varphi_1}^{\varphi_2} \int_{\theta_1}^{\theta_2} f(\tilde{\theta},\tilde{\varphi},\tilde{r})=1$. As an alternative to the Rayleigh fading model, the Rician distribution can be employed to characterize scenarios where both LoS and NLoS paths coexist between the UE and the BS. In such cases, the channel vector is expressed as:
\begin{align}\label{eq19}
\mathbf{h}_{\mathrm{NF}}^{\mathrm{Ric}}\sim \mathcal{CN}\left(\mathbf{h}_{\mathrm{NF}}^{\mathrm{LoS}},\mathbf{R}_{\mathrm{Ric}}\right),
\end{align}
where $\mathbf{h}_{\mathrm{NF}}^{\mathrm{LoS}}$ denotes the mean of the Rician model, while the correlation matrix is defined as
\begin{align}\label{eq20}
\mathbf{R}_{\mathrm{Ric}} =\mathbb{E}\left\{\left(\mathbf{h}_{\mathrm{NF}}^{\mathrm{Ric}}-\mathbf{h}_{\mathrm{NF}}^{\mathrm{LoS}}\right)\left(\mathbf{h}_{\mathrm{NF}}^{\mathrm{Ric}}-\mathbf{h}_{\mathrm{NF}}^{\mathrm{LoS}}\right)^{\mathrm{H}}\right\}.
\end{align}
As is seen, when $\mathbf{h}_{\mathrm{NF}}^{\mathrm{LoS}} = \mathbf{0}_N$, i.e., only NLoS paths exist between the UE and the BS, the Rician fading model in \eqref{eq19} reduces to the Rayleigh fading model in \eqref{eq17}. The above equations from \eqref{eq15} to \eqref{eq20} indicate that, in the MISO scenario, the array response vector at the BS is the dominant factor in characterizing the UE-BS channel. Hence, in the sequel we provide the array response vectors for various BS array geometries.

\subsubsection{ULA-based Channel Model}

\begin{figure}[t!]
\setlength{\abovecaptionskip}{-0.0cm}
\setlength{\belowcaptionskip}{-0.4cm}
\centering
\includegraphics[scale=0.27]{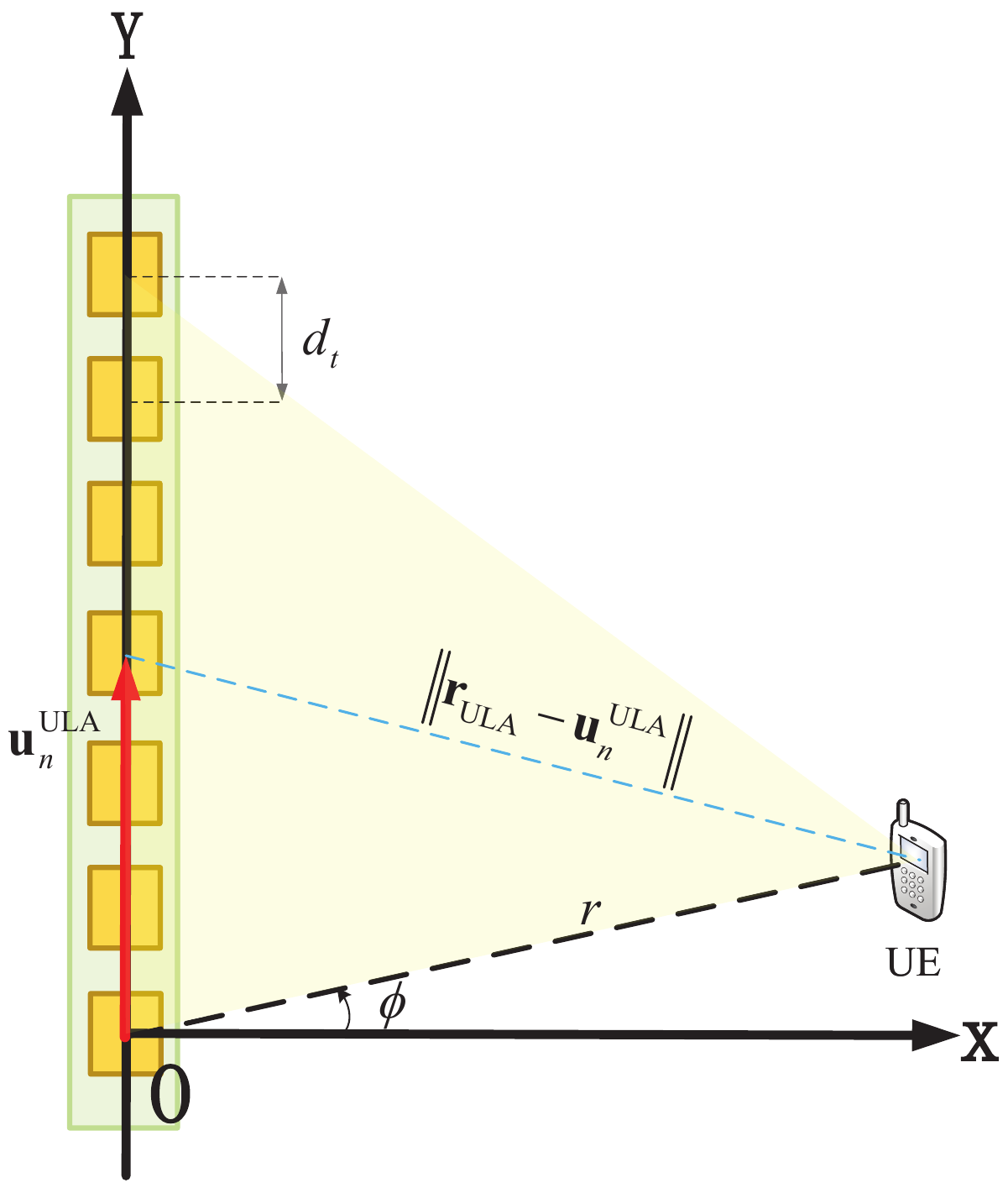}
\caption{ULA-based near-field channel model.}
\label{Fig.3}
\end{figure}

We first consider a MISO system where the BS is equipped with a ULA with inter-element spacing $d_t$. In such a case, we can always consider a coordinate system as shown in Fig. \ref{Fig.3}, such that all the ULA elements lie within the $\text{XOY}$ plane. Under this configuration, the $\text{Z}$-axis can be ignored and we can set $\theta = 0^{\circ}$. By placing the origin of the coordinate system at the reference element of the ULA, the coordinates of the $n$-th ULA element are expressed as $\mathbf{u}^{\text{ULA}}_{n}=\left[0,(n-1)d_t\right]^{\mathrm{T}}$, where $n=1,2,\cdots,N$. Furthermore, the coordinates of the UE antenna are given by $\mathbf{r}_{\text{ULA}}= r\mathbf{k}_{\text{ULA}} = r\left [\cos\phi,\sin\phi\right]^{\mathrm{T}}$, where $r$ and $\phi$ denote the distance and azimuth angle between the reference element of the ULA and the UE. In this
case, the propagation distance $\left \| \mathbf{r}_{\text{ULA}}-\mathbf{u}^{\text{ULA}}_{n} \right \|$ from the $n$-th ULA element to the UE can be calculated as\cite{dong2023near}:
\begin{equation}
\begin{aligned}\label{eq21}
&\left \| \mathbf{r}_{\text{ULA}}-\mathbf{u}^{\text{ULA}}_{n} \right \|\\
&=\sqrt{r^{2}+\left[(n-1)d_t\right]^{2}-2r(n-1)d_t\sin\phi}\\
&= r-(n-1)d_t\sin\phi+\frac{[(n-1)d_t]^{2}\cos^{2}\phi}{2r}+\cdots .
\end{aligned}
\end{equation}
Based on \eqref{eq21}, the difference between the path lengths from the UE to the $n$-th ULA element and from the UE to the ULA reference element is expressed by
\begin{equation}
\begin{aligned}\label{eq22}
\triangle r^{\text{ULA}}_n&=\left \| \mathbf{r_{\text{ULA}}}-\mathbf{u}^{\text{ULA}}_{n} \right \|-r\\
&= -(n-1)d_t\sin\phi+\frac{[(n-1)d_t]^{2}\cos^{2}\phi}{2r}+\cdots.
\end{aligned}
\end{equation}
Substituting \eqref{eq22} into \eqref{eq12}, we obtain the near-field array response vector for the ULA as
\begin{align}\label{eqAR}
\mathbf{a}_{\text{ULA}}(\phi,\!r)\!=\!\frac{1}{\sqrt{N}}\!
\left[1,\!\cdots,\!e^{-j\frac{2\pi}{\lambda}\triangle r^{\text{ULA}}_n}\!,\!\cdots\!,\!e^{-j\frac{2\pi}{\lambda}\triangle r^{\text{ULA}}_N }\!\right]^{\mathrm{T}}\!.
\end{align}
Finally, putting this array vector into \eqref{eq15}, \eqref{eq16}, \eqref{RhNF1}, and \eqref{eq20}, we can sequentially derive the ULA-based LoS channel model \cite{dong2022near,dong2023near}, the multipath channel model\cite{dong2022near,dong2023near}, the spatially correlated Rayleigh model\cite{liu2021non,liu2023spatially}, and the Rician channel model \cite{liu2021non,liu2023spatially} in the near-field case.

\begin{figure}[t!]
\setlength{\abovecaptionskip}{-0.0cm}
\setlength{\belowcaptionskip}{-0.4cm}
\centering
\includegraphics[scale=0.27]{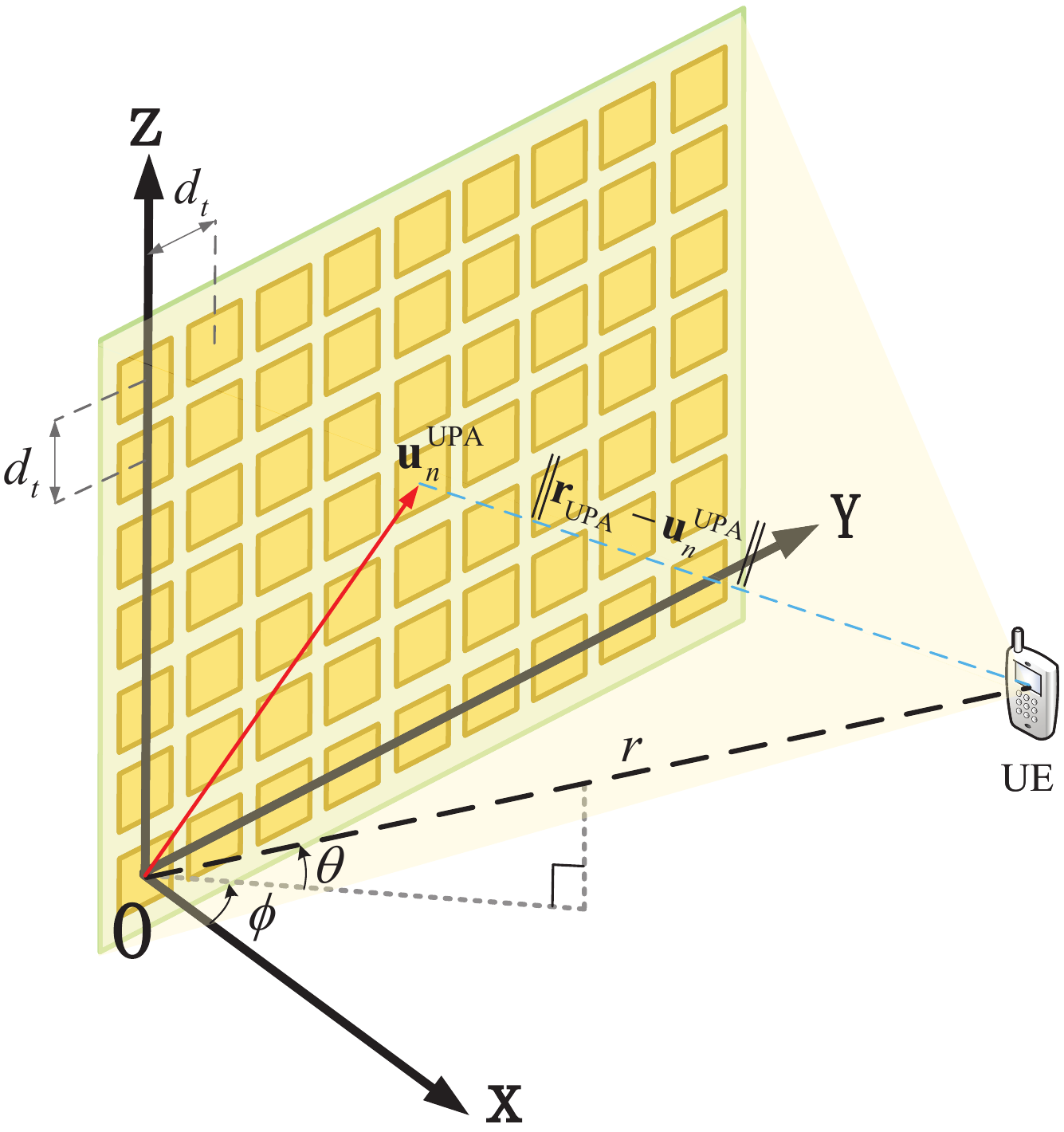}
\caption{UPA-based near-field channel model.}
\label{Fig.2}
\end{figure}

\subsubsection{UPA-based Channel Model}

We now consider a BS equipped with a 2-D UPA with inter-element spacing $d_t$, where the number of elements in each row and column are denoted by $N_H$ and $N_V$, respectively. As illustrated in Fig. \ref{Fig.2}, we employ a 3-D coordinate system at the BS, with the UPA lying on the ${\text{YOZ}}$ plane and the reference element placed in the origin ${\text{O}}$. When the UPA elements are indexed in a row-by-row manner with $n\in[1,N]$, where $N=N_H\times N_V$, the location of the $n$-th UPA element is expressed as
$\mathbf{u}^{\text{UPA}}_{n}=\left[0,i_nd_t, j_nd_t\right]^{\mathrm{T}}$, where $i_n=\text{mod}(n-1,N_H)$ and $j_n=\left\lfloor(n - 1)/N_H\right\rfloor$. On the other hand, the coordinates of the UE antenna are given by $\mathbf{r}_{\text{UPA}}=r\mathbf{k}_{\text{UPA}}=r\left[\cos\theta\cos\phi,\cos\theta\sin\phi,\sin\theta\right]^{\mathrm{T}}$, where $r$, $\theta$
and $\phi$ are the distance, elevation and azimuth angles between the UPA reference element and the UE, respectively. Thus, the difference in distance between the path from the UE to the $n$-th UPA element and that to the reference element is found to be as \cite{zhao2024performance}:
\begin{align}\label{eq30}
&\triangle r^{\text{UPA}}_{n}=\left\|\mathbf{r}_{\text{UPA}}-\mathbf{u}^{\text{UPA}}_{n}\right\|-r\nonumber\\
&=\sqrt{r^{2}\!+\left(i_n^{2}+j_n^{2}\right)d_t^{2}\!-2rd_t\left(i_n\cos\theta\sin\phi+j_n\sin\theta\right)}\!-\!r
\nonumber\\
&=-d_t(i_n\cos\theta\sin\phi+j_n\sin\theta)\!+\!\frac{i_n^{2}d_t^{2}(1\!-\!\cos^{2}\theta\sin^{2}\phi)}{2r}\nonumber\\
&\quad\ +\frac{j_n^{2}d_t^{2}\cos^{2}\theta}{2r}+\cdots .
\end{align}
Substituting the above distance difference into \eqref{eq12} provides the near-field array response vector of the UPA as \cite{lu2023near2}
\begin{align}\label{eq31}
\mathbf{a}_{\text{UPA}}\left(\theta,\phi,r\right)=\frac{1}{\sqrt{N}}
\begin{bmatrix}
 1,\cdots, e^{-j\frac{2\pi}{\lambda}\triangle r^{\text{UPA}}_{n}},\cdots ,e^{-j\frac{2\pi}{\lambda}\triangle r^{\text{UPA}}_{N}}
\end{bmatrix} ^{\mathrm{T}}.
\end{align}
Finally, substituting \eqref{eq31} into \eqref{eq15}, \eqref{eq16}, \eqref{RhNF1}, and \eqref{eq20}, we can derive the LoS channel model \cite{zhang2022beam,zhao2024performance}, the multipath channel model \cite{lu2023near2,guo2023compressed,demir2024spatial}, the spatially correlated Rayleigh model \cite{demir2024spatial}, and the Rician channel model \cite{wang2024geometry} for the near-field UPA~scenario.

\subsubsection{UCA-based Channel Model}

\begin{figure}[t!]
\setlength{\abovecaptionskip}{-0.0cm}
\setlength{\belowcaptionskip}{-0.4cm}
\centering
\includegraphics[scale=0.27]{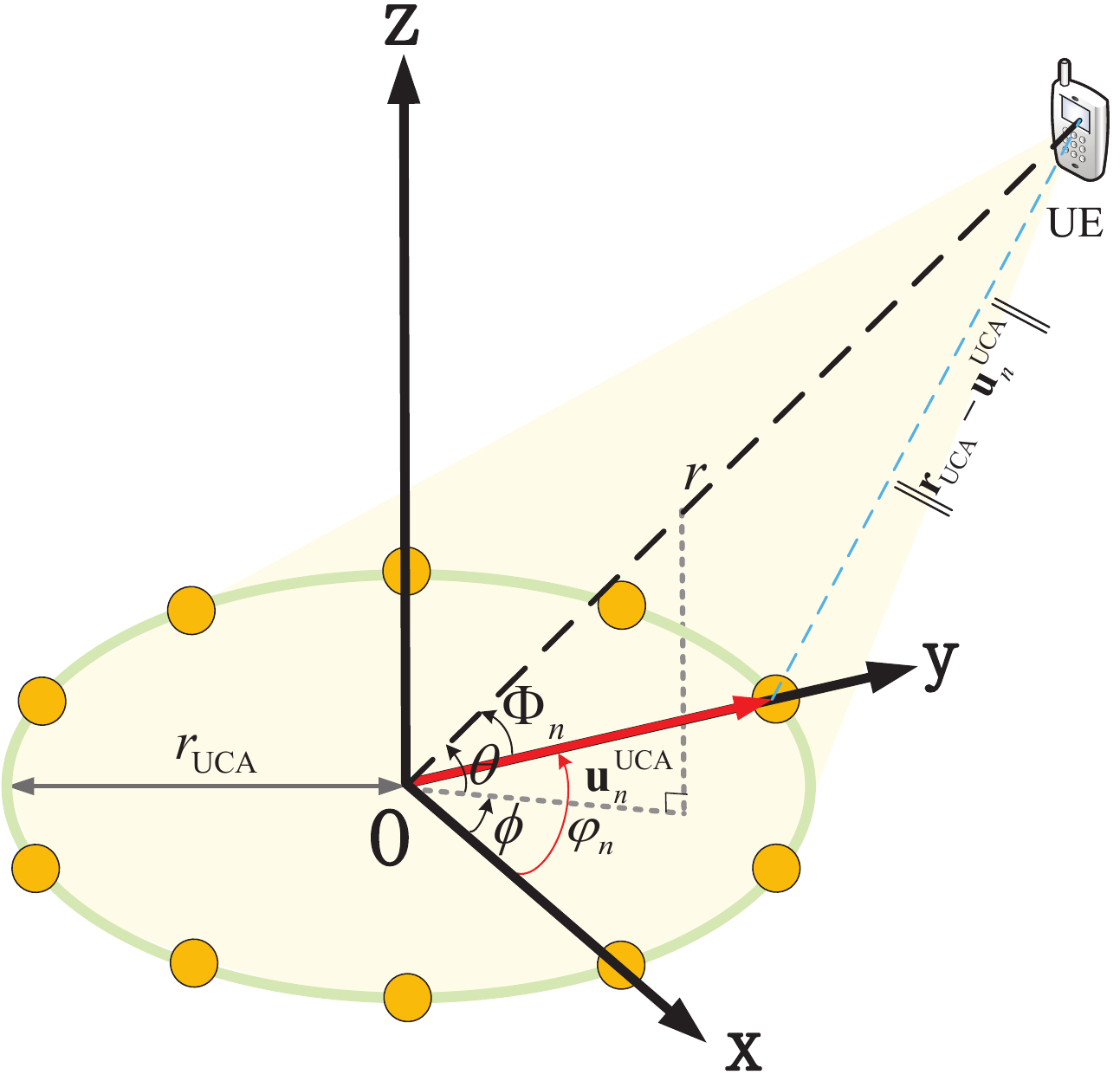}
\caption{UCA-based near-field channel model.}
\label{Fig.4}
\end{figure}

Fig. \ref{Fig.4} illustrates a BS equipped with an $N$-element UCA with a radius of $r_{\text{UCA}}$. In this case, it is convenient to use a coordinate system such that the UCA lies on the XOY plane and its center is located at the origin $\text{O}$. After placing the reference element at the position $(r_{\text{UCA}},0,0)$, the coordinates of the $n$-th UCA element is denoted as $\mathbf{u}^{\text{UCA}}_{n}=r_{\text{UCA}}\left[\cos\varphi_{n},\sin\varphi_{n},0\right]^{\mathrm{T}}$, where $\varphi_{n}=\frac{2\pi\left(n-1\right)}{N}$ is the azimuth angle of the $n$-th element, with $n\in \left \{1,\cdots,N\right\}$. The location of the UE is $\mathbf{r}_{\text{UCA}}=r\mathbf{k}_{\text{UCA}} = r \left [\cos\theta\cos\phi, \cos\theta\sin\phi,\sin\theta\right]^{\mathrm{T}}$,  where $(r,\theta,\phi)$ are the distance, elevation and azimuth angles between the origin $\text{O}$ and the UE, respectively. Thus, the path length difference between the UE and $n$-th UCA element and that to the reference element is given by
\cite{ji2016near,qin2024fast}:
\begin{align}\label{eq33}
\triangle r^{\text{UCA}}_{n} &= \left \| \mathbf{r}_{\text{UCA}} - \mathbf{u}^{\text{UCA}}_{n} \right \| - r \nonumber\\
&= \sqrt{r^{2} + r_{\text{UCA}}^{2} - 2r r_{\text{UCA}}\cos\Phi_{n}} - r \nonumber\\
&= -r_{\text{UCA}}\cos\Phi_{n} + \frac{r_{\text{UCA}}^{2}}{2r}\left(1 - \cos^{2}\Phi_{n}\right) + \cdots,
\end{align}
where $\Phi_{n}$ is the angle between the vectors $\mathbf{u}^{\text{UCA}}_{n}$ and $\mathbf{r}_{\text{UPA}}$, with $\cos\Phi_{n}=\cos\theta\cos\left(\phi-\varphi_{n}\right)$.
Substituting \eqref{eq33} into \eqref{eq12}, the near-field array response vector for the UCA is derived as \cite{wu2023enabling,qin2024fast}
\begin{align}\label{eq39}
\mathbf{a}_{\text{UCA}}\left(\theta,\phi,r\right)=\frac{1}{\sqrt{N}}
\begin{bmatrix}
 e^{-j\frac{2\pi}{\lambda}\triangle r^{\text{UCA}}_{1}},&\cdots&,e^{-j\frac{2\pi}{\lambda}\triangle r^{\text{UCA}}_{N}}
\end{bmatrix} ^{\mathrm{T}}.
\end{align}
Finally, inserting this array response vector into \eqref{eq15}, \eqref{eq16}, \eqref{RhNF1} and \eqref{eq20}, we can sequentially derive the UCA-based LoS channel model \cite{chen2024near}, the multipath channel model\cite{wu2023enabling,qin2024fast,peng2024near}, the spatially correlated Rayleigh model, and the Rician channel model \cite{he2023novel} in the near-field case.

\subsection{BS-UE Near-Field Channel Model for MIMO Systems}

Building on the above discussion, we can easily extend the MISO channel model to the MIMO channel model. In the MIMO case, the channel is influenced by the array response vectors at both the BS and UE. Assume that the BS is equipped with $N_t$ elements and the UE with $N_r$ elements. Hence, by treating the array at the UE as a point, the array response vector at the BS can be derived based on the position $(\theta_{\text{UE}},\phi_{\text{UE}},r_{\text{UE}})$ of the UE relative to the BS reference element, denoted as $\boldsymbol{a}_{T}(\theta_{\text{UE}},\phi_{\text{UE}},r_{\text{UE}})\in\mathbb{C}^{N_{t}}$ \cite{lu2024tutorial}. Similarly, the array response vector at the UE can be expressed as $\boldsymbol{a}_{R}(\theta_{\text{BS}},\phi_{\text{BS}},r_{\text{BS}})\in\mathbb{C}^{N_{r}}$, where $(\theta_{\text{BS}},\phi_{\text{BS}},r_{\text{BS}})$ is the position of the BS with respect to the UE reference element. Then, the LoS channel between the UE and the BS can be expressed as the outer product of $\boldsymbol{a}_{T}(\theta_{\text{UE}},\phi_{\text{UE}},r_{\text{UE}})$ and $\boldsymbol{a}_{R}(\theta_{\text{BS}},\phi_{\text{BS}},r_{\text{BS}})$ \cite{lu2024tutorial}, i.e.,
\begin{align}\label{eq43}
\mathbf{\bar{H}}_{\mathrm{NF}}^{\mathrm{LoS}}
=\bar{g}_{\mathrm{NF}}^{\mathrm{LoS}}\boldsymbol{a}_{T}(\theta_{\text{UE}},\phi_{\text{UE}},r_{\text{UE}})\boldsymbol{a}^{\mathrm{H}}_{R}(\theta_{\text{BS}},\phi_{\text{BS}},r_{\text{BS}}),
\end{align}
where $\bar{g}_{\mathrm{NF}}^{\mathrm{LoS}}$ is the attenuation factor, and the specific expressions for the array response vectors $\boldsymbol{a}_{T}(\theta_{\text{UE}},\phi_{\text{UE}},r_{\text{UE}})$ and $\boldsymbol{a}_{R}(\theta_{\text{BS}},\phi_{\text{BS}},r_{\text{BS}})$ can be derived from \eqref{eq12}, \eqref{eqAR}, \eqref{eq31} and \eqref{eq39} in the previous subsection. When multiple transmission paths exist between the UE and the BS, and the inter-path delays are neglected, the UE-BS channel model can further be described as \cite{tian2025analytical}
\begin{align}\label{eq44}
\mathbf{\bar{H}}_{\mathrm{NF}}^{\mathrm{NLoS}}=\sum_{l=1}^L\bar{g}_l\boldsymbol{a}_{T}(\theta^l_{\text{T}},\phi^l_{\text{T}},r^l_{\text{T}})\boldsymbol{a}_{R}^\mathrm{H}(\theta^l_{\text{R}},\phi^l_{\text{R}},r^l_{\text{R}}),
\end{align}
where $\bar{g}_l$ is the attenuation factor of the $l$-th path, $(\theta^l_{\text{T}},\phi^l_{\text{T}},r^l_{\text{T}})$ and $(\theta^l_{\text{R}},\phi^l_{\text{R}},r^l_{\text{R}})$ denote the positions of the $l$-th scatterer with respect to the BS reference element and the UE reference element, with $l\in\{1,\cdots, L\}$. As $L\rightarrow\infty$, the multipath model described in \eqref{eq44} converges to the Rayleigh fading model $\mathbf{H}_{\mathrm{NF}}^{\mathrm{Ray}}\sim \mathcal{CN}\left(\mathbf{0}_{N\times N},\mathbf{R}'_{\mathrm{Ray}}\right)$. Moreover, when the channel consists of a superposition of LoS and NLoS components, the channel model for the MIMO system follows a Rician distribution and is expressed by \cite{jiang2024high,tian2025analytical}
\begin{align}\label{eq45}
\mathbf{H}_{\mathrm{NF}}^{\mathrm{Ric}}\triangleq\sqrt{\frac{\kappa }{\kappa +1}}\mathbf{\bar{H}}_{\mathrm{NF}}^{\mathrm{LoS}}+\sqrt{\frac{1 }{\kappa +1}}\mathbf{\bar{H}}_{\mathrm{NF}}^{\mathrm{NLoS}},
\end{align}
where $\kappa$ is Rician factor. Furthermore, incorporating the Doppler factor $\exp\{j2\pi f_Dt\}$ into the quasi-static channel models in \eqref{eq15}-\eqref{eq20} and \eqref{eq43}-\eqref{eq45} allows for a natural extension to dynamic channel scenarios \cite{Wang2023A,Zheng2023Ultra,wang2025enhanced}.

By comparing the previously discussed near-field array response and channel models with their conventional far-field counterparts, it can be observed that the fundamental distinction lies in their phase structure and dominant parameter dimensions. Near-field channels are jointly characterized by angular and distance parameters, leading to a non-linear phase distribution across the array. This additional distance dependency introduces extra spatial degrees of freedom (DoF) for system design. In contrast, under the plane-wave assumption, far-field channels exhibit a linear phase variation, and their behavior is governed solely by the DoA. Moreover, as indicated by \eqref{eq14FF}, the far-field model can be interpreted as an asymptotic special case of the near-field formulation. When the propagation distance $r$ becomes sufficiently large such that the higher-order terms in the distance difference are negligible, the spherical-wave model reduces to the conventional plane-wave representation. It is worth noting that, due to the non-linear phase structure inherent in near-field propagation, the near-field LoS MIMO channel matrices typically exhibit a higher rank, resulting in increased DoF: $\text{DoF}^{\text{LoS}}_{\text{NF}}=\text{rank}\{\mathbf{\bar{H}}_{\mathrm{NF}}^{\mathrm{LoS}}\}\geq 1$. In contrast, the far-field LoS MIMO channel matrix always has rank one, leading to limited DoF, i.e., $\text{DoF}^{\text{LoS}}_{\text{FF}}=\text{rank}\{\mathbf{\bar{H}}_{\mathrm{FF}}^{\mathrm{LoS}}\}= 1$. For clarity, we summarize the characteristics of different channels in Table~\ref{TableIII}.



%
\setlength{\abovecaptionskip}{-0.0cm}
\setlength{\belowcaptionskip}{-0.6cm}
\begin{figure}[t!]
\centering
\includegraphics[width=5.3cm,height=6.8cm]{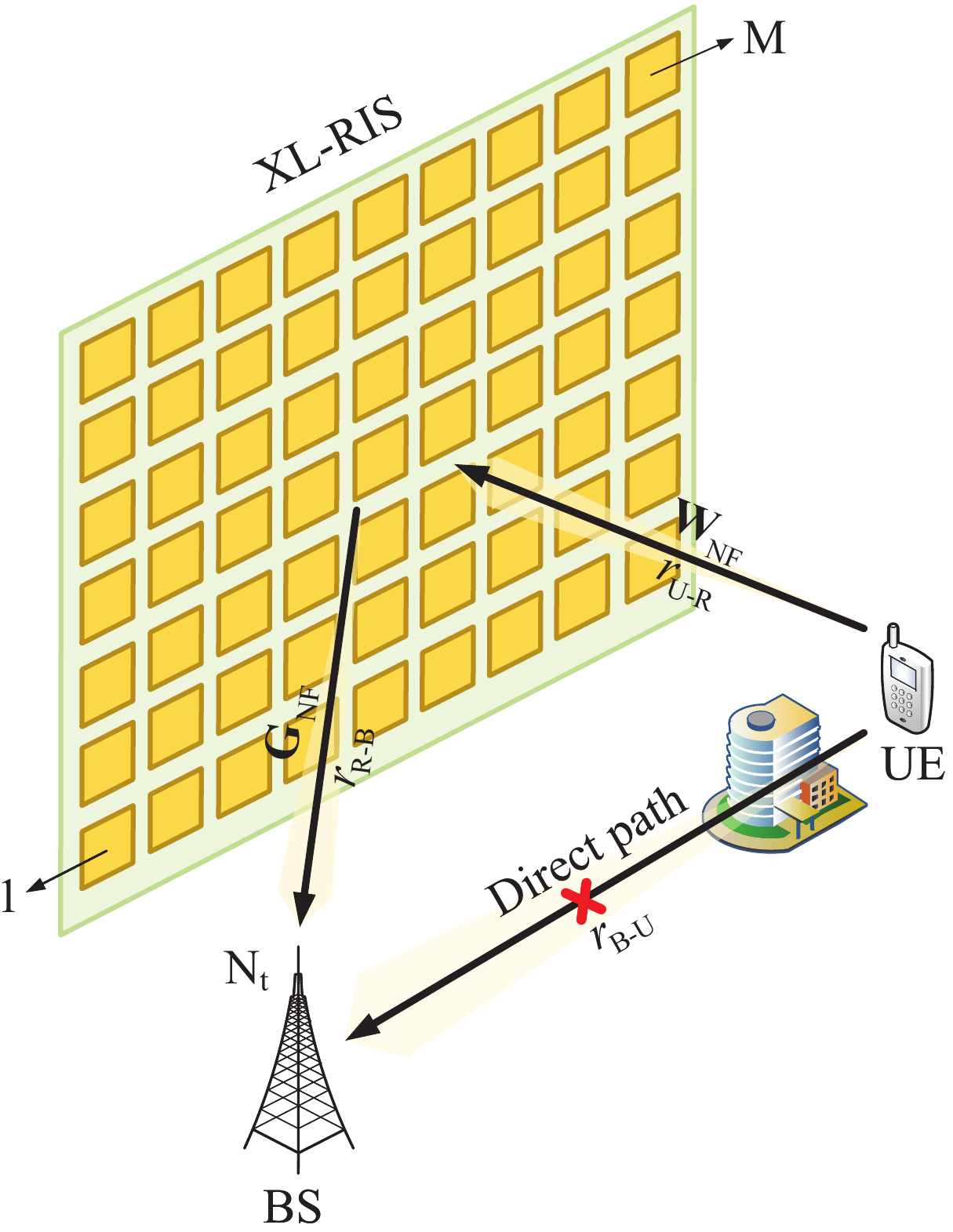}
\caption{The system model for the XL-RIS-aided communications.}
\label{Fig.5}
\end{figure}

\subsection{XL-RIS-aided Near-Field Cascaded Channel Model}

\begin{table*}[t!]
\captionsetup{skip=0pt}
\centering
\caption{Far-field vs. near-field channel comparison in single-user scenarios.}
\begin{small}
\begin{tabular}{|c|c|c|c|cc|c|}
\hline System & Channel Model & Characteristic & Main Factors & \multicolumn{2}{c|}{Type} & DoFs \\
\hline \multirow{8}{*}{MISO} & \multirow{4}{*}{Far-field} & \multirow{4}{*}{Linear phase} & \multirow{4}{*}{DoA} & \multicolumn{2}{c|}{LoS} & 1 \\
\cline{5-7} & & & & \multicolumn{1}{c|}{\multirow{2}{*}{NLoS}} & Multipath & 1 \\ \cline{6-7} & & & & \multicolumn{1}{c|}{} & Rayleigh & 1 \\
\cline{5-7} & & & & \multicolumn{2}{c|}{Rician} & 1 \\
\cline{2-7} & \multirow{4}{*}{Near-field} & \multirow{4}{*}{Non-linear phase} & \multirow{4}{*}{DoA and distance} & \multicolumn{2}{c|}{LoS} & 1 \\
\cline{5-7} & & & & \multicolumn{1}{c|}{\multirow{2}{*}{NLoS}} & Multipath & 1 \\
\cline{6-7} & & & & \multicolumn{1}{c|}{} & Rayleigh & 1 \\
\cline{5-7} & & & & \multicolumn{2}{c|}{Rician} & 1 \\
\hline \multirow{8}{*}{MIMO} & \multirow{4}{*}{Far-field} & \multirow{4}{*}{Linear phase} & \multirow{4}{*}{DoA} & \multicolumn{2}{c|}{LoS} & 1 \\
\cline{5-7} & & & & \multicolumn{1}{c|}{\multirow{2}{*}{NLoS}} & Multipath & $\geq1$ \\
\cline{6-7} & & & & \multicolumn{1}{c|}{} & Rayleigh & $\geq1$ \\
\cline{5-7} & & & & \multicolumn{2}{c|}{Rician} & $\geq1$ \\
\cline{2-7} & \multirow{4}{*}{Near-field} & \multirow{4}{*}{Non-linear phase} & \multirow{4}{*}{DoA and distance} & \multicolumn{2}{c|}{LoS} & $\geq1$ \\
\cline{5-7} & & & & \multicolumn{1}{c|}{\multirow{2}{*}{NLoS}} & Multipath & $\geq1$ \\
\cline{6-7} & & & & \multicolumn{1}{c|}{} & \multicolumn{1}{l|}{Rayleigh} & $\geq1$ \\
\cline{5-7} & & & & \multicolumn{2}{c|}{Rician} & $\geq1$ \\ \hline \end{tabular} \end{small}
\label{TableIII}
\end{table*}

When an XL-RIS is deployed in a scenario where the direct path between the BS and the UE is unobstructed, taking the uplink transmission as an example, the signal transmitted by the UE generally experiences lower attenuation along the direct path than along the RIS-reflected path (since $r_{\text{B-U}} < r_{\text{U-R}}+ r_{\text{R-B}}$). As the operating frequency increases, the propagation loss rises markedly, thereby further exacerbating the attenuation of the RIS-reflected link. Consequently, when the direct path is available, the performance gain achieved by introducing an XL-RIS is usually limited. As illustrated in Fig. \ref{Fig.5}, an XL-RIS is therefore more beneficial in scenarios where obstacles block the direct link between the BS and the UE; by intelligently adjusting the phase-shifts of its reflecting elements, it can effectively redirect the UE’s signal toward the BS and compensate for the performance degradation caused by direct path blockage.

When the UE is equipped with a single antenna, while the BS and RIS are equipped with $N_t$ and $M$ elements, respectively, the channel model of the RIS-aided system consists of two components: the channel $\mathbf{w}_{\mathrm{NF}}\in\mathbb{C}^{M}$ from the UE to the RIS, and the channel $\mathbf{G}_{\mathrm{NF}}\in\mathbb{C}^{N_t\times M}$ between the RIS and the BS. Thus, the cascaded channel from the UE to the BS, reflected by the RIS, can be expressed as \cite{dovelos2021intelligent,wu2022near,tang2023near,liu2023low,jin2023near,chen2024channel,rahal2024ris,yuan2024near,schroeder2024near,tian2024near}
\begin{align}\label{eq515}
\mathbf{x}_{\mathrm{NF}} &=\mathbf{G}_{\mathrm{NF}}\mathbf{\Theta} \mathbf{w}_{\mathrm{NF}},
\end{align}
where $\mathbf{\Theta}$ $\in$ $\mathbb{C}^{M\times M}$ is the controllable reflection-coefficient matrix of the RIS. It is worth noting that when the inter-element coupling of the RIS can be neglected, the RIS reflection-coefficient matrix $\mathbf{\Theta}$ can be further written in a diagonal form, i.e., $\mathbf{\Theta}=\mathrm{diag}(\boldsymbol{\phi})$, where $\boldsymbol{\phi}\in \mathbb{C}^{M}$ is the vector collecting the controllable RIS phase-shifts $\{{\phi}_m \in[0,2\pi); m=1,\ldots,M\}$. Accordingly, \eqref{eq515} can be simplified as
\begin{align}\label{eq51}
\mathbf{x}_{\mathrm{NF}} =\mathbf{G}_{\mathrm{NF}}\mathrm{diag}(\boldsymbol{\phi}) \mathbf{w}_{\mathrm{NF}}=\mathbf{G}_{\mathrm{NF}}\mathrm{diag}(\mathbf{w}_{\mathrm{NF}})\boldsymbol{\phi}=\mathbf{h}_{\text{cas}}\boldsymbol{\phi},
\end{align}
where $\mathbf{h}_{\text{cas}}=\mathbf{G}_{\mathrm{NF}}\mathrm{diag}(\mathbf{w}_{\mathrm{NF}})$. When the UE is equipped with $N_r$ elements, the cascaded channel model from the UE to the RIS and then to the BS is given by \cite{bartoli2023spatial,zhang2024ris,lee2025near,lv2024ris}
\begin{align}\label{RIS2}
\mathbf{X}_{\mathrm{NF}} =\mathbf{G}_{\mathrm{NF}}\mathbf{\Theta} \mathbf{W}_{\mathrm{NF}},
\end{align}
where $\mathbf{W}_{\mathrm{NF}}\in\mathbb{C}^{M\times N_r}$ is the UE-RIS channel. The specific modeling of channels $\mathbf{G}_{\mathrm{NF}}$ and $\mathbf{w}_{\mathrm{NF}}$ (or $\mathbf{W}_{\mathrm{NF}}$) is derived by taking into account the geometric configuration of the arrays employed at the UE, BS, and RIS, as outlined in Section III-A and Section III-B.

\subsection{Simulation Tools for Near-Field Channel Modeling}

A number of representative simulation platforms have been developed to provide effective support for near-field channel modeling. Specifically, the SEU-PML-6GPCS simulator \cite{SEUPML6GPCS}, jointly developed by Southeast University and Purple Mountain Laboratories, adopts a unified modeling framework that consistently accounts for both far-field plane-wave and near-field spherical-wave propagation, thereby enabling accurate near-field modeling of continuous-space channels across all candidate 6G frequency bands and a wide range of deployment scenarios. The QuaDRiGa channel simulator developed at the Fraunhofer Heinrich Hertz Institute is built upon GBSMs and incorporates an optional spherical-wave propagation mode, thereby enabling ELAA-driven near-field channel simulations \cite{QuaDRiGaChannelModel}. The BUPTCMCCCMG-IMT2030 platform \cite{BUPTCMCCCMG}, jointly released by Beijing University of Posts and Telecommunications and China Mobile, further extends the conventional GBSM framework by integrating ELAA-based near-field spherical-wave propagation and spatial non-stationarities, and can therefore be employed for channel modeling in key 6G near-field scenarios such as XL-MIMO, XL-RIS, and ISAC. For THz-band XL-MIMO systems, the TeraMIMO platform provides both plane-wave and spherical-wave propagation models \cite{TeraMIMO}, thereby accurately capturing near-field THz propagation effects such as wavefront curvature and wideband beam split effects. In parallel, NirvaWave \cite{NirvaWave}, developed at Princeton University, leverages scalar diffraction theory and Fourier optics to perform high-fidelity numerical simulations of near-field wavefronts for large-aperture arrays in complex electromagnetic environments, generating high-resolution near-field electromagnetic field distributions and equivalent channels, and thereby offering a fine-grained modeling tool for emerging near-field ISAC applications.

\setlength{\abovecaptionskip}{0.05cm}
\setlength{\belowcaptionskip}{-0.45cm}
\begin{figure}[t!]
\centering
\includegraphics[scale=0.39]{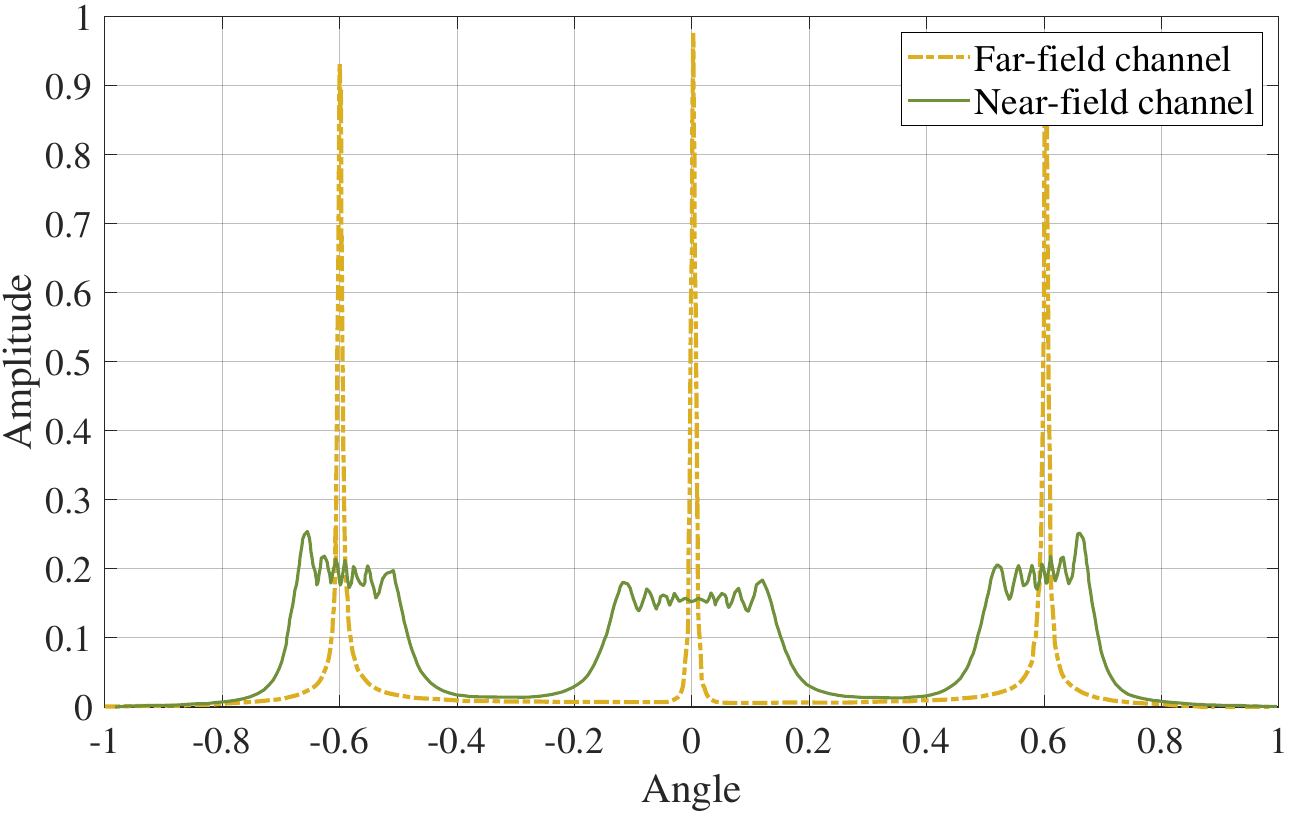}
\caption{Comparison of the angular-domain representations between the far-field and near-field channels~\cite{cui2022channel}.}
\label{Fig10}
\end{figure}

\subsection{Summary and Open Challenges}

In this Section, we review representative near-field channel models. Starting from the near-field array response vectors for different array configurations, we derive near-field LoS, multipath, spatially correlated Rayleigh fading, and Rician channel models, and compare these models with their classical far-field counterparts to elucidate their fundamental differences. Moreover, we introduce several representative near-field channel simulation platforms, which offer strong simulation support for signal processing tasks, especially the subsequent near-field channel estimation. Although both this section and the literature \cite{Liu2023Near, Gong2024Near} have established preliminary deterministic and statistical modeling frameworks for near-field channels, several critical challenges remain unresolved. The key challenges are summarized as follows \cite{Gong2024Near}:

\textbullet\ \textbf{Spatial non-stationarity awareness}: In near-field transmission with ELAAs, spatial non-stationarity is a fundamental characteristic that must be accurately incorporated into channel models \cite{Wu2014A}. Capturing this property often requires extensive scenario-specific measurement data, thereby motivating the development of non-stationarity-aware channel modeling approaches.

\textbullet\ \textbf{Concise channel model with controllable dimensionality}: Considering typical near-field scenarios such as XL-MIMO and XL-RIS, near-field channel models may become extremely high-dimensional, leading to significant computational overhead and costly parameter estimation. Hence, developing concise models with controllable dimensions is essential. However, achieving a balance between model simplicity and accuracy remains a key challenge.

\textbullet\ \textbf{Mutual coupling awareness}: In near-field systems with CAP arrays, mutual coupling between antenna elements is generally non-negligible. The radiated field from each element influences its neighboring elements, and these interactions propagate recursively across the array, altering the overall channel response. Therefore, incorporating mutual coupling into channel models is essential for accurate characterization and effective exploitation of near-field behavior.

\textbullet\ \textbf{Vectorial wavefront inclusion}: In addition to spherical wavefronts, vector wavefields are a fundamental characteristic of near-field channels and must be accurately captured in the modeling process. However, most existing models are built on scalar wavefield assumptions, which overlook vectorial effects. Furthermore, near-field channel models should account for polarization effects.

In summary, addressing the above challenges is of great significance for advancing near-field channel modeling theory and provides a clear roadmap for future research in this area.

\section{Channel Estimation Techniques for Near-Field Wireless Communications}

In the previous section, we systematically reviewed the typical near-field UE-BS direct channel models and the RIS-aided cascaded channel models. With the Rayleigh distance being extended in typical scenarios such as XL-MIMO and XL-RIS, the channel between the transmitter and receiver is more likely to exhibit spherical wavefront characteristics over the dominant propagation region, thereby introducing an additional distance dimension into the channel model. The inclusion of this distance dimension not only alters the spatial characteristics of the channel but also poses greater challenges for channel estimation, as detailed below:

\textbullet\ \textbf{Significantly increased computational overhead}: Due to the extremely large number of antennas in ELAAs, especially at high carrier frequencies where the antenna size shrinks, near-field channel estimation faces a severe scalability issue. In particular, conventional antenna-wise estimation approaches require  processing high-dimensional channel vectors whose dimension scales with the array size, resulting in significant computational burden.
Although LMMSE-based estimators \cite{LongMMSE2024, Long2025Near} can in principle be applied to near-field channels, their practical deployment becomes challenging, as the inversion of high-dimensional correlation matrices entails prohibitive computational complexity.

\textbullet\ \textbf{Non-sparse in the angular domain}: In a codebook designed based on near-field spherical-wave characteristics, each codeword must correspond to a spherical wave and be associated with a specific angle and distance. The spherical-wave nature of near-field transmission leads to significant energy dispersion in the angular domain, thereby weakening the angular sparsity \cite{cui2022channel}, as shown in Fig. \ref{Fig10}. Consequently, although compressive sensing algorithms \cite{tao2018regularized,wei2020deep,Qaisar2013Compressive} can effectively reduce computational complexity and pilot overhead, far-field schemes relying on angular sparsity become unsuitable for near-field channel estimation.

\setlength{\abovecaptionskip}{-0.1cm}
\setlength{\belowcaptionskip}{-0.4cm}
\begin{figure}[t!]
\centering
\includegraphics[scale=0.72]{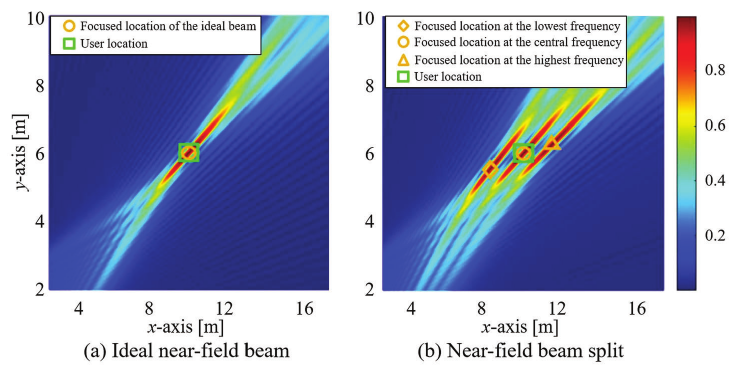}
\caption{The schematic diagram of the near-field beam split effect, reproduced from~\cite{cui2022near}, \textcopyright\ IEEE.}
\label{Fig9}
\end{figure}

\textbullet\ \textbf{Beam split under wideband operation}: In conventional phase-shifter-based communication systems, analog phase shifters are typically designed to be approximately frequency-flat within the communication bandwidth \cite{Cui2023Near2}. However, the actual phase of an electromagnetic wave depends on both the signal propagation delay and its frequency. Consequently, using such frequency-flat phase shifters for beam focusing in near-field wideband systems causes spherical waves at different subcarriers to focus at distinct spatial locations, leading to the so-called beam split or beam misfocus phenomenon \cite{Zhang20236G}, as illustrated in Fig. \ref{Fig9}. Notably, the near-field beam split effect is more pronounced than its far-field counterpart, as beams disperse simultaneously across the angular and distance domains \cite{lu2024tutorial}. This results in frequency-dependent variations in near-field channel responses, necessitating the joint estimation of angular and distance parameters across subcarriers, which substantially increases the model dimensionality and estimation computational complexity. In addition, beam split leads to phase inconsistencies in the signals received by antenna elements across subcarriers, thereby degrading the accuracy of near-field spherical-wave channel modeling and significantly increasing estimation errors. Moreover, due to frequency-dependent focal shifts, pilot signals may fail to cover all subcarriers’ energy peaks, leading to larger estimation errors for certain subcarriers.

\begin{figure}[t!]
\centering
\includegraphics[scale=0.47]{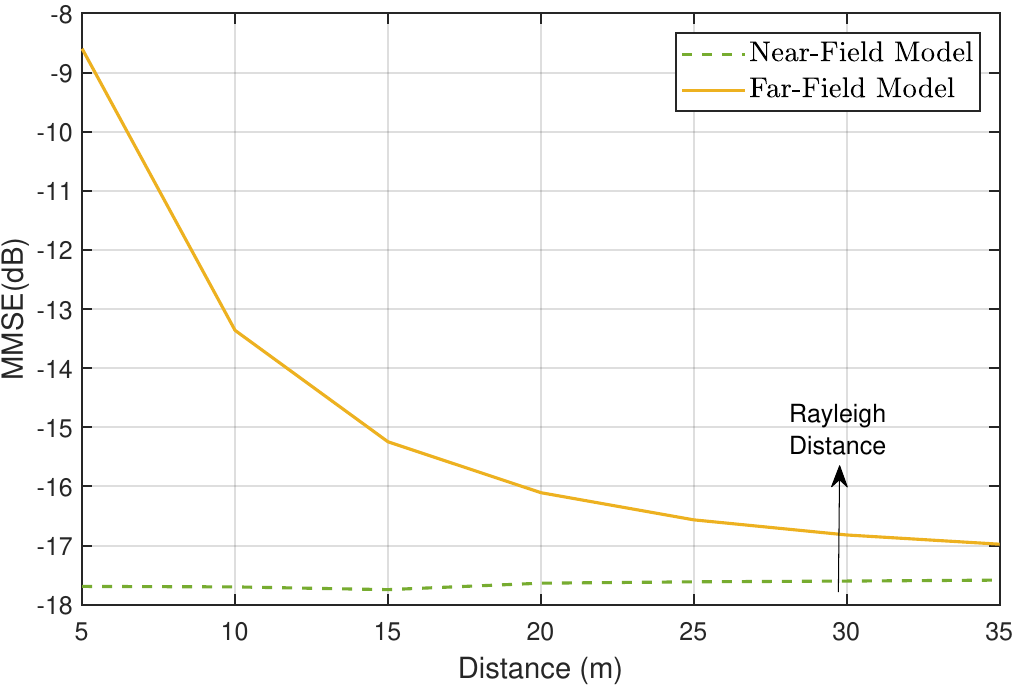}
\caption{Comparison of near-field channel estimation based on near-field and far-field models \cite{Long2025Near}.}
\label{Fig11}
\end{figure}

Building upon the above discussion, accurately characterizing near-field channel properties is a fundamental prerequisite for achieving high-accuracy channel estimation. As illustrated in Fig. \ref{Fig11}, we perform near-field channel estimation under both the conventional far-field assumption and the accurate near-field spherical-wave model. The results clearly demonstrate that incorporating the spherical-wave characteristics of near-field propagation is essential for improving estimation accuracy. Therefore, in this section, we will provide a comprehensive overview of existing near-field channel estimation methods. By considering various system configurations, we systematically analyze solutions tailored to single-user, multi-user, single-carrier, and multi-carrier scenarios. These methods aim to balance complexity and performance, offering technical support for the efficient implementation of near-field communication systems.

%
%

\subsection{BS-UE Near-Field Channel Estimation}

In general, when the UE is equipped with a single antenna, the received uplink pilot signal at the BS can be expressed as \cite{demir2024spatial,long2025parametric}
\begin{align}\label{pilot1}
   \mathbf{y} =\sqrt{\rho}\mathbf{h}_{\mathrm{NF}}+\mathbf{z},
\end{align}
where $\mathbf{h}_{\mathrm{NF}}$ denotes the UE-BS channel, which can follow an arbitrary channel model, $\rho$ is the pilot SNR and $\mathbf{z}\sim \mathcal{N}_{\mathbb{C}}(\mathbf{0}, \mathbf{I}_N)$. Conventional LS and LMMSE estimators can still be applied in the near-field case to obtain the corresponding channel estimates \cite{demir2024spatial,long2025parametric}, given by
\begin{align}\label{LS}
   \mathbf{\widehat{h}}^{\mathrm{LS}}_{\mathrm{NF}} =\frac{\mathbf{y}}{\sqrt{\rho}},
\end{align}
and
\begin{align}\label{LMMSE}
   \mathbf{\widehat{h}}^{\mathrm{LMMSE}}_{\mathrm{NF}} =\sqrt{\rho}\mathbf{R}_h(\rho\mathbf{R}_h+\mathbf{I}_N)^{-1}\mathbf{y},
\end{align}
where $\mathbf{R}_h$ is the spatial correlation matrix of $\mathbf{h}_{\mathrm{NF}}$.
%
%
However, on the one hand, LS essentially treats the channel as an unstructured high-dimensional vector to be recovered.
When the number of observations is limited, applying LS to ELAA-driven high-dimensional near-field channels often leads to large estimation errors.
On the other hand, although LMMSE is optimal when accurate second-order statistical priors are available, in near-field scenarios with ELAAs, channel statistics often exhibit spatial non-stationarity, which makes the acquisition of the spatial correlation matrix costly and prone to prior mismatch. Meanwhile, the matrix inversion required by LMMSE also incurs substantial computational overhead. Therefore, in near-field scenarios, the above conventional estimators are typically not the most performance-advantageous choices. Motivated by these observations, we will next discuss UE-BS channel estimation methods tailored to near-field characteristics and provide a systematic summary and analysis of existing studies.

\setlength{\abovecaptionskip}{-0.0cm}
\setlength{\belowcaptionskip}{-0.4cm}

\begin{table*}[t]
\captionsetup{skip=3pt}
\caption{Estimation methods for BS-UE near-field channels in single-user systems.}
\resizebox{\textwidth}{!}{
\begin{tabular}{|cc|c|cc|c|c|}
\hline
\multicolumn{2}{|c|}{\textbf{System Type}}                                                       & \textbf{Channel Type}      & \multicolumn{2}{c|}{\textbf{Algorithm}}                                                                                                                                                                                                               & \textbf{Features}                                                                                          & \textbf{Reference}                                                          \\ \hline
\multicolumn{1}{|c|}{\multirow{11}{*}{\textbf{Single-carrier}}} & \multirow{8}{*}{\textbf{MISO}} & \multirow{3}{*}{LoS}       & \multicolumn{1}{c|}{\multirow{3}{*}{Parametric estimation method}}                                                     & 3-D MUSIC algorithm                                                                                                          & High accuracy and high complexity                                                                          & \cite{long2025parametric}                                  \\ \cline{5-7}
\multicolumn{1}{|c|}{}                                          &                                &                            & \multicolumn{1}{c|}{}                                                                                                  & Two-stage MUSIC algorithm                                                                                                    & \multirow{2}{*}{High accuracy and low complexity}                                                          & \cite{He2012Efficient,Kosasih2023Parametric,Qu2024Two}     \\ \cline{5-5} \cline{7-7}
\multicolumn{1}{|c|}{}                                          &                                &                            & \multicolumn{1}{c|}{}                                                                                                  & SADCE method                                                                                                                 &                                                                                                            & \cite{huang2023low}                                        \\ \cline{3-7}
\multicolumn{1}{|c|}{}                                          &                                & \multirow{4}{*}{Multipath} & \multicolumn{1}{c|}{\multirow{2}{*}{Sparsity-aware method}}                                                            & \begin{tabular}[c]{@{}c@{}}Subarray-wise and the scattering-wise \\ estimation methods based on the refined OMP\end{tabular} & Enhanced estimation accuracy                                                                               & \cite{han2020channel}                                      \\ \cline{5-7}
\multicolumn{1}{|c|}{}                                          &                                &                            & \multicolumn{1}{c|}{}                                                                                                  & TPD framework                                                                                                                & \begin{tabular}[c]{@{}c@{}}Leakage-free parameter estimation \\ and low complexity\end{tabular}            & \cite{guo2023compressed}                                   \\ \cline{4-7}
\multicolumn{1}{|c|}{}                                          &                                &                            & \multicolumn{1}{c|}{\begin{tabular}[c]{@{}c@{}}Sparsity-aware +\\  deep learning method\end{tabular}}                  & SDL-LISTA algorithm                                                                     & Higher accuracy compared with OMP                                     & \cite{Zhang2023model} \\ \cline{4-7}
\multicolumn{1}{|c|}{}                                          &                                &                            & \multicolumn{2}{c|}{VNNCE method}                                                                                                                                                                                                                     & Estimation accuracy approaching the CRLB                                                                   & \cite{cao2023joint,cao2024newtonized}                      \\ \cline{3-7}
\multicolumn{1}{|c|}{}                                          &                                & Rayleigh                   & \multicolumn{2}{c|}{RS-LS estimator}                                                                                                                                                                                                                  & \begin{tabular}[c]{@{}c@{}}Validation of conventional estimators \\ in the near-field case\end{tabular}    & \cite{demir2024spatial}                                    \\ \cline{2-7}
\multicolumn{1}{|c|}{}                                          & \multirow{3}{*}{\textbf{MIMO}} & Multipath                  & \multicolumn{1}{c|}{\multirow{3}{*}{Sparsity-aware method}}                                                            & Eigen-dictionary based on DPSS                                                                                               & \begin{tabular}[c]{@{}c@{}}Substantial reductions \\ in baseband sampling and dictionary size\end{tabular} & \cite{liu2025sensing}                                      \\ \cline{3-3} \cline{5-7}
\multicolumn{1}{|c|}{}                                          &                                & \multirow{2}{*}{Rician}    & \multicolumn{1}{c|}{}                                                                                                  & OMP-based two-stage channel estimation                                                                                       & Performance approaching the CRLB                                                                           & \cite{lu2023near1}                                         \\ \cline{5-7}
\multicolumn{1}{|c|}{}                                          &                                &                            & \multicolumn{1}{c|}{}                                                                                                  & 3S-MMV framework based on unified OMP                                                                                        & Low complexity                                                                                             & \cite{shi2024double}                                       \\ \hline
\multicolumn{1}{|c|}{\multirow{8}{*}{\textbf{Multi-carrier}}}   & \multirow{6}{*}{\textbf{MISO}} & LoS                        & \multicolumn{2}{c|}{Time-frequency domain search}                                                                                                                                                                                                     & Reduced pilot overhead                                                                                     & \cite{cui2022near}                                         \\ \cline{3-7}
\multicolumn{1}{|c|}{}                                          &                                & \multirow{5}{*}{Multipath} & \multicolumn{1}{c|}{\multirow{2}{*}{Sparsity-aware method}}                                                            & Adaptive JSBL-CE algorithm                                                                                                   & Reduced codebook overhead                                                                                  & \cite{zhu2024sparse}                                       \\ \cline{5-7}
\multicolumn{1}{|c|}{}                                          &                                &                            & \multicolumn{1}{c|}{}                                                                                                  & AGSBL method                                                                                                                 & Enhanced estimation performance                                                                            & \cite{cheng2019adaptive}                                   \\ \cline{4-7}
\multicolumn{1}{|c|}{}                                          &                                &                            & \multicolumn{1}{c|}{\multirow{2}{*}{Deep learning method}}                                                             & CNN-based approach                                                                                                           & \begin{tabular}[c]{@{}c@{}}Enhanced BER performance \\ and reduced pilot overhead\end{tabular}             & \cite{lee2022intelligent}                                  \\ \cline{5-7}
\multicolumn{1}{|c|}{}                                          &                                &                            & \multicolumn{1}{c|}{}                                                                                                  & Mixed training strategy with DNN architecture                                                                                & Low complexity and high robustness                                                                         & \cite{gao2024deep}                                         \\ \cline{4-7}
\multicolumn{1}{|c|}{}                                          &                                &                            & \multicolumn{1}{c|}{\multirow{2}{*}{\begin{tabular}[c]{@{}c@{}}Sparsity-aware + \\ deep learning method\end{tabular}}} & A-RCE network combined with TFIST                                                                                            & \begin{tabular}[c]{@{}c@{}}Enhanced estimation accuracy \\ and reduced pilot overhead\end{tabular}         & \cite{Wang2025Near}                                        \\ \cline{2-3} \cline{5-7}
\multicolumn{1}{|c|}{}                                          & \multirow{2}{*}{\textbf{MIMO}} & LoS                        & \multicolumn{1}{c|}{}                                                                                                  & D-STiCE method                                                                                                               & Minimal pilot overhead                                                                                     & \cite{kim2024deep}                                         \\ \cline{3-7}
\multicolumn{1}{|c|}{}                                          &                                & Multipath                  & \multicolumn{1}{c|}{Parametric estimation method}                                                                      & Parametric SAGE algorithm                                                                                                    & --                                                                                                         & \cite{liu2025graph}                                        \\ \hline
\end{tabular}}
\label{TableIV}
\end{table*}

\subsubsection{Single-User Single-Carrier Systems}
In this part, we discuss channel estimation methods for the UE-BS link in single-user single-carrier systems, where the mainstream strategies include parametric estimation and compressive sensing techniques.

\textbf{Parametric Estimation Approaches}: For the LoS channel in MISO systems, when the channel can be characterized by a relatively small set of parameters, such as angles and distances, a promising strategy for the UE-BS channel estimation is to first recover the position of the UE, and subsequently incorporate these parameters into the parametric channel model to reconstruct the channel. This forms the basis of a parametric channel estimation approach. In this context, the authors of \cite{long2025parametric} effectively estimated the position of the UE in near-field sub-THz channels by employing a 3-D MUSIC algorithm. To further reduce the computational burden, a two-stage MUSIC algorithm was developed in \cite{He2012Efficient,Kosasih2023Parametric,Qu2024Two}. In this approach, the DoA of the UE is first estimated, followed by the derivation of the distance between the UE and the BS using the acquired angular information. Additionally, the authors of \cite{huang2023low} proposed a sequential angle-distance channel estimation (SADCE) method, which achieves position estimation accuracy comparable to that of the 3-D MUSIC algorithm, while substantially lowering computational complexity. Building upon these estimated position parameters, both \cite{long2025parametric} and \cite{Kosasih2023Parametric} incorporated them into a near-field channel model to ultimately derive accurate channel estimates. Since parametric methods operate directly in the continuous parameter space, they typically offer high estimation accuracy; however, this often comes at the cost of increased computational complexity.

\begin{figure}[t!]
\centering
\includegraphics[scale=0.55]{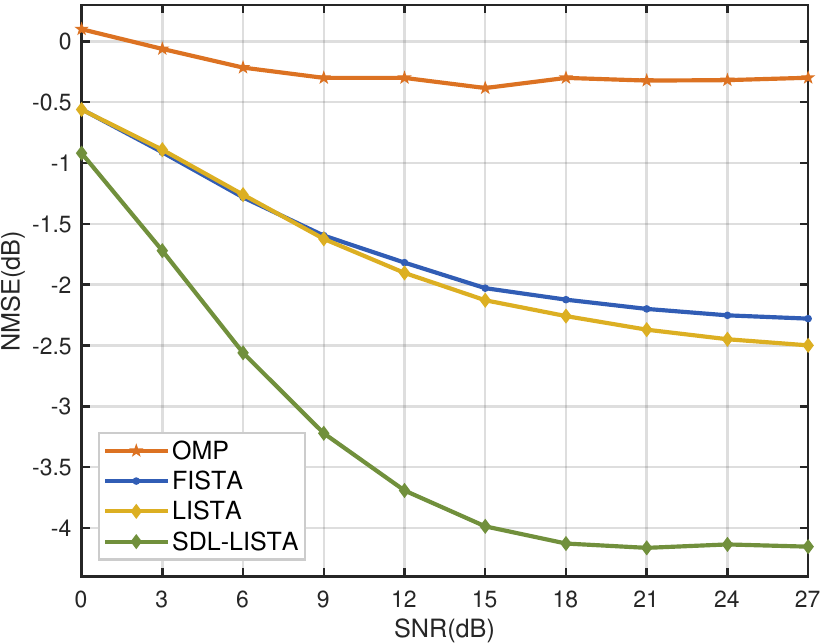}
\caption{The NMSEs vs. SNR using the SDL-LISTA and OMP approaches~\cite{Zhang2023model}.}
\label{Fig16}
\end{figure}

\textbf{Sparsity-Aware (Compressive Sensing) Approaches}: By reconstructing discretized sparse signals in a specific domain, compressive sensing algorithms avoid the high-dimensional search over continuous spaces required by parametric estimation methods, thereby reducing computational complexity. In typical far-field channel estimation, the inherent angular-domain sparsity enables the use of the classical OMP algorithm, which iteratively selects the dictionary codeword most correlated with the current residual and updates the corresponding sparse channel gains via the LS criterion until the termination condition is met. However, when traditional Fourier-based codebooks are used to transform near-field channels into the angular domain, the energy of each near-field path is dispersed across multiple angular bins, which weakens the angular sparsity \cite{Lei2025}. To address this issue, \cite{cui2022channel} first proposed a polar-domain codebook characterized jointly by angular and distance parameters, enabling both near-field and far-field channels to exhibit sparsity in the polar domain. Specifically, in the ELAA-based systems, the number of propagation paths is typically much smaller than the number of antenna elements, i.e., $L\ll N$. This implies that the number of channel parameters to be estimated remains limited, and therefore, the near-field channel remains compressible. For the sparse representation of the near-field channel, as shown in \eqref{eq16}, the multipath channel $\mathbf{h}_{\mathrm{NF}}^{\mathrm{NLoS}}$ is expressed as a weighted sum of a finite number of near-field array response vectors, where each vector $\mathbf{a}(\theta_l,\phi_l,r_l)$ is jointly determined by the angles $(\theta_l,\phi_l)$ and distance $r_l$. Based on this observation, \cite{cui2022channel} designed a novel polar-domain transform matrix $\mathbf{W}\in\mathbb{C}^{N\times Q}$, where $Q$ denotes the number of sampled near-field steering vectors in the polar domain, and each column of $\mathbf{W}$ is formed by a near-field array response vector discretely sampled from the joint angle-distance domain. When $Q\gg L$, this matrix can fully capture the angular and distance characteristics of each path component. Correspondingly, the channel $\mathbf{h}_{\mathrm{NF}}^{\mathrm{NLoS}}$ can be expressed as
\begin{align}\label{eqxs}
\mathbf{h}_{\mathrm{NF}}^{\mathrm{NLoS}}
\simeq\mathbf{W}\boldsymbol{\alpha}_{\mathrm{NF}}^{\mathrm{P}},
\end{align}
where $\boldsymbol{\alpha}_{\mathrm{NF}}^{\mathrm{P}}$
is the sparse channel gain corresponding to $\mathbf{W}$. As $L\ll N \ll Q$, $\boldsymbol{\alpha}_{\mathrm{NF}}^{\mathrm{P}}$ is highly sparse \cite{Zhang2023model}. Leveraging this sparsity, in \cite{han2020channel}, the subarray-wise and the scattering-wise channel estimation methods were proposed in MISO systems. The core idea is to partition the large-aperture array into multiple subarrays, each of which is processed independently using a refined OMP algorithm for accurate scatterer localization. Compared to conventional LS and standard OMP algorithms, this approach results in a significant improvement of the estimation accuracy. \cite{guo2023compressed} introduced a triple parametric decomposition (TPD) framework, which constructs separate covariance matrices for azimuth-elevation angles and propagation distances, so as to achieve an independent parameter decomposition. By employing both on-grid and off-grid techniques, this method facilitates sparse recovery of angle and distance information without the energy leakage effect, while also reducing the complexity associated to the gridding operation. In \cite{Zhang2023model}, a learning iterative shrinkage and thresholding algorithm (LISTA) was proposed for sparse multipath channel estimation. To address the limited estimation accuracy caused by the discretized sparsifying dictionary, the authors further introduced a sparsifying dictionary learning-LISTA (SDL-LISTA) framework that integrates deep learning algorithms with compressed sensing theory. Specifically, the sparsifying dictionary is represented as a neural network layer and jointly optimized within the LISTA architecture. As depicted in Fig.~\ref{Fig16}, the SDL-LISTA algorithm demonstrates superior estimation accuracy compared with conventional non-learning approaches such as OMP.

\setlength{\abovecaptionskip}{-0.0cm}
\setlength{\belowcaptionskip}{-0.4cm}

Extending to MIMO systems, \cite{liu2025sensing} first applied a time reversal algorithm to localize UEs and scatterers in near-field multipath channels, and then designed an eigen-dictionary based on the discrete prolate spheroidal sequences (DPSS) via eigenvalue decomposition (EVD) of the channel matrix. This sparsity-aware design reduces baseband sampling by 51\% and dictionary size by 66\% compared to conventional DFT and spherical dictionaries. For Rician channels, \cite{lu2023near1} proposed a two-stage channel estimation algorithm, where the LoS component is estimated via coarse grid search followed by iterative refinement, while the NLoS components are recovered using OMP by exploiting the sparsity in the polar domain. The Cram\'er-Rao lower bound (CRLB) was also derived to assess the algorithm’s estimation performance. In \cite{shi2024double}, a low-overhead unified OMP (UOMP) estimation method was presented and further extended with a three-stage multiple-measurement-vector (3S-MMV) framework to reduce computational complexity. In summary, compared to parametric estimation methods, compressive sensing approaches typically offer lower computational complexity, although their estimation accuracy is limited by the resolution of the codebook. However, it should be emphasized that the applicability of compressive sensing algorithms in near-field scenarios is limited. Although near-field channels often exhibit sparsity in the polar domain, such sparsity is not absolute. The sparsity in the polar domain strongly depends on the accuracy of the geometric model, such as array geometry, inter-element spacing, and calibration precision; model mismatches would cause energy leakage across multiple angle-distance bins, thereby degrading sparsity. Moreover, the sparsity assumption holds only when the number of propagation paths is much smaller than the number of antenna elements. In dense-scattering or rich-multipath environments, such as indoor reflections, this assumption may face challenges. Particularly, in continuously scattered or diffuse environments, the channel power spectrum is nearly continuous, making discretely sampled polar-domain codebooks inadequate for accurate channel representation. Consequently, compressive sensing algorithms are best suited for low-scattering or quasi-LoS conditions, while their performance deteriorates in complex multipath or statistical channels, such as Rayleigh fading scenarios.

\textbf{Other Approaches}: Apart from parametric and compressive sensing algorithms, works \cite{cao2023joint} and \cite{cao2024newtonized} proposed a variational Newtonized near-field channel estimation (VNNCE) algorithm for the multipath channel in the MISO systems, achieving an estimation accuracy that approaches the relevant CRLB. Moreover, for spatially correlated Rayleigh channels, \cite{demir2024spatial} validated the effectiveness of the reduced subspace least squares (RS-LS)  estimator under near-field conditions.

\subsubsection{Single-User Multi-Carrier Systems}

\setlength{\belowcaptionskip}{-0.2cm}
\begin{figure*}[t!]
    \centering
    \subfigure[]{\includegraphics[width=0.40\textwidth]{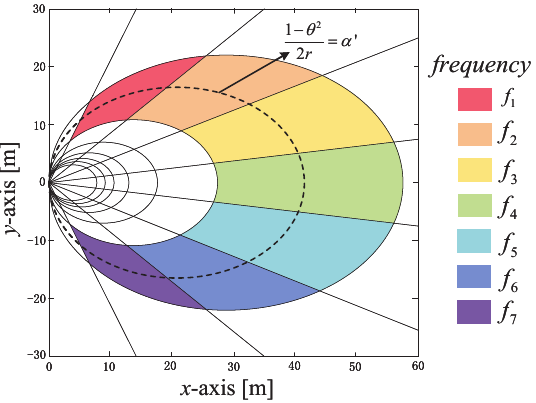}}
    \subfigure[]{\includegraphics[width=0.40\textwidth]{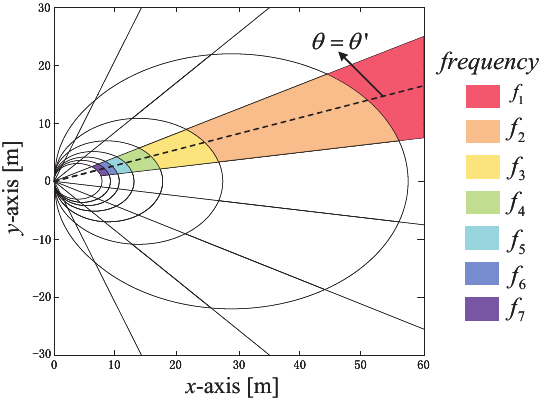}}
    \caption{The schematic diagrams of the near-field rainbow on (a) the angle dimension and (b) the distance dimension \cite{cui2022near}.}
    \label{Fig12}
\end{figure*}

Subsequently, we review channel estimation techniques for single-user multi-carrier systems, focusing on compressive sensing, deep learning-based approaches, and parametric estimation strategies.

\textbf{Time-Frequency Domain Search Approach}: As discussed earlier, due to the inherent limitations of frequency-flat phase shifters, spherical waves in wideband systems give rise to the near-field beam split effect. Regarding its impact on channel estimation, existing studies fall into two categories: beam split exploitation and beam split suppression techniques. As a representative work in the former category, \cite{cui2022near} first demonstrated that near-field time-delay (TD) beamforming can flexibly control the focusing of near-field beams at different frequencies onto multiple spatial angles and distances by adjusting the TD parameters in the angular and distance domains. This produces a rainbow-like multi-focal distribution, as illustrated in Fig. \ref{Fig12}, which cannot be achieved by conventional far-field beamforming. By analogy to the dispersion of white light through a prism, \cite{cui2022near} termed this controllable multi-focal distribution as the ``near-field rainbow effect''. Furthermore, \cite{cui2022near} proposed a fast channel estimation method tailored for wideband near-field LoS channels. By slightly adjusting the TD parameters across different time slots, the proposed scheme performs frequency-domain searches for the optimal angle and time-domain searches for the optimal distance, thereby enabling rapid acquisition of CSI.

\setlength{\abovecaptionskip}{-0.0cm}
\setlength{\belowcaptionskip}{-0.4cm}
\begin{figure}[t!]
\centering
\includegraphics[scale=0.47]{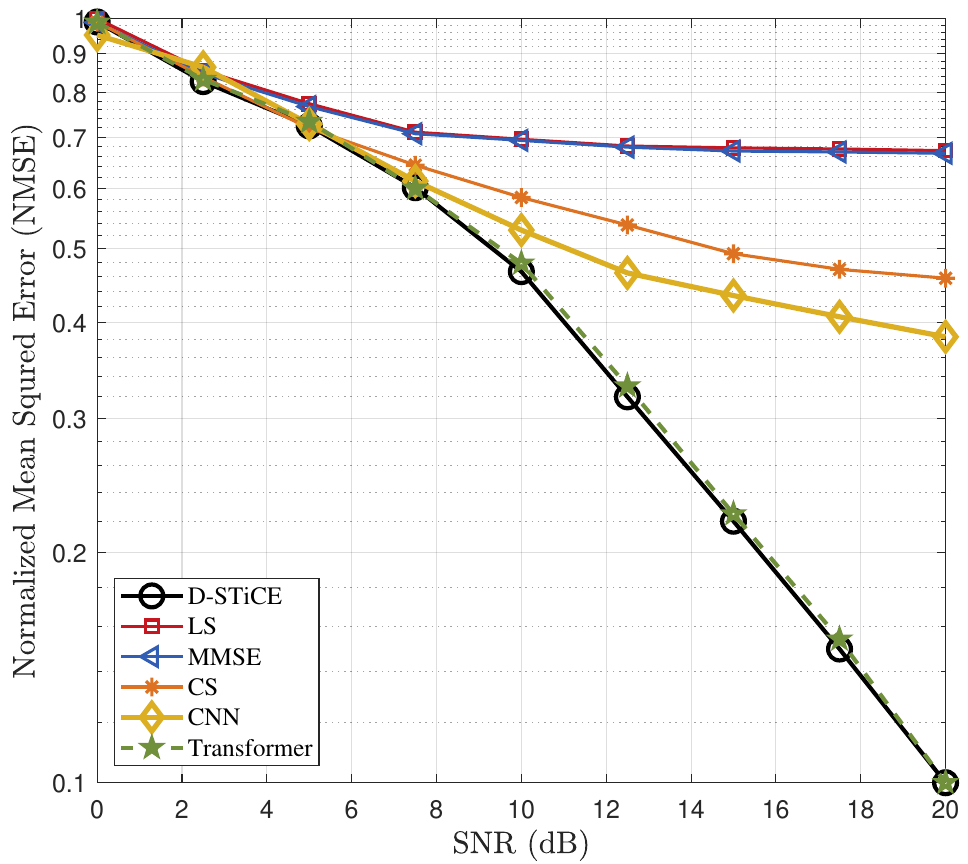}
\caption{The NMSEs vs. SNR using different channel estimation approaches~\cite{kim2024deep}.}
\label{Fig15}
\end{figure}

\textbf{Sparsity-Aware (Compressive Sensing) Approaches}: Moreover, sparsity-aware also provide an effective approach to alleviate the beam split effect in MISO-OFDM systems. The authors of \cite{zhu2024sparse} proposed an adaptive joint sparse Bayesian learning channel estimation (JSBL-CE) algorithm, which enables precise channel parameter estimation with reduced codebook overhead. In addition, \cite{cheng2019adaptive} introduced an adaptive group sparse Bayesian learning (AGSBL) method that incorporates a hierarchical prior model governed by tunable hyperparameters and employs a variational Bayesian algorithm. This method adaptively groups channels with similar sparsity patterns and exploits the partial joint sparsity within each group to achieve accurate channel estimation. However, the frequency dependence of channels in wideband systems increases the dimensionality of the parameters to be estimated, which may lead the aforementioned methods to encounter higher computational complexity when acquiring CSI for all active subcarrier channels.

\textbf{Deep Learning Approaches}: By constructing deep neural networks, deep learning algorithms can automatically learn the nonlinear mapping between jointly encoded features, such as angle, distance, delay, and frequency, and the channel coefficients from large amounts of pilot data. In this way, they can characterize the frequency-dependent properties of wideband near-field channels in an end-to-end manner with relatively low computational complexity, thereby offering a novel and efficient solution for wideband near-field channel estimation. In \cite{lee2022intelligent}, a convolutional neural network (CNN)-based approach was proposed to estimate  multiple channel parameters, including angles, distances, time delay and complex gains. This method outperforms conventional compressed sensing-based channel estimation algorithms in terms of bit error rate (BER) and pilot overhead. Furthermore, \cite{gao2024deep} proposed a deep unfolding-based algorithm specifically designed for channel estimation in multipath scenarios. This method mitigates the near-field beam split effect by employing frequency-dependent near-field dictionaries. A deep neural network (DNN) is also incorporated to learn optimal parameter updates in each iteration. Moreover, a mixed training strategy, combining the DNN architecture and loss function, is developed to enable low-complexity and highly robust estimation under various scenarios. To further leverage the complementary strengths of deep learning and sparsity-aware approaches, \cite{Wang2025Near} proposed a deep learning-based joint channel estimation and sparse reconstruction scheme. An attention mechanism-based residual channel estimation (A-RCE) network is designed to enhance estimation accuracy by leveraging the inherent spatial correlations of the channel matrix across subcarriers and antennas. In parallel, a trainable fast iterative shrinkage-thresholding (TFIST) network is introduced to exploit the polar-domain sparsity of near-field channels, enabling a low-dimensional sparse representation and thereby reducing the overhead associated with CSI feedback.
Extending to the MIMO-OFDM scenario, \cite{kim2024deep} combined the deep learning algorithm with compressed sensing theory and proposed a deep sparse time-varying channel estimation (D-STiCE) method for the THz LoS channel. By leveraging long short-term memory (LSTM), this approach effectively captures the temporally correlated features of sparse channel parameters, enabling accurate estimation with minimal pilot overhead. As illustrated in Fig. \ref{Fig15}, the proposed method demonstrates significantly higher estimation accuracy compared with conventional LS and MMSE estimators as well as compressive sensing-based approaches.

\textbf{Parametric Estimation Approach}: Apart from the above methods, the authors of \cite{liu2025graph} proposed a parametric space-alternating generalized expectation-maximization (SAGE) algorithm for MIMO-OFDM systems to effectively estimate the multi-bounce angular and range parameters in spatially non-stationary multipath channels.

\subsubsection{Multi-User Single-Carrier Systems}

This part surveys UE-BS channel estimation methods in multi-user single-carrier scenarios, where compressive sensing constitutes the predominant methodology.

\begin{table*}[]
\captionsetup{skip=3pt}
\caption{Estimation methods for BS-UE near-field channels in multi-user systems.}
\resizebox{\textwidth}{!}{
\begin{tabular}{|cc|c|cc|c|c|}
\hline
\multicolumn{2}{|c|}{\textbf{System Type}}                                                       & \textbf{Channel Type}       & \multicolumn{2}{c|}{\textbf{Algorithm}}                                                                                                                                                                        & \textbf{Features}                                                                                                                                                               & \textbf{Reference}                          \\ \hline
\multicolumn{1}{|c|}{\multirow{6}{*}{\textbf{Single-carrier}}} & \multirow{5}{*}{\textbf{MISO}}  & LoS                         & \multicolumn{1}{c|}{\multirow{4}{*}{Sparsity-aware method}}                                          & Belief-based OMP method                                                                                 & Superior accuracy at low SNR                                                                                                                                                    & \cite{tang2024joint}       \\ \cline{3-3} \cline{5-7}
\multicolumn{1}{|c|}{}                                         &                                 & \multirow{4}{*}{Multipath}  & \multicolumn{1}{c|}{}                                                                                & cS-VBL and dS-VBL approaches                                                                            & \begin{tabular}[c]{@{}c@{}}Higher accuracy and lower \\ complexity than \cite{cheng2019adaptive}\end{tabular}                                                  & \cite{pisharody2024near}   \\ \cline{5-7}
\multicolumn{1}{|c|}{}                                         &                                 &                             & \multicolumn{1}{c|}{}                                                                                & Turbo-CoSaMP algorithm                                                                                  & High estimation accuracy                                                                                                                                                        & \cite{xie2024massive}      \\ \cline{5-7}
\multicolumn{1}{|c|}{}                                         &                                 &                             & \multicolumn{1}{c|}{}                                                                                & \begin{tabular}[c]{@{}c@{}}Two-stage channel estimation \\ based on SOMP and 2D-OMP\end{tabular}        & \begin{tabular}[c]{@{}c@{}}32\% improvement in estimation \\ accuracy over \cite{xie2024massive}\end{tabular}                                                  & \cite{arai2024near}        \\ \cline{4-7}
\multicolumn{1}{|c|}{}                                         &                                 &                             & \multicolumn{1}{c|}{\begin{tabular}[c]{@{}c@{}}Sparsity-aware +\\ deep learning method\end{tabular}} & Grid-based polar-domain channel estimation                                                              & -                                                                                                                                                                               & \cite{nguyen2024channel}   \\ \cline{2-7}
\multicolumn{1}{|c|}{}                                         & \textbf{MIMO}                   & -                           & \multicolumn{1}{c|}{-}                                                                               & -                                                                                                       & -                                                                                                                                                                               & -                                           \\ \hline
\multicolumn{1}{|c|}{\multirow{16}{*}{\textbf{Multi-carrier}}} & \multirow{15}{*}{\textbf{MISO}} & LoS                         & \multicolumn{1}{c|}{\multirow{10}{*}{Sparsity-aware method}}                                         & Two-level P-OMP algorithm                                                                               & Superior performance over \cite{chen2023non}                                                                                                                   & \cite{zhang2024non}        \\ \cline{3-3} \cline{5-7}
\multicolumn{1}{|c|}{}                                         &                                 & \multirow{13}{*}{Multipath} & \multicolumn{1}{c|}{}                                                                                & P-SOMP and P-SIGW hybrid estimation framework                                                           & Superior performance in terms of NMSE                                                                                                                                           & \cite{cui2022channel}      \\
\cline{5-7}
\multicolumn{1}{|c|}{}                                         &                                 &                             & \multicolumn{1}{c|}{}                                                                                & GP-SOMP and GP-SIGW algorithms                                                                          & -                                                                                                                                                                               & \cite{chen2023non}         \\
\cline{5-7}
\multicolumn{1}{|c|}{}                                         &                                 &                             & \multicolumn{1}{c|}{}                                                                                & Inclusive P-SOMP method                                                                                 & \begin{tabular}[c]{@{}c@{}}Enhanced distance \\ accuracy over \cite{cui2022channel}\end{tabular}                                                               & \cite{jang2024near}        \\ \cline{5-7}
\multicolumn{1}{|c|}{}                                         &                                 &                             & \multicolumn{1}{c|}{}                                                                                & DL-OMP algorithm                                                                                        & Superior accuracy over P-OMP                                                                                                                                                    & \cite{zhang2024near}       \\
\cline{5-7}
\multicolumn{1}{|c|}{}                                         &                                 &                             & \multicolumn{1}{c|}{}                                                                                & Beam-focused SOMP algorithm                                                                             & -                                                                                                                                                                               & \cite{hussain2024near}     \\ \cline{5-7}
\multicolumn{1}{|c|}{}                                         &                                 &                             & \multicolumn{1}{c|}{}                                                                                & \begin{tabular}[c]{@{}c@{}}CCSAMP-PR algorithm based on \\ refined polar-domain dictionary\end{tabular} & \multirow{4}{*}{\begin{tabular}[c]{@{}c@{}}Mitigate the near-field \\ beam split effect \\ in wideband systems\end{tabular}}                                                    & \cite{ruan2024wideband}    \\ \cline{5-5} \cline{7-7}
\multicolumn{1}{|c|}{}                                         &                                 &                             & \multicolumn{1}{c|}{}                                                                                & NBA-OMP algorithm                                                                                       &                                                                                                                                                                                 & \cite{elbir2023nba}        \\ \cline{5-5} \cline{7-7}
\multicolumn{1}{|c|}{}                                         &                                 &                             & \multicolumn{1}{c|}{}                                                                                & BPD-based method                                                                                        &                                                                                                                                                                                 & \cite{cui2023near}         \\ \cline{5-5} \cline{7-7}
\multicolumn{1}{|c|}{}                                         &                                 &                             & \multicolumn{1}{c|}{}                                                                                & Segmented off-grid P-SIGW algorithm                                                                     &                                                                                                                                                                                 & \cite{chen2024segmented}   \\ \cline{4-7}
\multicolumn{1}{|c|}{}                                         &                                 &                             & \multicolumn{1}{c|}{\multirow{3}{*}{Deep learning method}}                                           & Neural network-based LIGW algorithm                                                                     & \begin{tabular}[c]{@{}c@{}}Enhanced distance and channel recovery \\ accuracy over \cite{cui2022channel}\end{tabular}                                          & \cite{jang2024neural}      \\ \cline{5-7}
\multicolumn{1}{|c|}{}                                         &                                 &                             & \multicolumn{1}{c|}{}                                                                                & \begin{tabular}[c]{@{}c@{}}Dictionary learning-based \\ bilevel optimization algorithm\end{tabular}     & \begin{tabular}[c]{@{}c@{}}Superior accuracy to P-SOMP \cite{cui2022channel} \\ and lower complexity than BPD \cite{cui2023near}\end{tabular} & \cite{zheng2024dictionary} \\ \cline{5-7}
\multicolumn{1}{|c|}{}                                         &                                 &                             & \multicolumn{1}{c|}{}                                                                                & FL-based framework                                                                                      & \begin{tabular}[c]{@{}c@{}}12x lower pilot overhead \\ compared to \cite{elbir2023nba}\end{tabular}                                                            & \cite{elbir2023near}       \\ \cline{4-7}
\multicolumn{1}{|c|}{}                                         &                                 &                             & \multicolumn{1}{c|}{\begin{tabular}[c]{@{}c@{}}Sparsity-aware +\\ deep learning method\end{tabular}} & \begin{tabular}[c]{@{}c@{}}Two-stage channel estimation \\ combining SOMP and GDM\end{tabular}          & Improved NMSE over LS and OMP                                                                                                                                                   & \cite{Jin2024A}            \\ \cline{3-7}
\multicolumn{1}{|c|}{}                                         &                                 & Rician                      & \multicolumn{1}{c|}{Sparsity-aware method}                                                           & Angular clustering-based estimation method                                                              & \begin{tabular}[c]{@{}c@{}}Outperforms P-SOMP in \cite{cui2022channel} \\ while reducing complexity\end{tabular}                                               & \cite{ruan2024channel}     \\ \cline{2-7}
\multicolumn{1}{|c|}{}                                         & \textbf{MIMO}                   & -                           & \multicolumn{1}{c|}{-}                                                                               & -                                                                                                       & -                                                                                                                                                                               & -                                           \\ \hline
\end{tabular}}
\label{TableV}
\end{table*}

\textbf{Sparsity-Aware (Compressive Sensing) Approaches}: In ELAA-enabled systems, considering the large array aperture, each UE may radiate pilot signals toward only a subset of the BS array. As a result, the overall channel in a multi-user MISO system tends to exhibit a block-sparse structure. Consequently, compressed sensing approaches can exploit this structured sparsity to perform hierarchical sparse modeling, thereby enabling efficient estimation of multi-user sparse channels. \cite{tang2024joint} proposed a two-stage joint visibility region (VR) detection and channel estimation method. In the first stage, the VR detection-oriented message passing (VRDO-MP) approach is employed to estimate antenna visibility using spatial correlation. In the second stage, leveraging both VR information and channel sparsity, the belief-based OMP (BB-OMP) method is applied to estimate the multi-user LoS channel, achieving superior performance at low SNR. Furthermore, by leveraging the polar-domain sparsity, \cite{pisharody2024near} proposed a decentralized channel estimation algorithm for multipath channels between the BS and multiple UEs. The algorithm first employs a centralized sub-array-based variational Bayesian learning (cS-VBL) approach and then extends it to a decentralized S-VBL (dS-VBL) implementation. Compared to the AGSBL algorithm in \cite{cheng2019adaptive}, the proposed approach achieves superior estimation accuracy with reduced computational complexity, thereby enhancing the overall SE. In \cite{xie2024massive}, a polar-domain sparse channel sampling technique was integrated with the turbo-type compressive sampling matching pursuit (Turbo-CoSaMP) algorithm for channel estimation in massive unsourced random access (URA) systems. By employing joint angle-distance sampling, this method enhances channel sparsity. Besides, the Turbo-CoSaMP, which incorporates joint activity detection and channel estimation, iteratively identifies active codewords and estimates channel parameters. Furthermore, a Newton-based optimization procedure is incorporated into Turbo-CoSaMP to mitigate basis mismatch issues arising from discrete polar-domain sampling, thereby enabling more accurate multipath channel estimation. In \cite{arai2024near}, a two-stage channel estimation algorithm was developed. The first stage employs sparse OMP (SOMP) to estimate angle and distance parameters, while the second stage applies 2D-OMP for UE-path association. Compared to the Turbo-CoSaMP algorithm in \cite{xie2024massive}, the proposed approach improves channel estimation accuracy by $32\%$ without increasing computational complexity. \cite{nguyen2024channel} proposed a grid-based polar-domain channel estimation method, which is founded on the gradient descent approach and exploits the polar-domain sparsity of near-field channels. The performance of this method is further enhanced through the application of a deep unfolding technique. However, the aforementioned methods primarily focus on channel estimation for multi-user MISO systems, while approaches tailored for multi-user MIMO systems remain largely unexplored.

\subsubsection{Multi-User Multi-Carrier Systems}

Subsequently, we summarize estimation methods for multi-user multi-carrier systems, where sparsity-aware and deep learning remain the primary strategies for achieving accurate estimation.

\textbf{Sparsity-Aware (Compressive Sensing) Approaches}: For multi-user MISO-OFDM systems, sparsity-aware remains one of the effective approaches for near-field channel estimation. \cite{wu2025near} proposed a dual-band near-field communication model to enhance the reconstructible sparsity of high-frequency near-field channels. By integrating structural characteristics derived from sparsity ambiguity with out-of-band spatial information from the low-frequency channel, this approach elevates the upper bound of sparsity reconstruction, thereby providing a foundation for the application of compressed sensing algorithms to channel estimation in multi-user and multi-carrier scenarios. Building on this, \cite{cui2022channel} introduced a hybrid estimation framework combining an on-grid polar-domain SOMP (P-SOMP) algorithm with an off-grid polar-domain simultaneous iterative gridless weighted (P-SIGW) algorithm, achieving superior performance in terms of the normalized mean square error (NMSE). Extending the approach of \cite{cui2022channel}, the authors of \cite{jang2024near} proposed an inclusive P-SOMP method for jointly estimating angles, distances, and gains in near-field multipath channels, resulting in a significant enhancement in distance estimation accuracy compared to \cite{cui2022channel}. In parallel, \cite{zhang2024near} developed a distance-parameterized angular-domain sparse near-field channel model and proposed a dictionary learning orthogonal matching pursuit (DL-OMP) estimation method, applicable to both LoS and multipath scenarios. By employing joint dictionary learning and sparse recovery, the proposed approach achieves estimation accuracy superior to that of the conventional polar-domain OMP (P-OMP) method. In \cite{hussain2024near}, a novel polar-domain codebook, independent of the user distance, was first designed to significantly reduce the codebook size. Using this optimized codebook, the authors proposed a beam-focused SOMP algorithm that effectively estimates near-field multipath channels. Furthermore, leveraging the fact that the multipath channels of different users share the same set of scatterers, the authors of \cite{ruan2024channel} first estimated scatterer positions relative to a single reference user and then enabled other users to share these estimates, thereby reducing the overall computational complexity. To further enhance the accuracy of scatterer positioning, they also designed an angular clustering algorithm, which outperforms the P-SOMP method in \cite{cui2022channel}. In \cite{chen2023non}, a signal extraction method based on group time block coding (GTBC) was illustrated to capture spatial-domain non-stationarity in multipath channels. Building on the extracted signals, an on-grid GTBC-based polar-domain simultaneous OMP (GP-SOMP) algorithm and an off-grid GTBC-based polar-domain simultaneous iterative gridless weighted (GP-SIGW) algorithm were developed to achieve accurate estimation of multipath channels. On this basis, \cite{zhang2024non} introduced a two-level P-OMP algorithm and a list-based channel estimation method, demonstrating superior performance over \cite{chen2023non}.

Furthermore, to mitigate the near-field beam split effect in wideband systems, the authors of \cite{ruan2024wideband} introduced a refined polar-domain dictionary for near-field multipath channels across different frequencies, enabling flexible adjustment of sampling angles and distances. They further proposed the correlation coefficient sparsity adaptive matching pursuit with parameter refinement (CCSAMP-PR) algorithm, which iteratively refines angle and distance estimates. This method achieves efficient near-field channel reconstruction with reduced complexity, even without exploiting any prior knowledge of channel sparsity. Similarly, in \cite{elbir2023nba}, a near-field beam-split aware OMP (NBA-OMP) algorithm was proposed for THz multipath channels. By constructing a dedicated NBA dictionary that accounts for angular and distance deviations, this approach effectively mitigates beam splitting over a 70 GHz bandwidth, achieving improved performance compared to \cite{cui2022channel}. Apart from these dictionary-based approaches, \cite{cui2023near} presented a bilinear pattern detection (BPD)-based method to mitigate beam-splitting effects. This approach leverages linear frequency-dependent variations in the angular and distance domains to estimate the DoA and multipath distances across frequencies. By incorporating compressed sensing, it enables the reconstruction of the entire wideband channel, delivering superior performance compared to existing methods reported in \cite{cui2022channel}. Following \cite{cui2023near}, \cite{chen2024segmented} further developed a segmented off-grid P-SIGW algorithm, which further improves wideband channel estimation accuracy, achieving performance superior to that of the methods proposed in \cite{cui2022channel} and \cite{cui2023near}.

\textbf{Deep Learning Approaches}: In addition to compressed sensing algorithms, deep learning techniques offer an effective solution for near-field channel estimation in multi-user and multi-carrier systems. The authors of \cite{jang2024neural} proposed a neural network-based localizer for initial near-field channel position estimation. The resulting channel estimate is further refined using a frequency-selectivity-based location-domain iterative gridless weighted (LIGW) method, which, compared to \cite{cui2022channel}, demonstrates a remarkable improvement in both distance estimation and overall channel recovery accuracy. Besides, in \cite{zheng2024dictionary}, a dictionary learning-based bilevel optimization algorithm was introduced, in which an unsupervised deep neural network learns an optimal dictionary at the upper level, while the lower level identifies the best sparse representation of the near-field channel. Numerical results show that this approach not only outperforms the P-SOMP algorithm in \cite{cui2022channel} but also achieves lower computational complexity than the BPD method in \cite{cui2023near}. Moreover, based on the model-based approach in \cite{elbir2023nba}, \cite{elbir2023near} developed a federated learning (FL)-based framework as a model-free alternative for multi-user channel estimation. By integrating labeled data from \cite{elbir2023nba}, this method reduces the training overhead by a factor of 12 compared to conventional techniques. Furthermore, by combining the complementary strengths of compressed sensing and deep learning techniques, \cite{Jin2024A} proposed a two-stage channel estimation method, where the SOMP is first used for coarse channel recovery. Then, treating the coarse estimate with errors as a noisy image, a generative deep model (GDM)-based approach is used to further refine the near-field channel estimate, achieving an improved NMSE over conventional LS and OMP methods. However, similarly, all the methods discussed in this subsection are designed for multi-user MISO-OFDM systems, while channel estimation approaches tailored for multi-user MIMO-OFDM systems remain largely unexplored.

\subsubsection{Summary and Lessons Learned}

In this subsection, we conduct a systematic survey of near-field BS-UE channel estimation methods under various system configurations. For ease of reference, the estimation methods for \textit{single-user} scenarios discussed in Sections IV-A(1) and IV-A(2), as well as those for \textit{multi-user} scenarios reviewed in Sections IV-A(3) and IV-A(4), are summarized in Tables \ref{TableIV} and \ref{TableV}, respectively. Based on these two tables, parametric channel estimation, sparsity-aware (or compressive sensing) methods, and deep learning-based approaches can be regarded as the three primary paradigms for near-field channel estimation under various system configurations. In practical deployments, these methods exhibit distinct advantages and trade-offs in terms of applicable channel types, estimation accuracy, online complexity, and robustness.

Specifically, parametric channel estimation methods are particularly suitable for geometrically structured channels with well-defined components, such as LoS channels or multipath channels containing only a limited number of dominant paths. In such cases, the channel can be accurately characterized by a small set of parameters, including distance, angle, and gain. However, when the channel exhibits stronger randomness and statistical behavior, as in Rayleigh fading or structurally complex Rician channels, a low-dimensional parametric description becomes inadequate, and the effectiveness of parametric estimation degrades significantly. In contrast, sparsity-aware methods are well suited to channels dominated by an LoS path or a limited number of significant multipath components. In these scenarios, the number of propagation paths is much smaller than the number of array antennas, so the channel exhibits a distinct sparse structure in an appropriate transform domain, enabling efficient estimation via sparse recovery algorithms. Deep learning-based methods, on the other hand, offer the broadest applicability. With sufficiently representative training data, they can flexibly model a wide range of channel types, such as LoS, multipath, as well as statistical models including Rayleigh and Rician fading. Moreover, deep learning can implicitly learn complex effects, such as frequency selectivity and spatial non-stationarity, in an end-to-end manner, making it particularly suitable for handling wideband beam split effects in multi-carrier systems and the high-dimensional parameter coupling inherent in ELAA deployments.

From the perspective of estimation accuracy, and taking the Cram\'er-Rao bound as a notional $10/10$ benchmark, the three paradigms can be roughly rated as $9/10$ for parametric estimation, $7/10$ for compressive sensing, and $9/10$ for deep learning approaches. Specifically, parametric estimation methods can fully exploit the spherical-wave propagation characteristics and array geometry when an accurate geometrical channel model is available. By operating directly in the continuous parameter space and employing high-precision parameter estimation algorithms, such as MUSIC and its variants, these methods can recover key parameters including angle, distance, and gain with very high accuracy, which in turn enables near-optimal channel reconstruction performance. In contrast, compressive sensing algorithms typically require constructing a channel codebook in a selected domain and then applying a sparse recovery algorithm, such as OMP, to reconstruct the channel. Since the codebook consists of discretized channel samples (i.e., on-grid points), the achievable accuracy is inherently constrained by the grid resolution, and even though finer grids or interpolation can mitigate this limitation to some extent, it is generally difficult to match the off-grid accuracy of parametric methods. Deep learning approaches, on the other hand, can attain an accuracy level of around $9/10$, provided that sufficiently representative and diverse training data are available. By learning both the underlying geometrical structure and the statistical properties of the channel in an end-to-end manner, they are able to achieve an estimation accuracy that is comparable to parametric methods for LoS and multipath channels, and can even surpass parametric estimation in more complex scenarios, such as statistical channel models.

\begin{table*}[]
\captionsetup{skip=3pt}
\caption{The performance comparison of mainstream estimation methods.}
\resizebox{\textwidth}{!}{
\begin{tabular}{|ccclcclccl|}
\hline
\multicolumn{1}{|c|}{\textbf{}}                                                                                    & \multicolumn{3}{c|}{\textbf{Parametric estimation}}                                                                                                                                                                                                                                                & \multicolumn{3}{c|}{\textbf{Sparsity-aware}}                                                                                                                                                                                                                                                                                                             & \multicolumn{3}{c|}{\textbf{Deep learning}}                                                                                                                                                                                                                                                           \\ \hline
\multicolumn{1}{|c|}{\textbf{Applicable channel}}                                                                  & \multicolumn{3}{c|}{\begin{tabular}[c]{@{}c@{}}Geometrically structured LoS/few-path\\ channels with well-defined components.\end{tabular}}                                                                                                                                                         & \multicolumn{3}{c|}{Sparse LoS/multipath channels.}                                                                                                                                                                                                                                                                                                       & \multicolumn{3}{c|}{\begin{tabular}[c]{@{}c@{}}Various channels, from LoS and sparse/rich \\ multipath to statistical channels; particularly \\ suitable for wideband near-field systems.\end{tabular}}                                                                                                \\ \hline
\multicolumn{1}{|c|}{\multirow{2}{*}{\textbf{\begin{tabular}[c]{@{}c@{}}Estimation \\ accuracy\end{tabular}}}}     & \multicolumn{1}{c|}{S}                                      & \multicolumn{2}{c|}{9/10}                                                                                                                                                                                                            & \multicolumn{1}{c|}{S}         & \multicolumn{2}{c|}{7/10}                                                                                                                                                                                                                                                                                               & \multicolumn{1}{c|}{S}                                                          & \multicolumn{2}{c|}{9/10}                                                                                                                                                                                           \\ \cline{2-10}
\multicolumn{1}{|c|}{}                                                                                             & \multicolumn{1}{c|}{R}                                      & \multicolumn{2}{c|}{\begin{tabular}[c]{@{}c@{}}Off-grid search in the continuous parameter \\ space under a known geometric channel model \\ ensures high reconstruction accuracy.\end{tabular}}                                     & \multicolumn{1}{c|}{R}         & \multicolumn{2}{c|}{\begin{tabular}[c]{@{}c@{}}Accuracy is limited by the resolution \\ of grid points in the channel codebook.\end{tabular}}                                                                                                                                                                           & \multicolumn{1}{c|}{R}                                                          & \multicolumn{2}{c|}{\begin{tabular}[c]{@{}c@{}}With sufficient training data, end-to-end \\ learning of channel structure and \\ statistics enables high estimation accuracy.\end{tabular}}                         \\ \hline
\multicolumn{1}{|c|}{\multirow{2}{*}{\textbf{\begin{tabular}[c]{@{}c@{}}Online\\complexity\end{tabular}}}} & \multicolumn{1}{c|}{S}                                      & \multicolumn{2}{c|}{3/10}                                                                                                                                                                                                            & \multicolumn{1}{c|}{S}         & \multicolumn{2}{c|}{5/10}                                                                                                                                                                                                                                                                                               & \multicolumn{1}{c|}{S}                                                          & \multicolumn{2}{c|}{8/10}                                                                                                                                                                                           \\ \cline{2-10}
\multicolumn{1}{|c|}{}                                                                                             & \multicolumn{1}{c|}{R}                                      & \multicolumn{2}{c|}{\begin{tabular}[c]{@{}c@{}}Multi-dimensional off-grid search in \\ the continuous parameter space leads\\ to the highest online complexity,\\ especially in wideband near-field systems.\end{tabular}}           & \multicolumn{1}{c|}{R}         & \multicolumn{2}{c|}{\begin{tabular}[c]{@{}c@{}}Sparsity-aware methods operate on low-dimensional\\  sparse representations, thus their complexity \\ mainly scales with the sparsity level, \\ dictionary size, and number of iterations rather than \\ with the total number of antennas or grid points.\end{tabular}} & \multicolumn{1}{c|}{R}                                                          & \multicolumn{2}{c|}{\begin{tabular}[c]{@{}c@{}}A single forward pass yields fixed \\ and typically lower online complexity\\  than sparsity-aware methods.\end{tabular}}                                            \\ \hline
\multicolumn{1}{|c|}{\multirow{2}{*}{\textbf{Robustness}}}                                                         & \multicolumn{1}{c|}{S}                                      & \multicolumn{2}{c|}{4/10}                                                                                                                                                                                                            & \multicolumn{1}{c|}{S}         & \multicolumn{2}{c|}{7/10}                                                                                                                                                                                                                                                                                               & \multicolumn{1}{c|}{S}                                                          & \multicolumn{2}{c|}{8/10}                                                                                                                                                                                           \\ \cline{2-10}
\multicolumn{1}{|c|}{}                                                                                             & \multicolumn{1}{c|}{R}                                      & \multicolumn{2}{c|}{\begin{tabular}[c]{@{}c@{}}Strongly relies on an accurate geometric \\ channel model and array calibration, so model\\ mismatch or complex scattering can \\ cause severe performance degradation.\end{tabular}} & \multicolumn{1}{c|}{R}         & \multicolumn{2}{c|}{\begin{tabular}[c]{@{}c@{}}Constrains the channel to a low-dimensional \\ sparse subspace and filters part of \\ the noise, but degrades when polar-domain \\ sparsity or dictionary matching is poor.\end{tabular}}                                                                                & \multicolumn{1}{c|}{R}                                                          & \multicolumn{2}{c|}{\begin{tabular}[c]{@{}c@{}}Can learn array errors, hardware \\ non-idealities, and statistical \\ variations from data and is thus robust\\ to noise and moderate model mismatch.\end{tabular}} \\ \hline
\multicolumn{10}{|l|}{S: Score, R: Reason.}\\ \hline
\end{tabular}
}
\label{TableX}
\end{table*}

From the perspective of online computational complexity, and taking a score of $10/10$ as an ideal benchmark corresponding to the lowest online complexity under the same system scale, the three paradigms can be approximately rated as $3/10$ for parametric estimation, $5/10$ for sparsity-aware methods, and $8/10$ for deep learning approaches. Specifically, parametric estimation operates directly in the continuous parameter space and typically relies on multi-dimensional off-grid searches, which leads to the largest online computational burden among the three categories. This drawback becomes more pronounced in wideband near-field systems, where the dimensionality of the unknown parameters increases dramatically with the number of subcarriers and the associated search space expands accordingly; consequently, parametric channel estimation methods are more suitable for narrowband or single-user scenarios where the channel is governed by a relatively small set of geometric parameters. By contrast, sparsity-aware methods exploit structured dictionaries and sparsity priors, and operate mainly on low-dimensional sparse representations. Their computational complexity depends primarily on the sparsity level, the dictionary size, and the number of iterations, rather than growing with the total number of antennas or grid points. Even when extended to wideband near-field channel estimation, the use of joint sparsity exploitation, frequency-compensated dictionary designs, or BPD allows these methods to maintain a substantially lower online complexity than parametric approaches that require exhaustive searches over the entire parameter space, so that they can be regarded as exhibiting a medium complexity level. Compared with the above two strategies, deep learning methods typically attain the lowest online computational cost: a single forward pass of the network suffices to produce the channel estimate, resulting in a fixed and implementation-friendly complexity which, in representative XL-MIMO and wideband configurations, is often lower than that of carefully engineered sparsity-aware algorithms. Hence, their online complexity can be rated at around $8/10$. It should be noted, however, that deep learning approaches rely on a potentially expensive offline training phase, which requires large training datasets and iterative optimization, and thus introduces a high, albeit typically one-time, computational cost.

With regard to the robustness of the three strategies, we mainly assess their performance in terms of sensitivity to model mismatch, susceptibility to noise and interference, and degradation under variations in channel statistics. Taking a score of $10/10$ as an ideal robustness benchmark, corresponding to only negligible performance loss under all these impairments and thus nearly preserving the ideal estimation performance, the three approaches can be roughly rated as $4/10$ for parametric estimation, $7/10$ for sparsity-aware methods, and $8/10$ for deep learning techniques. Specifically, parametric estimation exhibits relatively weak robustness, as it strongly relies on an accurate geometric channel model. Once discrepancies arise between the assumed and actual array geometry, RF chain amplitude-phase response, or scattering environment, the resulting modeling errors tend to be greatly amplified, so that the bias induced by model mismatch can dominate over the noise-induced error; such degradation becomes particularly pronounced in hybrid-field or richly scattering environments. Sparsity-aware methods, on the other hand, can achieve moderately high robustness when the dictionary and grid are properly designed. By leveraging sparse recovery algorithms such as OMP and sparse Bayesian learning, they constrain the channel to a low-dimensional sparse subspace and employ thresholding and sparsity priors to suppress unstable coefficients, thereby preserving the dominant paths while naturally filtering part of the noise and interference energy. At the same time, they can tolerate mild dictionary mismatch and array imperfections. However, when the polar-domain sparsity is severely impaired, for example, in rich multipath or nearly continuous scattering environments, or when the dictionary is strongly mismatched to the true channel structure, their performance degrades significantly, so their overall robustness can be regarded as moderately high. In contrast, deep learning methods typically exhibit the highest robustness. With sufficiently representative training data, neural networks can implicitly learn factors that are difficult to model explicitly, such as array errors, hardware non-idealities, hybrid-field effects, and variations in channel statistics, and thus tend to be more stable than the other two approaches under noise perturbations and moderate model mismatch. Nevertheless, their robustness still depends critically on the alignment between the training data and the deployment scenario; once the application scenario substantially departs from the training distribution, their performance may also experience noticeable degradation.

In summary, these three paradigms exhibit different trade-offs among estimation accuracy, computational complexity, and robustness, and their comparative characteristics are summarized in Table \ref{TableX}. In practical deployments, the choice among these strategies should be guided jointly by the channel characteristics and the availability, amount, and representativeness of training data. For multipath channels with a finite number of dominant components, deep learning-based methods are typically preferred when sufficiently large and representative datasets are available. When such data are limited or not well matched to the deployment scenario, the choice often shifts to parametric or sparsity-aware estimation. In particular, parametric methods are most suitable when the channel can be accurately captured by a geometric model and high accuracy is prioritized, whereas sparsity-aware approaches are attractive when the channel admits an effective sparse representation in an appropriate transform domain and a better accuracy-complexity trade-off is desired. Table \ref{TableX} provides a convenient reference summarizing these trade-offs.

Despite considerable progress has been made in near-field BS-UE channel estimation across various system configurations, existing research has primarily focused on structurally deterministic channel models, such as LoS and multipath scenarios. In contrast, channel estimation techniques for statistically characterized models, such as the Rayleigh fading model, remain underexplored. Moreover, near-field channel estimation techniques tailored for multi-user MIMO scenarios are still limited and warrant further investigation and development.

\subsection{XL-RIS-aided Near-Field Cascaded Channel Estimation}

Building upon the research advances in direct channel estimation between the BS and UE, the subsequent sections systematically review near-field cascaded channel estimation methods in XL-RIS-aided wireless communications. The introduction of XL-RIS results in an exponential increase in the dimensionality of the cascaded channel. When combined with the nonlinear coupling of multi-dimensional parameters (e.g., distance, angle, and phase) induced by spherical wave propagation, this gives rise to a twofold computational challenge. On the one hand, due to the passive nature of the RIS, it cannot transmit or receive pilot signals directly. As a result, the cascaded channel must be estimated indirectly at the BS or the UE by appropriately adjusting the RIS reflection patterns. On the other hand, the strong coupling between parameters further complicates cascaded channel estimation, often requiring the solution of high-dimensional non-convex optimization problems. This typically involves large-scale matrix inversion and iterative optimization, significantly increasing computational complexity.
In the following sections, we systematically summarize cascaded channel estimation methods applicable to various scenarios, with the aim of identifying effective solutions to the aforementioned challenges.

\subsubsection{Single-User Single-Carrier Systems}

In this part, we first provide an overview of estimation methods for the RIS-aided cascaded channel under the single-user single-carrier configuration, mainly including sparsity-aware and low-rank approaches.

\begin{table*}[]
\captionsetup{skip=3pt}
\caption{Estimation methods for RIS-aided near-field cascaded channels in single-user systems.}
\resizebox{\textwidth}{!}{
\begin{tabular}{|cc|c|cc|c|c|}
\hline
\multicolumn{2}{|c|}{\textbf{System Type}}                                             & \textbf{Channel Type}               & \multicolumn{2}{c|}{\textbf{Algorithm}}                                                                               & \textbf{Features}                                                                                       & \textbf{Reference}                                                  \\ \hline
\multicolumn{1}{|c|}{\multirow{5}{*}{\textbf{Single-carrier}}} & \multirow{4}{*}{MISO} & \multirow{5}{*}{Multipath} & \multicolumn{1}{c|}{\multirow{3}{*}{Sparsity-aware method}} & FSBL-based two-stage estimation strategy       & Superior to OMP                                                                                & \cite{Yu2023Channel}                      \\ \cline{5-7}
\multicolumn{1}{|c|}{}                                &                       &                            & \multicolumn{1}{c|}{}                                       & DNOMP-based separate channel estimation scheme & Reduced computational complexity                                                               & \cite{lu2024low}                          \\ \cline{5-7}
\multicolumn{1}{|c|}{}                                &                       &                            & \multicolumn{1}{c|}{}                                       & BESVR and TVR algorithms                       & Lower CPU time and NMSE than P-OMP                                                             & \cite{schroeder2024low,schroeder2024near} \\ \cline{4-7}
\multicolumn{1}{|c|}{}                                &                       &                            & \multicolumn{1}{c|}{Low-rank algorithm} & PW-CLRA-based two-timescale estimation famework                                         & \begin{tabular}[c]{@{}c@{}}Reduced pilot overhead \\ and complexity\end{tabular}               & \cite{lee2025near,lee2025two}                         \\ \cline{2-2} \cline{4-7}
\multicolumn{1}{|c|}{}                                & MIMO                  &                            & \multicolumn{1}{c|}{Sparsity-aware method}                  & Compressive sensing-based localizer            & --                                                                                             & \cite{rinchi2022compressive}              \\ \hline
\multicolumn{1}{|c|}{\multirow{4}{*}{\textbf{Multi-carrier}}}  & \multirow{2}{*}{MISO} & \multirow{4}{*}{Multipath} & \multicolumn{1}{c|}{\multirow{4}{*}{Sparsity-aware method}} & PF-RCE method                                  & Superior estimation accuracy over OMP                                                          & \cite{wu2022near}                         \\ \cline{5-7}
\multicolumn{1}{|c|}{}                                &                       &                            & \multicolumn{1}{c|}{}                                       & Polar bin design                               & 1.2 dB NMSE gain over \cite{wu2022near}                                       & \cite{wu2023parametric}                   \\ \cline{2-2} \cline{5-7}
\multicolumn{1}{|c|}{}                                & \multirow{2}{*}{MIMO} &                            & \multicolumn{1}{c|}{}                                       & 3-D P-SOMP algorithm                           & $96.9\%$ pilot overhead reduction                                                              & \cite{du2025three}                        \\ \cline{5-7}
\multicolumn{1}{|c|}{}                                &                       &                            & \multicolumn{1}{c|}{}                                       & 2D-OLS-based MMPSR framework                   & \begin{tabular}[c]{@{}c@{}}Outperforms all existing \\ 2-D reconstruction methods\end{tabular} & \cite{yang2024near}                       \\ \hline
\end{tabular}}
\label{TableVI}
\end{table*}

\textbf{Sparsity-Aware (Compressive Sensing) Approaches}:  Similar to BS-UE channel estimation, sparsity-aware methods serve as effective tools for recovering multipath channels in RIS-aided MISO systems. The authors of \cite{Yu2023Channel} modeled the RIS-aided cascaded channel as a combination of a near-field multipath channel from the UE to the RIS and a far-field multipath channel from the RIS to the BS. Due to partial blockages, different VRs were assumed for each scatterer between the RIS and the UE. Based on this model, a two-stage strategy was proposed for joint channel and VR estimation leveraging a fast sparse Bayesian learning (FSBL) framework, which demonstrated superior performance compared to conventional OMP methods. Besides, in \cite{lu2024low}, a hybrid RIS is configured in which the elements located in the central subarray, as long as many specially selected discrete elements, are active, while all the others remain passive. For this architecture, a separate channel estimation scheme based on the decoupling operation (SCEDO) was proposed to estimate both the UE-RIS and RIS-BS multipath channels. In this process, a novel damped Newtonized orthogonal matching pursuit (DNOMP) algorithm with reduced computational complexity was developed. This algorithm combines planar and spherical wave models to estimate angle parameters using the central subarray, while recovering path gains and distances through the discrete elements. Similarly, a hybrid RIS structure combining both passive and active elements was discussed in \cite{schroeder2024low}, enabling the RIS to demodulate received pilot signals during training. Subsequently, by exploiting the block sparsity of the near-field channel, \cite{schroeder2024low} proposed two low-complexity algorithms: the boundary estimation and sub-vector recovery (BESVR) algorithm and the linear total variation regularization (TVR) algorithm. Compared to the existing P-OMP method, both algorithms demonstrate significant advantages in terms of CPU time and NMSE. The work \cite{schroeder2024near} further investigates the impact of various parameters, such as carrier frequency and RIS configuration, on the TVR algorithm proposed in \cite{schroeder2024low}, validating its applicability in the millimeter-wave band ($f_c = 28$ GHz). Extending to RIS-aided MIMO systems, in \cite{rinchi2022compressive}, a location estimation algorithm based on compressive sensing is designed to effectively retrieve the DoA and distance of the UE in the cascaded multipath channel. Through iterative design of the RIS phase-shifts, the positioning performance can be further improved.

\textbf{Low-Rank Approach}: In addition to sparsity-aware methods, low-rank algorithms offer a promising alternative for cascaded channel estimation in RIS-aided systems, enabling a significant reduction in pilot overhead and computational complexity. These methods exploit the inherent redundancy of the channel matrix induced by limited scattering, spatial correlation, or structured array deployments. By modeling the channel as a low-rank or approximately low-rank structure, the original high-dimensional estimation task can be decomposed into low-dimensional subspace learning and coefficient recovery. This reformulation substantially reduces computational complexity and training overhead, while maintaining high estimation accuracy. In this context, the authors of \cite{lee2025two} derived the time-scaling property of the cascaded channel in RIS-aided MISO systems, demonstrating that the channel can be represented as the product of large-timescale and small-timescale channels. Building on this foundation, they propose a two-timescale channel estimation (2TCE) framework: the large-timescale channel is estimated using the existing piece-wise collaborative low-rank approximation (PW-CLRA) algorithm \cite{lee2025near}, followed by the estimation of the small-timescale channel using observations from beam training together and the previously large-timescale estimation. This approach significantly reduces pilot overhead and computational complexity in RIS-aided cascaded channel~estimation.

\subsubsection{Single-User Multi-Carrier Systems}

Subsequently, we review RIS-aided cascaded channel estimation methods for single-user multi-carrier systems, where compressive sensing constitutes the predominant technical route.

\textbf{Sparsity-Aware (Compressive Sensing) Approaches}: Sparsity-aware methods also provide an effective solution for cascaded channel estimation in wideband systems. Leveraging the polar-domain sparsity of near-field multipath channels and the common sparsity support property of THz wideband channels, the authors of \cite{wu2022near} proposed a novel polar-domain frequency-dependent RIS-aided channel estimation (PF-RCE) method for the THz MISO-OFDM system. Compared to existing OMP algorithms, PF-RCE demonstrates significantly improved estimation accuracy in wideband scenarios. Building upon this, the authors in \cite{wu2023parametric} introduced a novel polar bin design to further enhance the sparse recovery performance of the PF-RCE, resulting in an additional $1.2$ dB gain in terms of NMSE.
Extended to RIS-aided MIMO-OFDM systems, the authors of \cite{du2025three} proposed a 3-D P-SOMP algorithm for the cascaded multipath channel, which ensures accurate channel estimation while reducing the system pilot overhead by $96.9\%$. Moreover, the authors of \cite{yang2024near} characterized the beam split effect of wideband cascaded multipath channels in both the angular and distance domains and, based on this characterization, constructed a wideband spherical wave dictionary. Subsequently, a multi-frequency parallel subspace recovery (MMPSR) framework was developed, which transforms the 2-D compressed sensing problem into multiple sparse vector recovery problems and performs multi-frequency joint processing. Based on this framework, an atom matching method based on correlation coefficients was designed to replace the inner-product-based atom matching used in conventional OMP algorithms. In addition, the authors proposed a 2-D oracle least squares (2D-OLS) estimator, which ultimately achieved estimation accuracy surpassing all existing 2-D channel reconstruction~methods.

\subsubsection{Multi-User Single-Carrier Systems}

In this part, we shift our focus to multi-user single-carrier systems, where parametric estimation, sparsity-aware, low-rank algorithms, and deep learning can all be employed to enable RIS-aided cascaded channel estimation.

\begin{table*}[]
\captionsetup{skip=3pt}
\caption{Estimation methods for RIS-aided near-field cascaded channels in multi-user systems.}
\resizebox{\textwidth}{!}{
\begin{tabular}{|cc|ccccc|}
\hline
\multicolumn{2}{|c|}{\textbf{System Type}}                                             & \multicolumn{1}{c|}{\textbf{Channel Type}}      & \multicolumn{2}{c|}{\textbf{Algorithm}}                                                                                                                                                                                 & \multicolumn{1}{c|}{\textbf{Features}}                                                                                                                          & \textbf{Reference}                          \\ \hline
\multicolumn{1}{|c|}{\multirow{9}{*}{\textbf{Single-carrier}}} & \multirow{6}{*}{MISO} & \multicolumn{1}{c|}{\multirow{5}{*}{Multipath}} & \multicolumn{1}{c|}{Parametric estimation method}           & \multicolumn{1}{c|}{\begin{tabular}[c]{@{}c@{}}VR identification based on accumulation\\ function and sliding window\end{tabular}}                        & \multicolumn{1}{c|}{\begin{tabular}[c]{@{}c@{}}Centimeter-level UE localization and\\ VR identification success rate $>97\%$\end{tabular}}                      & \cite{han2022localization} \\ \cline{4-7}
\multicolumn{1}{|c|}{}                                         &                       & \multicolumn{1}{c|}{}                           & \multicolumn{1}{c|}{\multirow{2}{*}{Sparsity-aware method}} & \multicolumn{1}{c|}{PDS-OMP algorithm combined with LS}                                                                                                   & \multicolumn{1}{c|}{Low pilot overhead}                                                                                                                         & \cite{jin2023near}         \\ \cline{5-7}
\multicolumn{1}{|c|}{}                                         &                       & \multicolumn{1}{c|}{}                           & \multicolumn{1}{c|}{}                                       & \multicolumn{1}{c|}{Two-step estimation protocol}                                                                                                         & \multicolumn{1}{c|}{--}                                                                                                                                         & \cite{chen2023channel}     \\ \cline{4-7}
\multicolumn{1}{|c|}{}                                         &                       & \multicolumn{1}{c|}{}                           & \multicolumn{1}{c|}{\multirow{2}{*}{Deep learning method}}  & \multicolumn{1}{c|}{\begin{tabular}[c]{@{}c@{}}Two-stage estimation strategy \\ based on DnCNN and ISTA\end{tabular}}                                     & \multicolumn{1}{c|}{Superior NMSE over OMP}                                                                                                                     & \cite{tang2023near}        \\ \cline{5-7}
\multicolumn{1}{|c|}{}                                         &                       & \multicolumn{1}{c|}{}                           & \multicolumn{1}{c|}{}                                       & \multicolumn{1}{c|}{DRN-NFCE algorithm}                                                                                                                   & \multicolumn{1}{c|}{\begin{tabular}[c]{@{}c@{}}Higher accuracy and $80\%$ pilot overhead \\ reduction over LS and MMSE estimators\end{tabular}}                 & \cite{wang2024enhanced}    \\ \cline{3-7}
\multicolumn{1}{|c|}{}                                         &                       & \multicolumn{1}{c|}{Rayleigh}                   & \multicolumn{2}{c|}{LMMSE-based three-stage estimation framework}                                                                                                                                                       & \multicolumn{1}{c|}{Reduced pilot overhead}                                                                                                                     & \cite{wang2020channel}     \\ \cline{2-7}
\multicolumn{1}{|c|}{}                                         & \multirow{3}{*}{MIMO} & \multicolumn{1}{c|}{\multirow{2}{*}{Multipath}} & \multicolumn{1}{c|}{Sparsity-aware method}                  & \multicolumn{1}{c|}{\begin{tabular}[c]{@{}c@{}}Two-stage estimation protocol \\ based on M-LAOMP and 3D-D-CS\end{tabular}}                                & \multicolumn{1}{c|}{Superior NMSE over OMP}                                                                                                                     & \cite{yang2023channel}     \\ \cline{4-7}
\multicolumn{1}{|c|}{}                                         &                       & \multicolumn{1}{c|}{}                           & \multicolumn{1}{c|}{Low-rank algorithm}                     & \multicolumn{1}{c|}{Two-step estimation framework based on CLRA-JO}                                                                                       & \multicolumn{1}{c|}{$80\%$ training overhead reduction}                                                                                                         & \cite{lee2024near}         \\ \cline{3-7}
\multicolumn{1}{|c|}{}                                         &                       & \multicolumn{1}{c|}{Rician}                     & \multicolumn{1}{c|}{Low-rank algorithm}                     & \multicolumn{1}{c|}{Two-stage PW-CLRA algorithm}                                                                                                          & \multicolumn{1}{c|}{\begin{tabular}[c]{@{}c@{}}Higher accuracy and $70\%$ training \\ overhead reduction over \cite{lee2024near}\end{tabular}} & \cite{lee2025near}         \\ \hline
\multicolumn{1}{|c|}{\multirow{3}{*}{\textbf{Multi-carrier}}}  & \multirow{2}{*}{MISO} & \multicolumn{1}{c|}{LoS}                        & \multicolumn{1}{c|}{Parametric estimation method}           & \multicolumn{1}{c|}{\begin{tabular}[c]{@{}c@{}}Fresnel-based decoupled parametric estimation \\ with a 2-D subspace method and a 1-D search\end{tabular}} & \multicolumn{1}{c|}{Reduced complexity}                                                                                                                         & \cite{pan2023ris}          \\ \cline{3-7}
\multicolumn{1}{|c|}{}                                         &                       & \multicolumn{1}{c|}{Multipath}                  & \multicolumn{1}{c|}{Sparsity-aware method}                  & \multicolumn{1}{c|}{P-SOMP combined with P-SIGW}                                                                                                          & \multicolumn{1}{c|}{Improved accuracy over LS estimator}                                                                                                        & \cite{chen2023novel}       \\ \cline{2-7}
\multicolumn{1}{|c|}{}                                         & MIMO                  & \multicolumn{5}{c|}{--}                                                                                                                                                                                                                                                                                                                                                                                                                                                                   \\ \hline
\end{tabular}}
\label{TableVII}
\end{table*}

\textbf{Parametric Estimation Approach}: Similar to BS-UE channel estimation, the parametric estimation techniques provide an efficient approach for recovering cascaded channels in RIS-aided multi-user MISO systems. The authors of \cite{han2022localization} proposed a low-overhead parametric channel reconstruction method for cascaded multipath channels. This approach introduces VR identification methods based on the accumulation function and sliding window by leveraging channel statistics, achieving centimeter-level joint localization for multiple users with a VR identification success rate exceeding $97\%$. The estimated user positions are subsequently incorporated into the cascaded channel model to accurately reconstruct the multipath channels of different~users.

\textbf{Sparsity-Aware (Compressive Sensing) Approaches}: Sparsity-aware techniques also offer a promising solution for estimating cascaded channels in RIS-aided multi-user MISO systems. In \cite{jin2023near}, a polar domain dual-sparsity orthogonal matching pursuit (PDS-OMP) algorithm was combined with the LS estimator. This approach exploits the unique row-column sparsity of multi-user cascaded multipath channels in the polar domain, enabling accurate channel reconstruction with low pilot overhead. Besides, building on the row-column sparsity of multipath channels across users, \cite{chen2023channel} proposed a two-step joint multi-user channel estimation protocol. First, the received signal containing cascaded channel information from multiple users is projected onto a common column subspace by exploiting the shared column-block sparsity of the BS-RIS channel, reducing the channel matrix to one with only row-block sparsity. Then, an alternating optimization and iterative reweighted algorithm is applied to recover the joint sparse matrix, enabling more accurate estimation of the cascaded channels from different users. Extended to MIMO systems,the authors of \cite{yang2023channel} proposed a two-stage channel estimation protocol based on compressed sensing. In the first stage, the 3-D and 2-D multiple measurement vector look-ahead orthogonal matching pursuit (M-LAOMP) algorithms sequentially estimate the common direction of departure (DoD) of the BS-RIS channel and the DoAs of different users. Based on these estimates, the 3-D distributed compressed sensing (3D-D-CS) framework is integrated with a compressed sensing approach based on the polar-domain dictionary in the second stage, to jointly recover the angle and distance information of the cascaded channels. Compared to conventional OMP algorithms, this approach significantly improves the NMSE~performance.

\textbf{Deep Learning Approaches}: Beyond sparsity-aware methods, deep learning has also emerged as a competitive approach for estimating cascaded channels across different users. The authors of \cite{tang2023near} proposed a deep unfolding-based two-stage cascaded channel estimation strategy. In the first stage, a denoising convolutional neural network (DnCNN) is employed to estimate the angle and distance of the common path between the RIS and the BS. Subsequently, a network based on the iterative shrinkage-thresholding algorithm (ISTA) is designed to recover the angles, distances, and complex gains of the UE-RIS channels. Compared to existing OMP methods, the proposed approach demonstrates superior performance in terms of NMSE. In \cite{wang2024enhanced}, a deep residual network-driven near-field cascaded channel estimation (DRN-NFCE) algorithm was developed. By leveraging the powerful feature-learning capability of deep learning techniques, the proposed DRN-NFCE network achieves significantly higher estimation accuracy while reducing the pilot overhead by four-fifths compared to conventional LS and MMSE estimators.

\textbf{Low-Rank Approaches}: Moreover, low-rank algorithms are also a potential approach for estimating the cascaded channels in RIS-aided multi-user MIMO systems. In \cite{lee2024near}, a two-step channel estimation framework based on the CLRA and joint optimization (CLRA-JO) was developed. The cascaded multipath channels between the UEs and the RIS are first modeled as the product of a common column space matrix of the BS-RIS channel and user-specific coefficient matrices. Based on this model, the CLRA algorithm is employed to estimate the column space matrix, followed by the JO method to estimate the user-specific coefficient matrices. This approach ensures estimation accuracy while reducing the training overhead by approximately $80\%$. Moreover, the authors of \cite{lee2025near} partitioned the cascaded channels from different users into multiple piece-wise channels that share low-rank subspaces and developed a two-stage PW-CLRA algorithm. In the first stage, training observations from different users are used to estimate the shared subspace, while in the second stage, the user-specific coefficient matrices are jointly optimized. Compared to the CLRA-JO method in \cite{lee2024near}, this approach improves estimation accuracy while reducing the training overhead by $70\%$.

\textbf{Other Approach}: In addition to the aforementioned methods, \cite{wang2020channel} proposed a three-stage cascaded channel estimation framework that exploits the fact that the cascaded channels of different users share a common RIS-BS channel. In the first two stages, a conventional LMMSE estimator is employed to estimate the UE-BS and UE-RIS-BS channels of a typical user, respectively. In the third stage, the UE-RIS-BS channels of the remaining users are recovered by leveraging the strong correlation between these channels and that of the typical user. This framework effectively reduces pilot overhead while maintaining high channel estimation accuracy.

\subsubsection{Multi-User Multi-Carrier Systems}

Finally, RIS-aided cascaded channel estimation in multi-user multi-carrier configurations can be realized using either parametric methods or compressive sensing techniques.

\textbf{Parametric Estimation Approach}: Parametric channel estimation offers a competitive solution for recovering LoS channels in RIS-aided multi-user multi-carrier systems. In \cite{pan2023ris}, a second-order Fresnel approximation was employed to decouple the angular and distance domains of the LoS channel, thereby reducing the complexity of cascaded channel estimation. Based on this decoupling, a 2-D subspace method and a 1-D search were applied to estimate the DoAs and distances of different UEs, respectively. Subsequently, a conventional LS method was used to recover the channel attenuation coefficients across the subcarriers, yielding a complete CSI.

\textbf{Compressive Sensing Approach}: Moreover, compressive sensing methods are also commonly employed for estimating cascaded multipath channels in this scenario. In \cite{chen2023novel}, the P-SOMP algorithm is combined with the P-SIGW algorithm to perform joint channel estimation across multiple subcarriers, offering improved accuracy compared to the conventional LS estimator.

\subsubsection{Summary and Lessons Learned}

In this subsection, we provide a detailed discussion of existing cascaded channel estimation methods for RIS-aided near-field communications under various scenarios. For ease of reference, the cascaded channel estimation methods for \textit{single-user} scenarios, as presented in Sections IV-B(1) and IV-B(2), and \textit{multi-user} scenarios, as discussed in Sections IV-B(3) and IV-B(4), are summarized in Tables \ref{TableVI} and \ref{TableVII}, respectively.

It is worth noting that parametric estimation, compressive sensing, and deep learning techniques, originally developed for BS-UE channel estimation, can be naturally extended to the RIS-aided cascaded channel estimation scenario. Beyond these mainstream approaches, the intrinsic redundancy of cascaded channels across the temporal or user domain makes low-rank algorithms a promising alternative. Compared with parametric estimation or sparsity-aware schemes, low-rank approaches offer comparable estimation accuracy, while substantially reducing pilot overhead and computational complexity by exploiting low-dimensional subspace structures. Moreover, similar to BS-UE channel estimation, existing studies on RIS-aided cascaded channel estimation have primarily focused on scenarios with structurally deterministic channels, such as LoS or multipath models. In contrast, the investigation of cascaded channel estimation methods tailored for statistical channel models remains in its early stages and requires further exploration and refinement.

\section{Conclusion and Further Discussion}

\subsection{Conclusion}

Under the evolution from massive MIMO to ELAAs, the array aperture may span tens or even hundreds of operating wavelengths, making spherical-wavefront-based near-field transmission a fundamental paradigm in 6G wireless communications, where wireless channels must be jointly characterized by both angular and distance parameters. This joint dependency complicates the channel structure, posing new and unique challenges for accurate channel estimation. This survey has provided a comprehensive overview of near-field channel estimation solutions, encompassing a variety of system configurations and propagation scenarios. We first introduced the criteria for delineating far-field and near-field regions in antenna arrays. Building on this, we review mainstream antenna architectures and wavefront control techniques for spherical-wave generation, representative near-field application scenarios, as well as ongoing standardization activities related to near-field communications. A detailed examination of existing modeling approaches was then presented, including near-field array response vectors, LoS and multipath channels, as well as spatial correlation-based models tailored to near-field characteristics. We further reviewed and analyzed various channel estimation techniques developed for both BS-to-UE direct channels and RIS-aided cascaded channels, covering a broad range of system configurations. It is hoped that this survey provides timely and valuable references to support and inspire future research and development in near-field channel estimation for 6G wireless~communications.

\subsection{Challenges and Open Issues}

Although a variety of preliminary advances have been reported in the literature, existing works on near-field channel estimation still face several critical challenges and open issues when viewed against the practical requirements and large-scale deployment of future 6G networks. The main ones can be summarized as follows:

    \textbullet\ \textit{Accurate modeling and robust estimation of non-stationary and non-uniform near-field channels}: In ELAA architectures based on XL-MIMO and XL-RIS, near-field channels often exhibit pronounced spatial non-stationarity and statistical non-uniformity. Even within the same aperture, different subarrays may observe different sets of effective scattering clusters, and their corresponding power-delay profiles as well as angle-distance distributions can vary along the array. As a result, the conventional assumptions of spatial stationarity and statistical homogeneity underlying classical channel models are no longer valid. Consequently, employing a single global statistical model or a unified angle-distance dictionary over the entire aperture often leads to model mismatch and performance degradation. A key open issue in near-field channel estimation is thus the development of near-field channel models that explicitly capture segment-wise variations along the array, together with estimation algorithms that can adapt to the distinct channel statistics across subarrays while remaining robust under inaccurate or partially unknown priors.


    \textbullet\ \textit{Severe pilot overhead in high-frequency near-field scenarios}: Pilot overhead is a primary bottleneck for the high-frequency near-field channel estimation, especially under wideband and multi-user operation.
    As the operating frequency moves toward the sub-THz and THz bands, the effective beamwidth of the channel becomes significantly narrower, enabling the near-field channel energy to be precisely focused on a specific spatial location of the UE rather than being characterized solely by the DoA, as in far-field channels. Under such conditions, higher estimation accuracy of channel parameters is required, and the estimation error typically needs to be controlled within a sub-wavelength scale.
    At the same time, severe path loss and pronounced hardware impairments (e.g., phase noise) in the high-frequency communications further strengthen the reliance on highly accurate beam focusing,
    often leading to more frequent updates of high-dimensional CSI. Under these conditions, conventional pilot design and estimation procedures can incur prohibitively high training overhead. Wideband OFDM transmission and multi-user operation further aggravate this issue. On the one hand, as the operating frequency increases, the coupling between the spatial and frequency domains becomes stronger.
    As a result, if channel estimation is performed independently on each subcarrier, the required pilot overhead tends to scale roughly linearly with the number of active subcarriers. On the other hand, in multi-user scenarios, if a conservative orthogonal pilot allocation strategy is adopted to strictly avoid inter-user pilot interference, the resulting pilot overhead grows rapidly with the number of users. Despite the practical importance of addressing these issues, the existing literature still lacks systematic, in-depth analyses of the interplay among channel-parameter dimensionality, wideband beam-split effects, and pilot overhead in high-frequency near-field scenarios, and of its implications for channel-estimation design and overhead-performance trade-offs.

    \textbullet\ \textit{Near-field channel estimation under hardware limitations, channel aging, and real-time constraints}: In near-field systems operating at typical 6G frequency bands,  hardware limitations become especially pronounced. Representative impairments include finite-resolution phase shifters and ADCs, phase noise, amplitude and phase imbalances, mutual coupling, and array calibration errors. Meanwhile, channel aging is more severe in high-mobility and densely deployed scenarios, so that the acquired CSI would quickly become outdated. In spite of this, most existing near-field channel estimation algorithms are still developed under idealized assumptions, such as perfect hardware and quasi-static channels, and are therefore not directly suited for practical implementations. Moreover, many high-accuracy approaches rely on computationally intensive iterative optimization, eigen-decomposition, or deep neural network inference, which poses significant real-time implementation challenges in ELAA-based systems with stringent latency and power constraints. Designing near-field channel estimation frameworks that are hardware-aware, robust to channel aging, computationally efficient under practical processing budgets, and ideally supported by recent experimental results or testbed validations, remains an important and largely open research issue.

    \textbullet\ \textit{A unified framework for channel parameter estimation and performance evaluation in multi-scenario near-field applications}: It should be emphasized that channel parameter estimation in the near-field is not exclusively a problem of ``near-field communications'', but rather a common core requirement across a wide range of near-field applications. For example, in near-field sensing, the same class of spatial parameters, such as the distance, angle, velocity, and scattering characteristics of targets or terminals, must be reconstructed with high accuracy from array observations. In near-field ISAC scenarios, parameter estimation is further required to simultaneously support reliable target localization and maintain high communication link quality. However, most existing works on near-field channel estimation remain primarily communication-centric: the adopted models, pilot and beam designs, and performance metrics are mainly tailored to CSI accuracy and throughput, while the requirements of near-field sensing or ISAC are rarely addressed within a unified framework. This mismatch highlights the need for a unified channel parameter estimation and performance evaluation framework built upon a common angle-distance-delay-Doppler near-field model. Such a framework should be able to characterize communication CSI and sensing parameters jointly, which is particularly important for pushing near-field ISAC technologies toward practical deployment.

    \textbullet\ \textit{Lack of measurement data, experimental validation, and prototype platforms for practical deployment}: Most existing studies on near-field channel estimation are still confined to theoretical analysis and simulation-based validation, typically relying on idealized or semi-empirical channel models with limited connection to real propagation environments and hardware constraints. Systematic measurements and open datasets explicitly targeting near-field scenarios are still scarce, and experimental platforms or prototype systems capable of supporting representative near-field architectures such as XL-MIMO and XL-RIS systems are also relatively limited. This gap makes it difficult to thoroughly assess the practical applicability of current near-field channel models and estimation algorithms, and it constrains the generalization capability and reliability of data-driven and deep learning-based approaches in realistic near-field environments. Therefore, designing dedicated measurement systems and establishing open data resources tailored to near-field applications, and using them to refine channel models and validate algorithms, remains an urgent open issue on the path toward practical deployment.

\subsection{Future Research Directions}

Building on the above challenges and open issues, near-field channel estimation should be further investigated along the following directions, in order to better support the practical needs of diverse near-field applications in 6G networks.

    \textbullet\ \textit{Segment-wise modeling and adaptive estimation for non-stationary and non-uniform near-field channels}: Due to the pronounced spatial non-stationarity and statistical non-uniformity of near-field channels in ELAA architectures, it is no longer adequate to rely on conventional models that assume a single set of shared statistics over the entire array. Future research should instead move toward more refined segment-wise or subarray-level descriptions. On the modeling side, simulation and measurement data can be jointly exploited to identify and group the scattering components observed by different subarrays into region-specific clusters, leading to segment-wise statistical models or multi-cluster hybrid models that explicitly capture how channel statistics vary along the array aperture. On the estimation side, there is a clear need for adaptive algorithms capable of automatically inferring the local statistical characteristics of each subarray and adjust the associated priors and dictionaries accordingly, thereby maintaining robust performance even when the underlying statistics are inaccurate or only partially known. Furthermore, hybrid model-driven/data-driven approaches, such as deep unfolding or meta-learning frameworks that learn which parts of the array should share common statistics and which should be modeled separately, represent particularly promising directions for future work.

    \textbullet\ \textit{Low-overhead pilot design and beam-split-aware estimation for high-frequency near-field systems}: To cope with the stringent estimation-accuracy requirements and the resulting pilot overhead in high-frequency near-field scenarios, especially in wideband and multi-user settings, future work should develop more accurate near-field channel models in the joint angle-distance-delay-frequency domain, explicitly accounting for the wideband beam split effect. On the modeling and beamforming side, promising directions include the construction of frequency-dependent polar-domain dictionaries, array-response models learned from measurement or simulation data, and beam control schemes incorporating TD structures, so as to more accurately characterize how high-frequency near-field focal points vary across subcarriers. On the training side, for multi-carrier and multi-user scenarios, it is desirable to design structured and reusable pilot or training sequences that fully exploit correlations across time, frequency and space, thereby enabling pilot sharing and hierarchical training across subcarriers and users and significantly reducing the overall training overhead. Furthermore, a deeper theoretical understanding of the fundamental coupling among channel parameter dimensionality, beam split effects, and pilot overhead, together with the derivation of pilot-overhead lower bounds, Cram\'er-Rao bounds, and complexity-performance tradeoffs for high-frequency wideband near-field systems, represents another important direction for future research.

    \textbullet\ \textit{Hardware-aware, time-variation-robust, and real-time-implementable near-field channel estimation}: In near-field systems operating under pronounced hardware impairments, rapidly time-varying channels, and stringent processing constraints, it is crucial to develop channel estimation frameworks that are hardware-aware, robust to temporal variations, and suitable for real-time implementation. On the modeling side, a natural direction is to explicitly incorporate practical imperfections, such as finite-resolution phase shifters and ADCs, phase noise, amplitude and phase imbalances, mutual coupling, and array calibration errors, into tractable equivalent channel and noise models that remain amenable to analysis and estimation. On the algorithmic side, in high-mobility and densely deployed scenarios, there is a clear demand for low-complexity tracking and prediction schemes, for example near-field CSI tracking based on state-space models and Kalman filtering, or channel aging prediction that leverages historical CSI and underlying geometric relationships. From an implementation perspective, particular attention should be paid to computational complexity and latency in ELAA-based systems, motivating the design of lightweight solutions co-optimized with hardware architectures, such as deep unfolding networks with controllable number of iterations and low-rank update algorithms. These solutions should ultimately be validated through end-to-end real-time experiments on prototype platforms and testbeds representative of future 6G deployments.

    \textbullet\ \textit{A unified parameter estimation and performance evaluation framework for multi-scenario near-field applications}: Given that near-field channel parameter estimation plays a central role in a wide range of near-field applications, such as communications, sensing, ISAC, and WPT, future research should aim to establish a unified framework for parameter estimation and performance evaluation across these scenarios. On the one hand, it is desirable to build upon a common near-field angle-distance-delay-Doppler model to systematically characterize the intrinsic relationships between communication CSI and sensing-related parameters. Within such a unified model, the parameter estimation requirements of near-field sensing and ISAC can be explicitly incorporated into the channel estimation problem. On the other hand, it is important to investigate joint pilot and beamforming design within this unified framework, so that the same set of array observations can simultaneously support reliable near-field communication and high-precision target localization and environmental sensing. Based on this unified view, a joint performance evaluation strategy should be developed that encompasses both communication metrics (e.g., data rate and reliability) and sensing metrics (e.g., localization accuracy and detection probability). Using this strategy to guide algorithm design and system optimization will be a key step toward evolving near-field channel estimation from a purely communication-centric perspective to a truly integrated multi-scenario solution.

    \textbullet\ \textit{Validation of near-field channel estimation via measurements, datasets, and testbeds}: To narrow the gap between theoretical research and practical deployment, future research should place greater emphasis on measurement campaigns, dataset construction, and prototype platform development specifically tailored to near-field applications. On the one hand, dedicated near-field measurement systems should be designed for representative XL-MIMO and XL-RIS in candidate 6G frequency bands, enabling systematic measurements across diverse scenarios and array configurations, thereby facilitating the construction of open, reusable near-field channel datasets. On the other hand, these measurement data can be exploited to calibrate and refine existing near-field channel models, extract statistical characteristics and structural priors that more accurately represent real propagation conditions, and support the experimental validation of representative near-field channel estimation algorithms on hardware prototypes and 6G testbeds. Establishing such a closed loop linking modeling, data, and prototype-based validation will enable more objective assessments of algorithm performance and robustness in realistic near-field scenarios, provide quantitative evidence to support ongoing standardization and trials (e.g., within 3GPP, ITU, ETSI, and RIS-related alliances), and lay a solid foundation for generalizing and deploying data-driven and deep learning-based methods in near-field systems.

\section*{Acknowledgment}
The authors would like to thank the Editor and the anonymous Reviewers for their careful reading and valuable suggestions that helped to improve the quality of this manuscript.

\bibliographystyle{IEEEtran}
\bibliography{IEEEabrv,reference_b}

\begin{IEEEbiography}[{\includegraphics[width=1in,height=1.25in,clip,keepaspectratio]{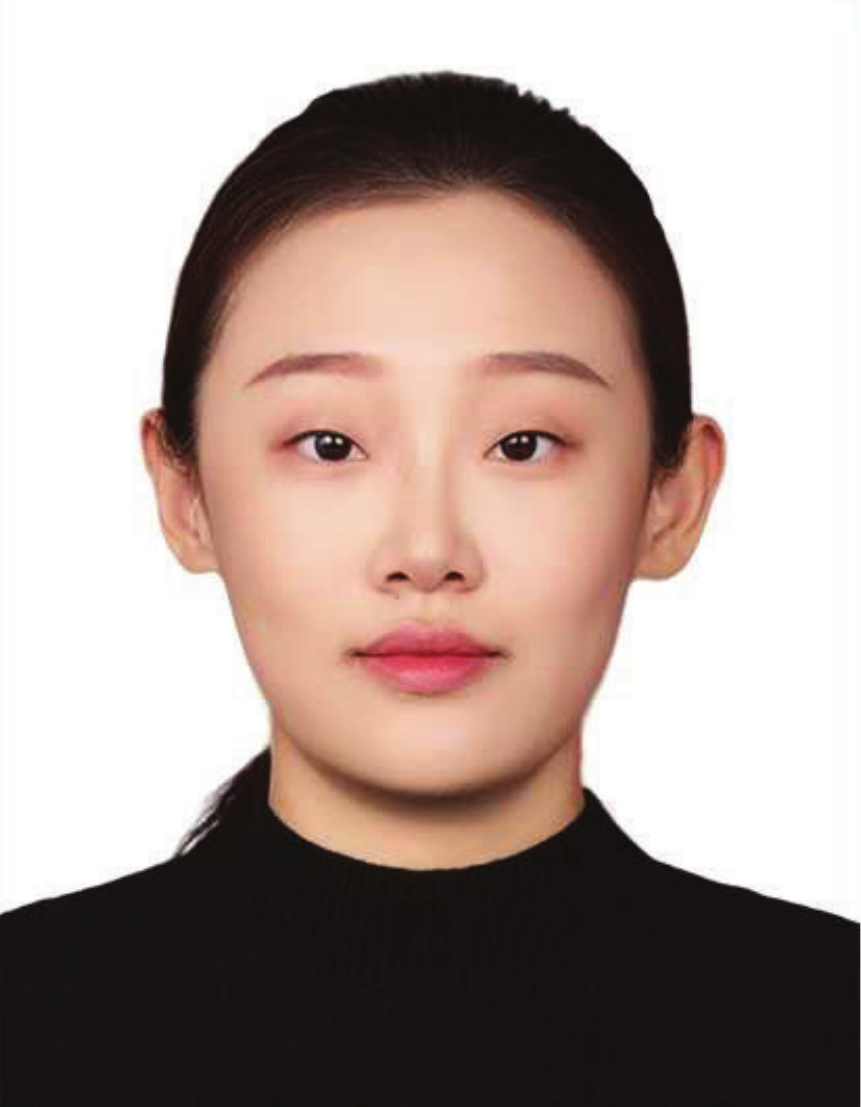}}]{Wen-Xuan Long}
(Member, IEEE) received her B.S. degree in Rail Transit Signal and Control from Dalian Jiaotong University, Dalian, China, in 2017. In 2023, she earned her Ph.D. degrees (cum laude) in Information and Communication Engineering from Xidian University, Xi'an, China, and in Information Engineering from the University of Pisa, Pisa, Italy. She is currently a postdoctoral research fellow in the Department of Information Engineering at the University of Pisa, Pisa, Italy. Her expertise encompasses wireless communications and signal processing, as well as estimation and detection theory. Her current research interests focus on channel estimation methods for 5G/B5G wireless communications, including RIS-aided cascaded channel estimation and near-field channel estimation suitable for THz communications. She has published more than 30 papers in academic journals and international conferences and co-authored a textbook: \textit{Circular Array-Based Radio Frequency OAM Communications} (2023). From 2020 to 2023, Dr. Wen-Xuan Long served as a member of the IMT-2030 (6G) promotion group. She is currently a member of IEEE VTS Propagation Committee.
\end{IEEEbiography}

\begin{IEEEbiography}[{\includegraphics[width=1in,height=1.25in,clip,keepaspectratio]{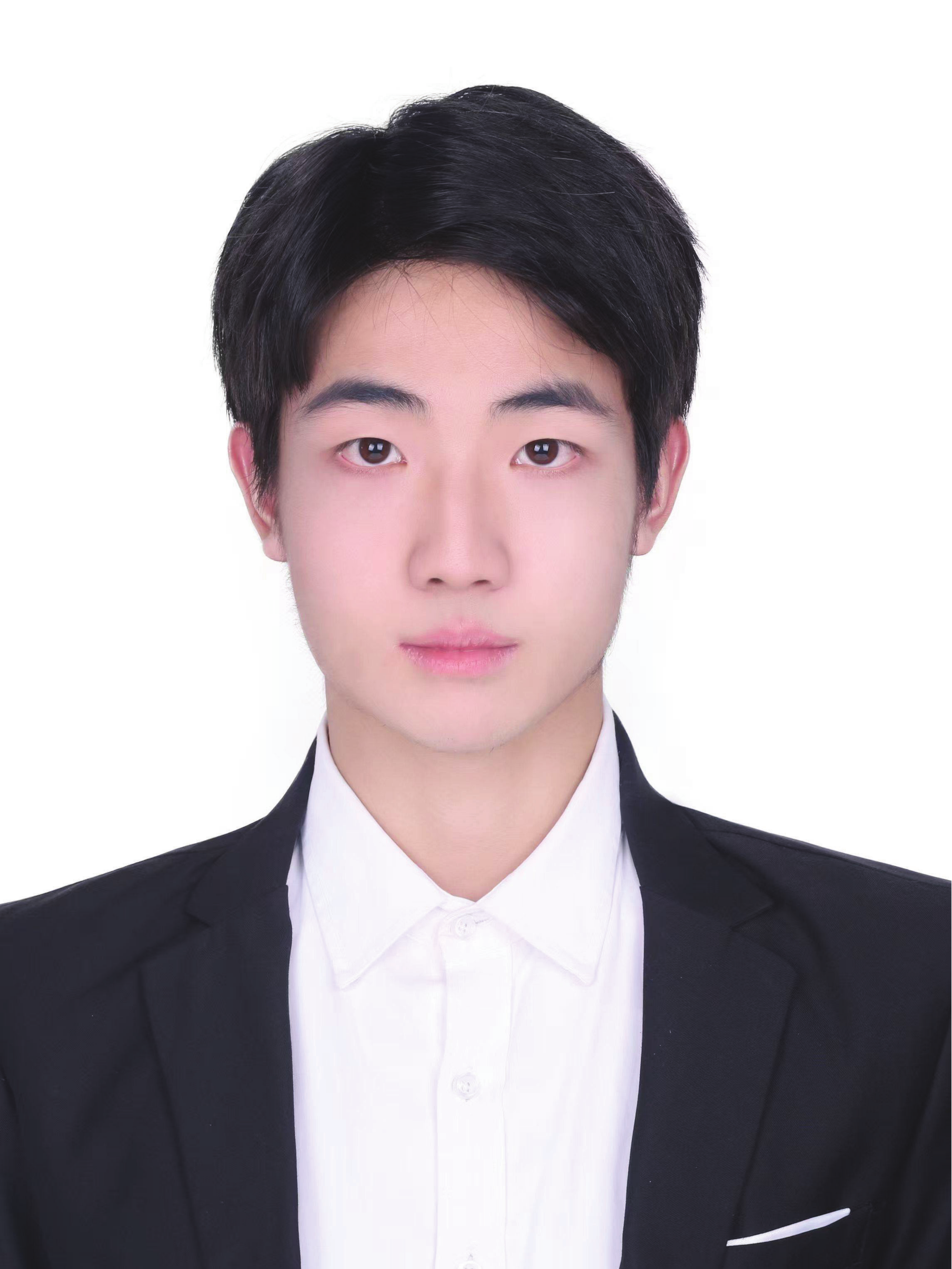}}]{Shengyu Ye}
(Graduate Student Member, IEEE) received the B.S. degree in communications engineering from Anhui University, Hefei, China, in 2024. He is currently pursuing the M.S. degree in information and communication engineering at Xidian University, Xi’an, China. His research interests include near-field communications and channel estimation.
\end{IEEEbiography}

\newpage

\begin{IEEEbiography}[{\includegraphics[width=1in,height=1.25in,clip,keepaspectratio]{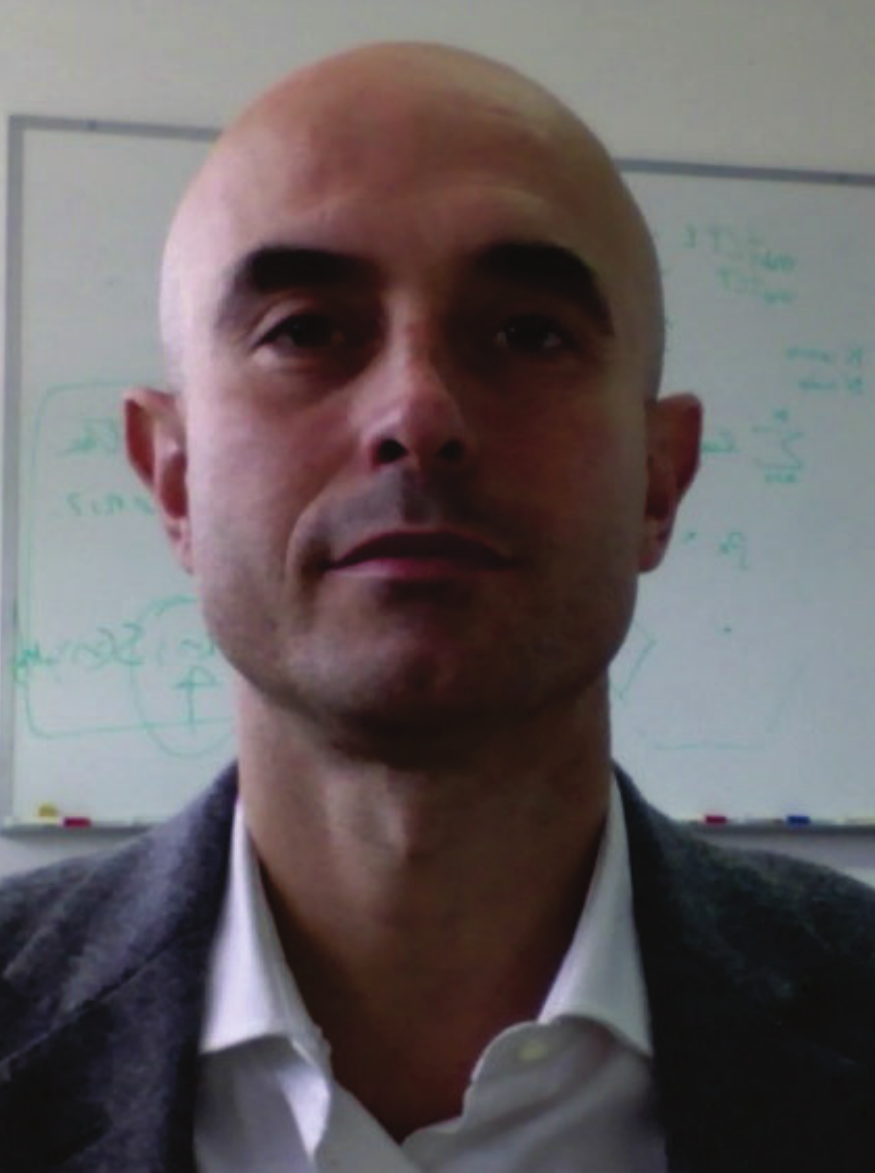}}]{Marco Moretti}
(Member, IEEE) graduated in Electronic Engineering at the University of Florence, Florence, Italy, in 1995, and received the Ph.D. degree from Delft University of Technology, Delft, Netherlands, in 2000. From 2000 to 2003, he was a Senior Researcher at Marconi Mobile. He is currently an Associate Professor at the University of Pisa, Pisa, Italy. His research interests include optimization algorithms and artificial intelligence for wireless communications, synchronization, channel estimation, and satellite communications. He serves as an Area Editor for IEEE TRANSACTIONS ON SIGNAL PROCESSING.
\end{IEEEbiography}

\begin{IEEEbiography}[{\includegraphics[width=1in,height=1.25in,clip,keepaspectratio]{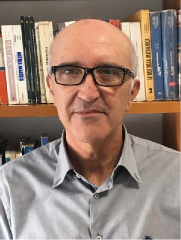}}]{Michele Morelli}
(Senior Member, IEEE) received the Laurea degree (cum laude) in electrical engineering
from the University of Pisa, Italy, in 1991, and the Ph.D. degree in telecommunications engineering from the Department of Information Engineering, University of Pisa. From 1992 to 1995, he was with the Department of Information Engineering, University of Pisa. In 1996, he joined Italian National Research Council (CNR), where he held the position of a Research Fellow for five years. Since 2002, he has been with the Department of Information Engineering, University of Pisa, where he is currently a Professor of digital transmissions and telecommunications. His research interests include digital communications, with emphasis on synchronization methods, equalization schemes, and precoding techniques. He is a member of the Communication Theory Committee. He was a co-recipient of the Best Student Paper Award at the IEEE Vehicular Technology Conference VTC ’06, Fall. He served
as an Editor for IEEE TRANSACTIONS ON WIRELESS COMMUNICATIONS
from 2007 to 2011, IEEE WIRELESS COMMUNICATIONS LETTERS
from 2011 to 2016, and IEEE TRANSACTIONS ON COMMUNICATIONS
from 2016 to 2018.
\end{IEEEbiography}

\begin{IEEEbiography}[{\includegraphics[width=1in,height=1.25in,clip,keepaspectratio]{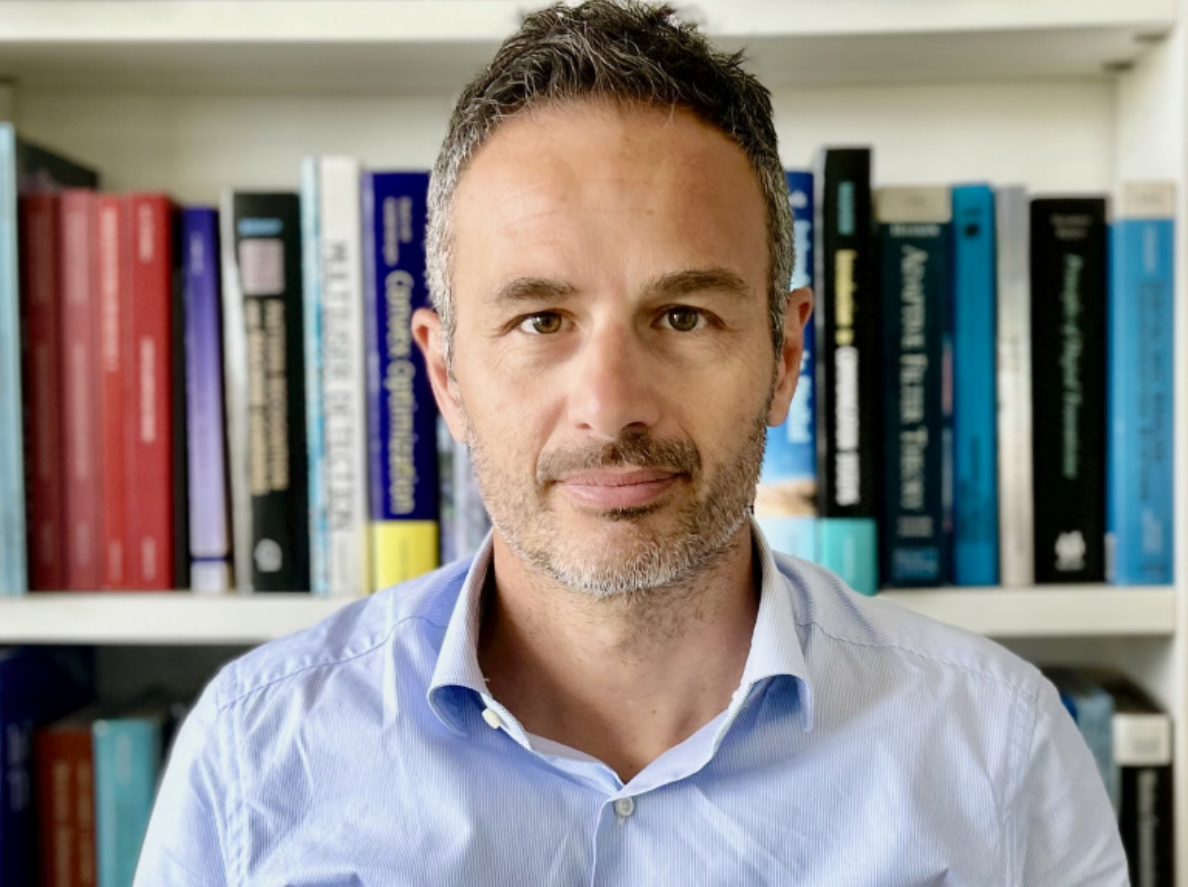}}]{Luca Sanguinetti} (Fellow, IEEE) received the Laurea Telecommunications Engineering degree (cum laude) and the Ph.D. degree in information engineering from the University of Pisa, Pisa, Italy, in 2002 and 2005, respectively. From June 2007 to June 2008, he was a Postdoctoral Associate with the Department of Electrical Engineering, Princeton University, Princeton, NJ, USA. From July 2013 to October 2017, he was with the Large Systems and Networks Group, Centrale Supelec, France. He is currently a Full Professor with the Department of Information Engineering, University of Pisa. He has co-authored two textbooks \textit{Massive MIMO Networks}: \textit{Spectral, Energy, and Hardware Efficiency} (2017) and \textit{Foundations of User-Centric Cell-Free Massive MIMO} (2021). His expertise and general interests include wireless communications and signal processing for communications.

Prof. Sanguinetti was the recipient of the 2018 and 2022 Marconi Prize Paper Awards in Wireless Communications and the 2023 Outstanding Paper Award of the IEEE Communications Society, and co-authored a paper that received the Young Best Paper Award from the ComSoc/VTS Italy Section. He was also the recipient of the FP7 Marie Curie IEF 2013, ``Dense Deployments for Green Cellular Networks." He was an Associate Editor of the IEEE TRANSACTIONS ON WIRELESS COMMUNICATIONS, IEEE TRANSACTIONS ON COMMUNICATIONS and IEEE SIGNAL PROCESSING LETTERS, the Lead Guest Editor of the IEEE JOURNAL ON SELECTED AREAS OF COMMUNICATIONS Special Issue on Game Theory for Networks, and an Associate Editor of the IEEE JOURNAL ON SELECTED AREAS OF COMMUNICATIONS, series on Green Communications and Networking. He is currently a member of the Executive Editorial Committee of the IEEE TRANSACTIONS ON WIRELESS COMMUNICATIONS.
\end{IEEEbiography}

\begin{IEEEbiography}[{\includegraphics[width=1in,height=1.25in,clip,keepaspectratio]{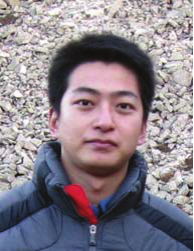}}]{Rui Chen}
(Member, IEEE) received the B.S., M.S., and Ph.D. degrees in communications and information systems from Xidian University, Xi’an, China, in 2005, 2007, and 2011, respectively. From 2014 to 2015, he was a Visiting Scholar with Columbia University, New York, NY, USA. He is currently a Full Professor and a Ph.D. Supervisor with the School of Telecommunications Engineering, Xidian University. He is also the Director of the Guangzhou Key Laboratory of Multimodal Traffic Information Perception Processing Technology and Intelligent Equipments. He has published one book and over 100 papers in international journals and conferences and held 50 patents. His research interests include broadband wireless communication systems, array signal processing, and intelligent transportation systems. He serves as an Associate Editor for International Journal of Electronics, Communications, and Measurement Engineering (IGI Global), an Editor for Information Countermeasure Technology and Electronics Optics and Control, and a committee member for several conferences. He is a member of the IMT-2030 Promotion Group.
\end{IEEEbiography}

\begin{IEEEbiography}[{\includegraphics[width=1in,height=1.25in,clip,keepaspectratio]{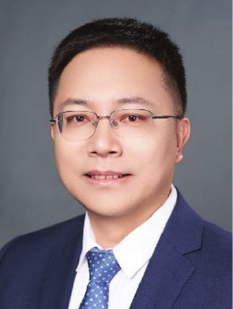}}]{Cheng-Xiang Wang}
(Fellow, IEEE) received the B.Sc. and M.Eng. degrees in communication and information systems from Shandong University, China, in 1997 and 2000, respectively, and the Ph.D. degree in wireless communications from Aalborg University, Denmark, in 2004.

He was a Research Assistant with the Hamburg University of Technology, Hamburg, Germany, from 2000 to 2001, a Visiting Researcher with Siemens AG Mobile Phones, Munich, Germany, in 2004, and a Research Fellow with the University of Agder, Grimstad, Norway, from 2001 to 2005. He was with Heriot-Watt University, Edinburgh, U.K., from 2005 to 2018, where he was promoted to a professor in 2011. He has been with Southeast University, Nanjing, China, as a professor since 2018, and he is now the Dean of the School of Information Science and Engineering. He is also a professor with Purple Mountain Laboratories, Nanjing, China. He has authored 4 books, 3 book chapters, and over 690 papers in refereed journals and conference proceedings, including over 270 IEEE journal/magazine papers and 32 highly cited papers. He has also delivered 39 invited keynote speeches and 24 tutorials in international conferences. His current research interests include wireless channel measurements and modeling, 6G/B6G ubiquitous intelligent connectivity networks, and electromagnetic information theory.

Dr. Wang is a Member of the Academia Europaea (The Academy of Europe) and European Academy of Sciences and Arts (EASA), a Fellow of the Royal Society of Edinburgh (FRSE), IEEE, IET and China Institute of Communications (CIC), an IEEE Communications Society Distinguished Lecturer in 2019 and 2020, and a Highly-Cited Researcher recognized by Clarivate Analytics in 2017-2020 and 2025. He was an Executive Editorial Committee Member of the IEEE TRANSACTIONS ON WIRELESS COMMUNICATIONS. He has served as an Editor for over ten international journals, including the IEEE TRANSACTIONS ON WIRELESS COMMUNICATIONS, from 2007 to 2009, the IEEE TRANSACTIONS ON VEHICULAR TECHNOLOGY, from 2011 to 2017, and the IEEE TRANSACTIONS ON COMMUNICATIONS, from 2015 to 2017. He was a Guest Editor of the IEEE JOURNAL ON SELECTED AREAS IN COMMUNICATIONS, the IEEE TRANSACTIONS ON BIG DATA, and the IEEE TRANSACTIONS ON COGNITIVE COMMUNICATIONS AND NETWORKING. He has served as a TPC Chair and General Chair for more than 30 international conferences. He received IEEE Neal Shepherd Memorial Best Propagation Paper Award in 2024. He also received 20 Best Paper Awards from international conferences.
\end{IEEEbiography}

\end{document}